\newif\ifCONDITION
\CONDITIONtrue
\documentclass[aps,prl,preprint,groupedaddress,longbibliography]{revtex4-2}

\usepackage{amsmath,amssymb}
\usepackage{color}
\usepackage{graphicx}
\usepackage{bm}
\usepackage{subfig}
\newcommand{\hglght}{black}

\newcommand{\ang}[1]{\langle{#1}\rangle}
\newcommand{\dd}[1]{\times10^{#1}}

\newcommand{\Cm}{C_{\mu}}

\newcommand{\St}{St}
\newcommand{\Fp}{F^{+}}

\newcommand{\figs}{pdf}

\newcommand{\size}{0.2}
\newcommand{\sizepng}{0.475}

\newcommand{\func}{u}
\newcommand{\casesjaa}{cmu2d-3-F01}
\newcommand{\casesjab}{cmu2d-3-F06}
\newcommand{\casesjac}{cmu2d-3-F10}

\newcommand{\casesjba}{cmu2d-5-F01}
\newcommand{\casesjbb}{cmu2d-5-F06}
\newcommand{\casesjbc}{cmu2d-5-F10}

\newcommand{\hcasesjaa}{$C_\mu=2.00\times 10^{-3},\Fp=1.0$ }
\newcommand{\hcasesjab}{$C_\mu=2.00\times 10^{-3},\Fp=6.0$ }
\newcommand{\hcasesjac}{$C_\mu=2.00\times 10^{-3},\Fp=10$  }

\newcommand{\hcasesjba}{$C_\mu=2.00\times 10^{-5},\Fp=1.0$ }
\newcommand{\hcasesjbb}{$C_\mu=2.00\times 10^{-5},\Fp=6.0$ }
\newcommand{\hcasesjbc}{$C_\mu=2.00\times 10^{-5},\Fp=10$  }

\newcommand{\del}{\partial}
\bmdefine{\br}{r}

\begin{document}


\title{Flow instability and momentum exchange in separation control by a synthetic jet}


\author{Yoshiaki Abe}
\email{yoshiaki.abe@tohoku.ac.jp.}
\altaffiliation[Present affiliation: ]{Insitute of Fluid Science, Tohoku University.}
\affiliation{%
Department of Aeronautics and Astronautics, University of Tokyo.
}%
\author{Taku Nonomura}
 \altaffiliation[Present affiliation: ]{Department of Aerospace Engineering, Tohoku University.}
\author{Kozo Fujii}%
 \altaffiliation[Present affiliation: ]{Department of Information and Computer Technology, Tokyo University of Science.}
\affiliation{ 
Institute of Space and Astronautical Science, Japan Aerospace Exploration Agency.
}%


\date{\today}

\begin{abstract}
This study investigates a mechanism of controlling separated flows around an airfoil using a synthetic jet (SJ). 
A large-eddy simulation (LES) was performed for a leading-edge separation flow around a NACA0015 airfoil at the chord Reynolds number of $63,000$ and the angle of attack of $12^\circ$. The present LES resolves a turbulent structure inside a deforming SJ cavity by a sixth-order compact difference scheme with a deforming grid.
An optimal actuation-frequency band is identified between $F^+=6.0$ and $20$ (normalised by the chord length and the freestream velocity), which suppresses the separation and drastically improves the lift-to-drag ratio.
It was found that in the controlled flows, the laminar separation bubble near the leading edge periodically releases multiple spanwise-uniform vortex structures, which diffuse and merge to generate a single coherent vortex in the period of $F^+$.
Such a coherent vortex plays a significant role in exchanging a chordwise momentum between a near-wall surface and the freestream away from the wall. It also entrains smaller turbulent vortices and eventually enhances the turbulent component of the Reynolds stress throughout the suction surface.
\textcolor{\hglght}{Linear stability theory (LST) was subsequently compared with the LES result, which clarifies the limitations and applicability of the LST to controlled flows with the present SJ condition.}
It is also revealed that in the optimal $F^+$ regime, both linear and nonlinear modes are excited in a well-balanced manner, where the first mode \textcolor{\hglght}{is associated with the Kelvin-Helmholtz instability and} contributes to a quick and smooth turbulent transition, while the second mode \textcolor{\hglght}{shows a frequency lower than that of the linear mode and} encourages a formation of the coherent vortex structure that eventually entrains smaller turbulent vortices.
\textcolor{\hglght}{The foregoing mechanism which relates the optimal $F^+$ to an efficient momentum exchange helps us construct a strategy to identify an optimal $F^+$ for separation control around an airfoil at a low angle of attack and a relatively low Reynolds number.}
\end{abstract}


\maketitle


\section{Introduction}\label{sec:intro}
An improvement of the ability to control fluid flows brings a great benefit to systems that have fluid flow in or around them.
The target of flow control widely spreads in vehicles, turbomachinery, chemical, and biomedical technologies, by means of a jet vectoring~\cite{Pack2001,Smith2002}, control of external flows~\cite{Chen2000,Mittal2005,Amitay2002,Tensi2002}, heat transfer~\cite{Kercher2003,Mahalingam2010,Chaudhari2010}, and mixing~\cite{Wang2001}. 
Among these targets, separation control often drastically improves their performance, e.g., avoiding a ``stall'' condition around an airfoil that may lead to a fatal accident of aircraft, while it is so far challenging mainly due to the nature of nonlinear dynamics that is typically involved in turbulent flows.
To date, various methodologies have been proposed for separation control.
In this study, we focus on so-called ``active control devices'' that have an ability to change the operation according to the surrounding unsteady flows:
for example, acoustic excitation \cite{Nishioka1989}, sweeping jet \cite{Vatsa2012, Graff2013}, plasma actuator (PA in Ref.~\onlinecite{Corke2010,Fujii2014}), and synthetic jet (SJ in Refs.~\onlinecite{Glezer2002,Greenblatt2008}).
Among these devices, the SJ and PA are normally categorized into ``microdevices,'' which are of microscale ($\mathcal{O}(10)$[mm]), lightweight, more simple structure, and lower energy consumption, and thus these are considered to be one of the most promising and realizable devices so far.
Indeed, several groups have already successfully applied the SJ (which is also denoted as ``periodic active flow control'' by Ref.~\onlinecite{Nagib2004}) to real-scale testings \cite{Nagib2004, Martin2014}.
The present study adopts the SJ to control a separated flow around an airfoil, while our broad interests are primarily in a separation control mechanism applicable to general microdevices including the PA~\textcolor{\hglght}{\cite{Enloe2004,Visbal2006,Patel2007a,Sidorenko2007,Corke2010,WangChoi2013,Sato2015PoF,Yakeno2015,Yakeno2017,Yarusevych2017,Ziade2018,Benton2019,Sato2020}}.
\subsection{Basic characteristics of induced flows from the SJ}
The SJ consists of a cavity and an orifice connected to the cavity whose bottom oscillates with a small amplitude, which produces weak and periodic flow from the orifice exit with zero mean.
The bottom of the cavity is typically a piezoelectric material bonded to membranes driven by an electric alternative current (AC) signal.
The width of the orifice exit (or slot) is generally $\mathcal{O}(10)$[mm] \cite{Rizzetta1999a,You2008}, and the velocity of induced flows can be widely changed from $\mathcal{O}(1)$ to $\mathcal{O}(10)$[m/s] \cite{Greenblatt2008} (the induced flow over $100$[m/s] has been also reported by Ref.~\onlinecite{Crowther2008}.)
The first attempt to use an SJ-like actuator has been made by Ref.~\onlinecite{Ingard1950}, where acoustic waves were used to generate sinusoidal flow motion.
The concept of today's SJ has been initiated by Ref.~\onlinecite{Dauphinee1957} as an air-jet generator, and subsequent works by Refs.~\onlinecite{Smith1997,Smith1998,Smith2002} has followed.
Smith and Glezer have investigated basic characteristics of induced flows by the SJ through an experimental study with a velocity measurement, spectrum analysis, and visualization of induced vortices.
Ref.~\onlinecite{Chen2000} has later experimentally revealed that the induced flow is initially a pair of vortices that is transient to a turbulent jet as evolving in the vertical direction.
As such, experimental studies have reported that the induced flow becomes turbulent away from the orifice exit depending on the flow conditions, although the time- and spanwise-averaged velocity field still shows a two-dimensional vertical flow.
On the other hand, Ref.~\onlinecite{Kral1997} has performed a Reynolds-averaged Navier-Stokes simulation (RANS), which agrees well with the experimental results in the time-averaged vertical velocity;
however, the other turbulent statistics such as velocity fluctuations were not in good agreement.
This is probably due to the use of
(1) the RANS that often fails to capture unsteady turbulent statistics;
(2) and the computational model of the SJ that adopts a boundary condition of a top-hat velocity distribution on the airfoil surface, where the flow inside the cavity was not resolved.
Accordingly, Ref.~\onlinecite{Rizzetta1999a} has carried out a direct numerical simulation (DNS) of the SJ, where the SJ is still modeled by the two-dimensional velocity boundary condition, thereby resulting in disagreement with the experimental plot, especially in its turbulent statistics.
Ref.~\onlinecite{Okada2012} has conducted a three-dimensional large-eddy simulation (LES) resolving the flows inside the orifice and cavity, where the velocity fluctuation qualitatively agrees well with the experimental results.
Therefore, generally, both a high-fidelity model of the SJ and the unsteady flow simulation, e.g., LES, are necessary for an accurate prediction of the induced flows by the SJ.
The present study follows Ref.~\onlinecite{Okada2012a} for the modeling of the SJ using the orifice and deforming cavity with a uniform shape in the spanwise direction.
\textcolor{\hglght}{Although the present SJ shape is not easy to realize in experiments due to geometrical uniformity in the spanwise direction, such a simple geometry can effectively remove the shape effect on resultant flow structures inside the SJ; furthermore, the jet properties of the present shape have been well investigated by former studies Refs.~\onlinecite{Rizzetta1999a,Okada2012,Okada2012a}. 
More recently, Ref.~\onlinecite{Ziade2018} has also performed numerical and experimental analysis of the cavity-shape influence on flow properties injected from the SJ. They have reported a cavity shape that introduces the strongest momentum at the exit as well as an importance of the high-fidelity flow simulation inside the cavity.}
\subsection{Optimal actuation frequency for separation control}
Many studies have been devoted to identifying an optimal actuation frequency for separation control, where separated flows over an airfoil and a backward-facing step have received considerable attention due to their simple geometries and well-known characteristics of the base flow.
In many cases, an actuation frequency of the SJ ($f^*$) is normalized by the reference length ($L^*$) and freestream velocity ($U^*$) as $\Fp=f^* L^*/U^*$.
An extensive review of experimental study has been reported in Ref.~\onlinecite{Greenblatt2008}, where the optimal frequency mostly exists within the range of $0.3<\Fp_{e}<4.0$ ($\Fp_{e}$ is based on a separation length), which generally corresponds to $\Fp=\mathcal{O}(1)$.
Indeed, Ref.~\onlinecite{Seifert1996} has shown that the actuation frequency of $\Fp=\mathcal{O}(1)$ is optimal for improving the lift in high-Reynolds-number ($Re=\mathcal{O}(10^5)$ to $\mathcal{O}(10^7)$) flows around a NACA0015 airfoil.
Note that they have also demonstrated that the input momentum by the SJ can be one hundred times smaller than that of the steady jet that is one of the most conventional devices for separation control so far.
The computational study of the same condition has been carried out by Ref.~\onlinecite{Donovan1998} using a two-dimensional RANS at $Re=\mathcal{O}(10^6)$, which shows that $\Fp=1$ performs better than $\Fp=10$.
More recently, in the separation control around a NACA0025 airfoil at a relatively low-Reynolds-number regime ($Re=\mathcal{O}(10^4)$), Ref.~\onlinecite{Feero2015} has experimentally shown that the better suppression can be achieved by the low-frequency control ($\Fp=\mathcal{O}(1)$), which corresponds to the wake instability.
Similarly, the separation control around a NACA0015 airfoil at $Re=\mathcal{O}(10^4)$ has been investigated by Ref.~\onlinecite{Zhang2015} using LES, where the low-frequency control performs better than the high-frequency control.
On the other hand, the advantage of the high-frequency control ($\Fp=\mathcal{O}(10)$) has also been reported by Ref.~\onlinecite{Glezer1999}, where the boundary-layer separation around a circular cylinder is controlled at a low-Reynolds-number regime ($Re=7.55\times 10^{4}$).
Accordingly, Refs.~\onlinecite{Amitay2002,Glezer2005} have reported that the high-frequency control is more effective in the separation control around a NACA0015 airfoil with its leading edge replaced by a cylinder at $Re=\mathcal{O}(10^5)$.
Ref.~\onlinecite{Okada2012a} has also conducted the LES of separation control around a backward-facing step flow, where the actuation frequency of the shear-layer instability provides a better control capability, which corresponds to the high-frequency control introduced above and the result is consistent with the preceding experimental work~\cite{Yoshioka2001}.
The effects of the high-frequency control ($\Fp=4$) has also been investigated by Ref.~\onlinecite{Dandois2007}, which identified an {\it acoustic-dominated mode} on the basis of a linear inviscid stability analysis, although the low-frequency actuation ($\Fp=0.5$) enables the better separation control via a {\it vorticity-dominated mode} under their conditions.
In this way, there are two different frequency regimes, $\Fp=\mathcal{O}(1)$ and $\mathcal{O}(10)$ (low- and high-frequency control), for achieving a better separation control in both of low- and high-Reynolds-number regimes.
Finally, although the focus of this paper is confined to the separation control by the SJ, it is worth noting that in the separation control using the PA, two distinct frequency regimes bands, i.e., $\Fp=\mathcal{O}(1)$ and $\Fp=\mathcal{O}(10)$, have been similarly reported as an optimal value \cite{Corke2004,Visbal2006,Patel2007a,Sidorenko2007,Sato2015PoF, Aono2017,Sekimoto2017}, which implies an analogy of the separation control mechanism by the SJ and PA, and possibly by the other microdevices.
Although several physical mechanisms have been proposed for each frequency regimes as will be described in the next subsection, the consistent explanation of optimal actuation frequency has not yet been adequately provided, which would be an obstacle to develop and construct universal design criteria for individual control parameters of the microdevices including the SJ.
\subsection{Mechanism of separation control}
Many of the literature have attempted to describe a mechanism of separation control, mainly from the viewpoint of a relationship between the optimal actuation frequency and flow instabilities.
There are typically two characteristic frequencies that are associated with the wake and shear-layer instabilities, respectively:
the first one corresponds to such a frequency that a large-scale vortex is shedding in the wake \cite{Wu1998};
the second one appears in the separated shear layer typically behind the separation point.
Most of the previous studies have reported that the low-frequency control utilizes the wake instability \cite{Smith1997,Amitay2002,Zhang2015}.
Meanwhile, the high-frequency control utilizes several flow phenomena including the shear-layer instability \cite{Yoshioka2001,Dandois2007,Okada2012a,Feero2015}.

\subsubsection{Shear-layer instability}
The advantage of utilizing the shear-layer instability has been investigated extensively for controlling the size of a laminar-separation-bubble (LSB) \cite{Gaster1966,Watmuff1999,Spalart2000,Rist2002,Rist2006,Marxen2011}. 
In many cases, the most unstable frequency is compared with that predicted by the linear stability theory (LST), which assumes the inviscid and parallel flow approximation, based on the Rayleigh equation, i.e., the convective Kelvin Helmholtz (KH) instability in the separated shear layer.
Ref.~\onlinecite{Nishioka1989} has investigated the receptivity and instability of the LSB in the separation control around an airfoil using an acoustic excitation.
They have performed the LST analysis and demonstrated that the most unstable frequency measured from the experiments agrees well with that predicted by the LST, which suggests that the disturbance in the LSB grows \textcolor{\hglght}{exponentially (linearly in the log-plot)} with distance downstream (linear growth regime).
This result is consistent with the previous studies of Refs.~\onlinecite{Ahuja1984,Zaman1987}, which suggested the existence of some appropriate frequencies in the input flow disturbance for minimizing the LSB size.
When the disturbance is sufficiently amplified via the linear instability, it grows more rapidly (nonlinear growth regime) afterwards, thereby resulting in the formation of coherent vortices that are periodically released from the LSB \cite{Marxen2004}.
The relationship between the coherent vortex emitted from the LSB and the large-scale vortex shedding from the trailing edge, i.e., the wake of an airfoil, has been investigated in Ref.~\onlinecite{Yarusevych2009} at a low-Reynolds-number flow, including the effect of a feedback loop in control parameters.
More precise discussion on the spatial growth of disturbance has been also carried out through experimental studies as well as numerical studies using both DNS and LES \textcolor{\hglght}{\cite{Rist2006,Dandois2007,Avdis2009,Boutilier2012b,Marxen2015,Yarusevych2017JFM,Yarusevych2017,Yarusevych2019}}.
Ref.~\onlinecite{Marxen2015} has conducted two-dimensional DNS around a flat-plate-like airfoil to control the size of the LSB, where a small disturbance is introduced by modelling an SJ-like suction and blowing device.
They have reported that the most unstable frequency obtained from the LST, i.e., the inviscid linear-instability frequency, maximally suppresses the size of the LSB;
on the other hand, based on the Orr-Sommerfelt equations, the lower and higher frequencies can be found unstable due to the viscous effect and the convective instability, respectively, which are however not effective for controlling the LSB size.
\textcolor{\hglght}{
More recently, Ref.~\onlinecite{Yarusevych2017JFM} performed an experimental study for steady and transient response of a laminar separation bubble using a DBD plasma actuator. They employed the LST analysis to relate the shear layer stability and the bubble dynamics. In the LSB, the unstable disturbances in the separated shear layer is amplified to form shear layer roll-up vortices, which are locked onto a excitation frequency with strong coherence structure if the frequency is sufficiently close to the unstable frequency.
Furthermore, Ref.~\onlinecite{Yarusevych2017} represented an applicability of the LST analysis to controlled flows where the mean flow deformation by flow control is significant and also reported that the LST analysis shows a good agreement with experimental measurements even in the aft portion of the LSB where disturbances are significantly amplified. Note that the existence of shear layer instability and the effectiveness of the LST analysis in the controlled LSB has been represented also by Ref.~\onlinecite{Postl2011}.}
Finally, it should be emphasized that these studies have mostly focused on controlling the LSB size in the attached flows, which might not straightforwardly adapt to the discussion on the control of massively separated flows through the SJ.
Nevertheless, the receptivity and instability in the LSB are expected to be strictly relevant to the optimal actuation frequency~\cite{Dejoan2004,Dandois2007,Avdis2009} in conjunction with the mechanism of momentum exchange as is described in the next sub-subsection.

\subsubsection{Momentum injection near the airfoil surface}
Ref.~\onlinecite{Glezer2005} has experimentally shown that the high-frequency actuation promotes a turbulent mixing through the Reynolds stress, thereby achieving the effective separation control around an airfoil.
The similar argument that the turbulent mixing plays an essential role in the momentum exchange has been given in Ref.~\onlinecite{You2008} based on the LES of separation control around a NACA0015 airfoil.
Later on, Ref.~\onlinecite{Sato2015PoF} has conducted the LES around a NACA0015 airfoil although the PA is adopted for the separation control.
They have more precisely reevaluated the momentum injection near the airfoil surface in the separation-controlled flows by means of a phase decomposition of the Reynolds stress, whose mechanism has been classified into several categories.
One of those categories is the ``freestream momentum entrainment,'' which indicates that the momentum in the freestream direction is transferred from an external flow (freestream) away from the airfoil surface to the bottom of the boundary layer.
The similar mechanism has been partly referred in Ref.~\onlinecite{You2008} as well.
The freestream momentum entrainment is considered significant given that the momentum directly introduced from the device is not strong and not comparable to that of the freestream, because the momentum in the freestream direction needs to be constantly injected to maintain the attached flow by balancing with an adverse pressure gradient.
Such a momentum transfer mechanism has been further categorized into the mechanisms by a ``coherent (large-scale) vortex'' and ``turbulent vortex.''
The existence of the coherent vortex and its formation by a vortex merging process have been similarly reported in \textcolor{\hglght}{Refs.~\onlinecite{Smith1997,Glezer2005,Zhang2015,Yarusevych2017JFM,Yarusevych2017,Yarusevych2019,Lambert2019}}, and the consecutive large-scale vortex emerges in the period of $\Fp$ that is then periodically released from the trailing edge.
However, the relationship between the shear-layer instability and the promotion of momentum exchange has not been adequately discussed, and moreover, it is not yet clear how the linear instability frequency resultantly relates to the optimal actuation frequency for separation control.
It is worth noting that in Ref.~\onlinecite{Sato2015PoF}, the optimal actuation frequency is lower than the linear instability frequency based on the velocity profile in the LSB.
The similar discrepancy between the linear instability frequency and the optimal actuation frequency can be seen in Ref.~\onlinecite{Nishioka1989}, which implies that the optimal actuation frequency for the separation control cannot be straightforwardly determined from the LST, unlike controlling the size of an LSB.
\textcolor{\hglght}{
Ref.~\onlinecite{Yarusevych2017} also reported the similar mechanism that the actuation frequencies close to the linear instability frequency enhances a coherence of the shed vortices which are locked onto the actuation frequency. The formation of coherent vortices delays a breakdown into smaller turbulent vortices, and thus the LSB size is shortened.}

\subsection{Objectives and structure of the present paper}\label{sec:objectives}
To this end, it has not yet been adequately clarified 1) how the coherent vortex is generated as a consequence of a disturbance that the SJ introduces to an external flow, and 2) how the momentum injection is maintained near the airfoil surface, especially in terms of the relationship with the optimal actuation frequency for separation control.
\textcolor{\hglght}{
The present study primarily aims at identifying the optimal actuation frequencies of the SJ to suppress the separation and maintain the attached flow, based on simulations with a three-dimensinal flow disturbance from the SJ.
Then, we focus on clarifying the mechanism how the SJ effectively maintains attached flows with those optimal actuation frequencies using the time- and phase-averaged analysis with the aid of LST.
In specific, this study performs a phase decomposition of the Reynolds stress that characterizes the momentum exchange by the coherent vortex and turbulent vortex structures as well as comparing a spatial growth rate of the SJ disturbance with the LST results on the separated shear layer.}
To avoid complex multifactorial effects, we confine our target to the separation control around an NACA0015 airfoil at the chord-Reynolds-number of $63,000$.
The SJ is modelled by a two-dimensional (spanwise-uniform) shape that is embedded in the airfoil surface and consists of a small orifice with a deforming cavity.
At this Reynolds number, the flow separates in the vicinity of the leading edge at an angle-of-attack (AoA) of 12$^\circ$, and thus the location of the SJ is fixed to the leading edge so that the spatial growth of the flow disturbance can be adequately observed in the separated shear layer.
Note that the location of the SJ (or PA) has been found significant for the improvement of a separation-control ability, which is however not investigated in this paper and left to the future study.
In this study, the LES is performed for various operating conditions of the SJ including an input momentum, $\Cm$, and actuation frequency, $\Fp$.
The present LES adopts a sixth-order compact scheme for spatial discretization, which resolves turbulent flows even inside the SJ cavity using deforming grids.
\textcolor{\hglght}{
Finally, it should be mentioned that the use of LST to clarify the {\it transient} mechanism from the separated flow to the attached flow with control is not appropriate due to a drastic change of the flow field and difficulty in a definition of mean flows. Therefore, in the context of the LST, this study only focuses on the mechanism of how the SJ effectively maintains attached flows with the optimal actuation frequencies after the transient period, which does not fully answer the question of how the identified optimal frequencies realize a drastic change from the separated flow to the attached flow. The application of the LST to the controlled flow with LSBs has been often reported because a quasi-steady flow field can be identified and regarded as a mean flow even with a control if the input disturbance is sufficiently small compared to the mean-flow velocity~\cite{Postl2011,Marxen2015,Yarusevych2017}. Further relevant studies using the LST for controlled flows can be also found in Refs.~\onlinecite{Lin1996,Sato2015PoF}. 
The mechanism of a transient process from the separated (noncontrolled) to attached (controlled) flow can be seen in the other studies~\cite{Amitay2002,Asada2015,Fukumoto2016}.
}

The rest of this paper is organized as follows.
Section~\ref{sec:formulation} describes a formulation of the problem including the SJ and the separated flow over an airfoil, then Sec.~\ref{sec:method} represents the computational cases, flow solver, and post-processing tools.
Section~\ref{sec:basicflow} discusses the fundamental characteristics of the separation controlled flow mainly from time-averaged flow fields, which include an ability of separation control.
Section~\ref{sec:momentum} evaluates the momentum injection in the chordwise direction that is exchanged between the near-wall surface and freestream, through a phase decomposition of the Reynolds stress.
Based on the classification above, the mechanism of the optimal $\Fp$ is investigated in Sec.~\ref{sec:spatialgrowth}, based on a spatial growth of the input disturbance from the SJ.
Finally, Sec.~\ref{sec:conclusion} concludes this paper.

\section{Formulation of the problem}\label{sec:formulation}
We describe conditions of the separated flow and the SJ in this section.
The airfoil is a spanwise-uniform NACA0015 airfoil, which is a simple symmetric airfoil and has frequently been used for fundamental studies through a lot of experimental and numerical approach.
We consider the flow at the chord Reynolds number of $63,000$.
At this Reynolds number, the flow remains attached up to $AoA=11^\circ$ with the LSB in the vicinity of the leading edge \cite{Asada2009}.
This study aims to control the separated flow at a post-stalled angle, i.e., $AoA=12^\circ$, where the flow separates from approximately $2.3\%$ of the chord location and becomes turbulent via the KH instability.
This suggests that during the separation control, even a small disturbance from the SJ could be amplified in the downstream direction through the turbulence transition process, which has been regarded as one of the crucial phenomena in the separation control at this Reynolds number.
Thus, the present conditions help us understand the mechanism of how such small disturbance from the SJ is amplified and eventually modifies the entire flow.
Finally, the fluid is assumed to be air, and the specific heat ratio and the Prandtl number are defined as $1.4$ and $0.72$, respectively.

\begin{figure}
\centering
\ifCONDITION
  \includegraphics[width=1.0\textwidth,clip]{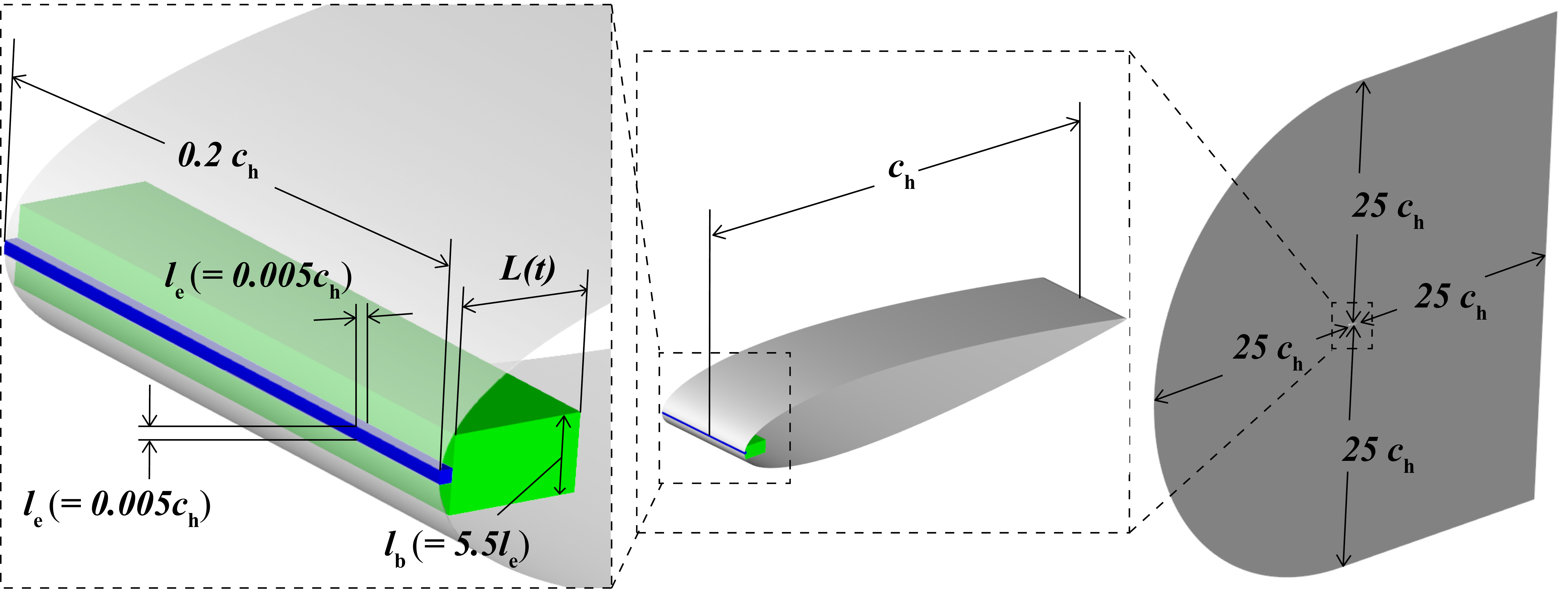}
\else
\fi
  \caption{Schematic illustration of the computational domain and the SJ installed to the leading edge of a NACA0015 airfoil.}
\label{fig:SJconfig}
\end{figure}
For simplicity, the SJ is modeled by a simple two-dimensional (spanwise-uniform) shape that is embedded in the airfoil surface, which consists of a small orifice and deforming cavity (the similar shape is adopted by Refs.~\onlinecite{Rizzetta1999a,Okada2012a}.)
The SJ is installed at the leading edge ($0\%$ of the chord length), where the direction of its orifice is normal to the airfoil surface to avoid direct injection of the chordwise momentum and more clearly characterize the turbulent mixing effects on separation control. 
Figure~\ref{fig:SJconfig} shows the computational domain and schematic illustration of the SJ, where green and blue colored regions indicate those of the airfoil surface and the SJ, respectively.
The width and height of the orifice are set to be $0.5\%$ of the chord length ($l_{\text{e}}=0.005c_{\text{h}}$, where $c_{\text{h}}$ indicates the chord length), which has often been used in the previous studies~\cite{You2008,Okada2012a}.
The depth of the cavity is expressed as $L(t)$ ($t$ denotes the nondimensional time), which is initially set as $L(0)=L_{0}=10l_\text{e}$;
the width of the cavity bottom is set to be $l_{\text{b}}=5.5l_\text{e}$.
The nondimensional actuation frequency is denoted as $F^+$ according to the previous studies~\cite{Greenblatt2000,Glezer2002}, which is defined as
\begin{align}
F^+\equiv \frac{ f^*c_h^*}{u^*_\infty},\label{Fpdef}
\end{align}
which is the same normalization as the Strouhal number based on the freestream velocity and chord length: $\St$.
The bottom of the cavity oscillates in a translational motion, where the amplitude of oscillation is denoted by $A$ (normalized by $c_\text{h}$), and the depth of the cavity, $L(t)$, is defined as follows:
\begin{align}
L(t)
=L_{0}+A\cos(2\pi F^+t).\label{F+}
\end{align}
The momentum coefficient $C_\mu$ is defined as
\begin{align}
C_{\mu}\equiv\frac{\rho u_{\text{max}}^2 l_\text{e}}{\rho u_{\infty}^2 c_{\text{h}}},\qquad u_{\text{max}} l_\text{e}\equiv l_{\text{b}}\text{max}\left(\frac{{\rm d} L(t)}{{\rm d}t}\right)=2\pi l_{\text{b}}AF^+,\label{uj}
\end{align}
which is the ratio of the momentum induced by the SJ and the freestream per unit time.
In this definition, the fluid in the cavity is assumed to be incompressible, and the maximum momentum ($\rho u_{\text{max}}$) induced by the SJ is adopted to define the momentum coefficient $C_{\mu}$.
Note that the amplitude $A$ is changed according to $F^+$ when $C_\mu$ is kept constant as $A=u_{\infty}\sqrt{c_{\text{h}} C_{\mu}l_\text{b}}/(2\pi F^+l_\text{e})$.

The computational cases are summarized in Table~\ref{chap1-cases}.
The actuation frequencies are set to be $\Fp=1.0,6.0,10,15,20,$ and $30$.
The momentum coefficient, $\Cm$, is similar to or smaller than that in the previous studies,
e.g., $\Cm=3.5\dd{-3}$ by Ref.~\onlinecite{Amitay2002},
$\Cm=1.23\dd{-3}$ by Ref.~\onlinecite{You2008},
$\Cm=2.00\dd{-3}$ by Ref.~\onlinecite{Okada2012a},
and $\Cm=2.13\dd{-4}$ by Ref.~\onlinecite{Zhang2015}.
The present study adopts a small $\Cm$ so that a contribution from the direct momentum addition is suppressed and the effects of freestream momentum entrainment can be clarified in the separation control mechanism.

\begin{table}
  \begin{center}
\def~{\hphantom{0}}
  \begin{tabular}{ccc}\hline\hline
case description & input momentum ($C_\mu$)  & $\Fp$  \\\hline
Noncontrolled & ---  & ---    \\
Controlled (strong input)&$2.0\times 10 ^{-3}$  & $1.0,\quad 6.0,\quad 10,\quad 15,\quad 20,\quad 30$  \\
Controlled (weak input)&$2.0\times 10 ^{-5}$  & $1.0,\quad 6.0,\quad 10,\quad 15,\quad 20,\quad 30$  \\\hline\hline
  \end{tabular}
\caption{Computational cases.}\label{chap1-cases} 
  \end{center}
\end{table}
%
%

\section{Methodology}\label{sec:method}
\subsection{Flow solver}\label{subsec:flowsolver}
In the present study, LANS3D~\cite{Fujii1990lans3d} is employed for the series of computations.
LANS3D is a high-order compressible flow solver for structured grids, which has been developed at the ISAS/JAXA and applied to a considerable number of engineering problems~\cite{Fujii2005,Okada2012,Okada2012a,Asada2015,Sato2015PoF,Fukumoto2016,Aono2017,Sato2020} as well as fundamental problems~\cite{Fujii2008,Yakeno2015,Yakeno2017}, and the capability of the code has been sufficiently verified through the literature above.
The followings are an overview of the present code employed for the LES of the separated flow control.
The governing equations are the three-dimensional compressible Navier-Stokes equations in body-fitted coordinates.
The spatial derivatives of the convective and viscous terms are evaluated by a sixth-order compact finite-difference scheme~\cite{Lele1992}.
The metrics and Jacobian for the coordinate transformation are evaluated by the symmetric conservative forms~\cite{Abe2013b} which can avoid a freestream preservation error on moving and deforming grids even with the high-order compact scheme.
To suppress a numerical oscillation, a tenth-order filtering~\cite{Gaitonde2000} is used with a filtering coefficient of $\alpha_f=0.495$.
For a time integration, a backward second-order difference formula converged by the five subiterations~\cite{Visbal2002} of the lower-upper symmetric alternating direction implicit and symmetric Gauss-Seidel (ADI-SGS)~\cite{Nishida2009} is employed, and the second-order of accuracy in time is ensured.

\subsection{Computational grids and boundary conditions}\label{subsec:grids}
\begin{figure}
\centering
\ifCONDITION
\includegraphics[width=1.0\textwidth,clip]{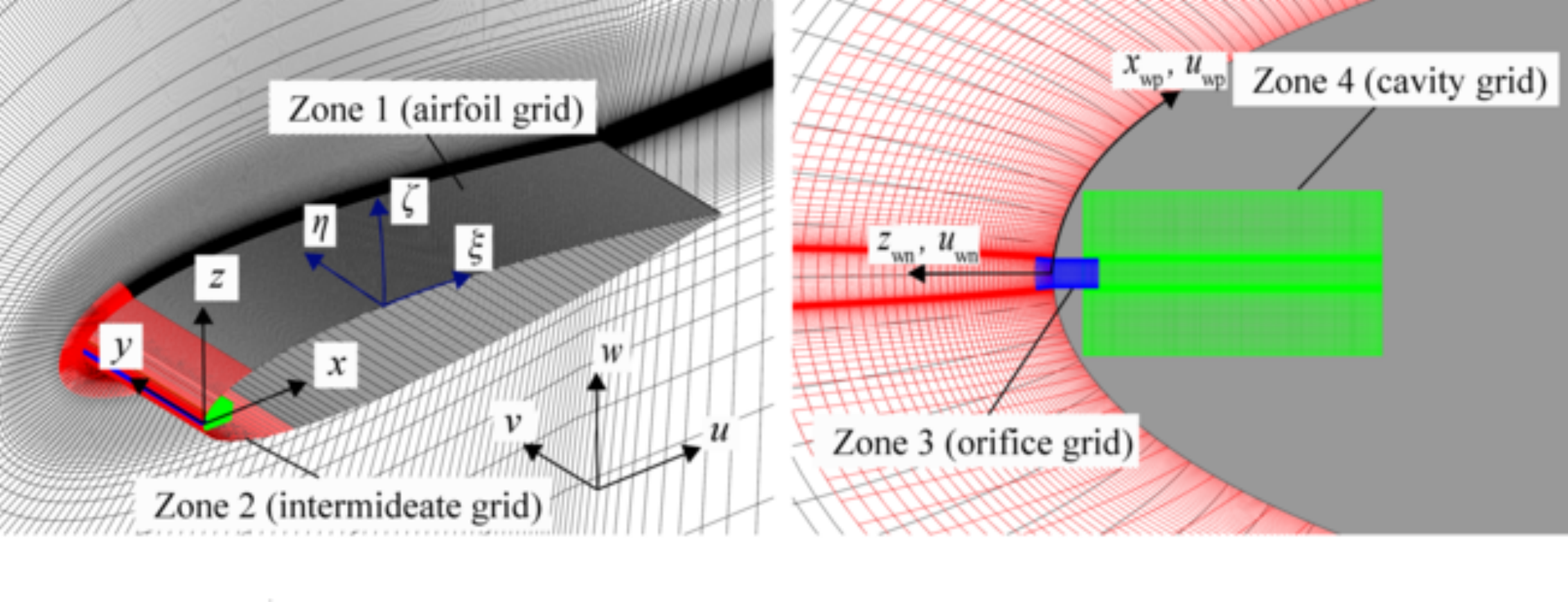}
\else
\fi     
  \caption{Computational grids (every 5 points are visualized) and coordinates}
  \label{fig:grid}
\end{figure}

The computational grid is constructed based on the zonal grid approach~\cite{Fujii1995a}:
background grid around an airfoil (Zone 1), intermediate region (Zone 2), the cavity of the SJ (Zone 4), and the orifice of the SJ (Zone 3) are generated separately, as shown in Fig.~\ref{fig:grid}.
The C-type grid is adopted around the airfoil, and the outer boundary is located at $25c_{\text{h}}$ away from the leading edge.
The size of the computational domain in the spanwise direction ($y$ direction) is $0.2c_{\text{h}}$.
The boundary-fitted coordinate system $(\xi,\eta,\zeta)$ is employed as shown in Fig.~\ref{fig:grid};
the minimum grid size in the wall-normal direction ($\zeta$ direction) is $0.12\%$ of the chord length $c_{\text{h}}$ (or $0.03/\sqrt{Re}$).
The grid of the cavity (Zone 4) is deformed periodically in time, where the grid points are determined in the algebraic manner given by Refs.~\onlinecite{Melville1997,Okada2012a}.
On the boundaries where the zonal grids are connected with each other, approximately 20 grid points are overlapped and the flow variables are exchanged with small errors~\cite{Fujii1995a}.
The number of total grid points is approximately 30 million (Table~\ref{tab:Gridpoints}).
At the outflow boundary, all variables are extrapolated from the points next to the outflow boundary.
A periodic boundary condition is applied to the spanwise direction.
An adiabatic no-slip condition is adopted on the surface of the airfoil and the walls in the SJ. 
\textcolor{\hglght}{
The nondimensionalized computational time step is $\Delta t^* u_\infty / c_{\text{h}} =4.0\times 10^{-5}$ and the corresponding maximum Courant number becomes approximately 1.6.
The time step and grid size normalized by the wall-unit size approximately satisfy $\Delta t^+\leq 0.02$ and $(\Delta\xi^+,\Delta\eta^+,\Delta\zeta^+)\leq (10,9,1)$, respectively.
These are sufficient for a wall-resolved LES of a turbulent boundary layer as suggested in ~\cite{Larsson2016,Kawai2008,Choi1994}.
A series of convergence study in Appendix~\ref{sec:validation} indicates that the present resolution is almost comparable to DNS criteria used in some references (see Appendix~\ref{sec:validation} in detail), however, which are not rigorously examined in terms of a comparison between grid spacing and Kolmogorov scale. 
Therefore, the present simulation is defined as a wall-resolved LES, which provides sufficiently converged results in terms of turbulence transition and prediction of separation for this study.
In Appendix~\ref{sec:validation}, we performed a comparison of $C_{\rm p}$ distribution on the airfoil surface with experimental data as well as a series of grid convergence studies. Although the experimental result for reference was available only for the noncontrolled flows in this study, the same flow solver has been utilized to perform separation-control simulations at the same Reynolds number using the DBD plasma actuator~\cite{Sato2015PoF, Aono2017, Sato2020}, where the simulation results have been well validated in the controlled cases. 
Therefore, the present flow solver and simulation set up are believed to provide sufficiently accurate results considering the similar separation-control simulations performed as above, although full experimental validation for the controlled flow was not available due to a spanwise-uniform SJ geometry which simplifies the jet characteristics.}

\begin{table}
  \begin{center}
\def~{\hphantom{0}}
  \begin{tabular}{cccccc}
\hline\hline
zone name &description  &  $(N_\xi,N_\eta,N_\zeta)$& number of total grid points \\\hline
Zone 1    &airfoil grid &   $( 795 , 134, 179)$    & 19,068,870\\
Zone 2    &intermediate grid  & $ (253 ,134 , 91)$    & 3,085,082\\
Zone 3    &orifice grid &    $( 45 , 134 , 75)$    & 452,250\\
Zone 4    &cavity grid &    $(157 , 134 ,214)$    & 4,502,132\\\hline\hline
  \end{tabular}
\caption{Number of grid points is summarized.
$(N_\xi,N_\eta,N_\zeta)$ represents the number of grid point in each direction.}\label{tab:Gridpoints}
  \end{center}
\end{table}
%
%

\subsection{Post-processing methods for mean flow characteristics}\label{subsec:numerical-tools}

\subsubsection{Averaging operators}
In this part, we describe the definitions of averaging operators.
Let $f$ represent an instantaneous physical quantity, which is the function of time $t$ and space $\br=(x,y,z)$.
The total and phase averaging operators are defined as $\overline{\bullet}$ and $\ang{\bullet}_\varphi$, respectively:
\begin{align}
\overline{f}&=\lim_{T_{\text{all}} \to \infty}\frac{1}{T_{\text{all}}}\int_{t_0}^{t_0+T_{\text{all}}}f(t,\br){\rm d}t\simeq\frac{1}{K}\sum_{i=0}^{K-1}f(t_0+i\Delta t,\br),\quad (\Delta t= T_{\text{all}}/K )\label{eq:totalave}\\
\ang{f}_\varphi&=\lim_{N_\varphi \to \infty}\frac{1}{N_\varphi}\sum_{n=0}^{N_\varphi-1}f(t_\varphi+n T,\br)
\simeq \frac{1}{N_\varphi}\sum_{n=0}^{N_\varphi-1}f(t_\varphi+n T,\br), \notag\\
&\qquad\qquad\qquad\qquad\qquad
(T=1/\Fp, \quad 0\leq \varphi < 2\pi, \quad t_0\leq t_\varphi < t_0+T)\label{eq:phaseave}
\end{align}
where $t_0$ represents a start time of the averaging procedure;
$T_{\text{all}}$ denotes the averaging period;
$\Delta t$ is identical to the computational time step;
$\varphi$ denotes the phase angle between $0$ and $2\pi$ (in this study, $\varphi$ is associated with the actuation frequency of the SJ), and $N_\varphi$ indicates the maximum number of ensembles;
$t_\varphi$ denotes the time that corresponds to the phase angle $\varphi$;
$T$ is the time period that is characterized by a specific flow motion, e.g., the actuation period $T=1/\Fp$ in this study.
Accordingly, the averaging operator $\overline{\bullet}^\varphi$ is also defined as,
\begin{align}
\overline{\ang{f}_\varphi}^{\varphi}&=\frac{1}{2\pi}\int_{0}^{2\pi}\ang{f}_\varphi {\rm d}\varphi
\simeq\frac{1}{M}\sum_{m=0}^{M-1}\langle{f}\rangle_{\varphi_m},
\end{align}
where $M$ denotes the total number of discrete phases (the present paper adopts $20$ segments for each period.)
Note that the total averaging operator $\overline{\bullet}$ is theoretically (but not numerically) the same as the total phase averaging operator $\overline{\bullet}^\varphi$, although they are separately used in this study.

When a spatially homogeneous direction can be defined, the ensemble average should also be taken with respect to that spatial direction.
In this study, the spanwise direction ($y$) is regarded as a homogeneous direction, and thus we introduce the spanwise averaging operator $[\bullet]$ as follows:
\begin{align}
[f(t,\br)]=\lim_{L \to \infty}\frac{1}{L}\int_{y_0}^{y_0+L}f(t,\br){\rm d}y
\simeq\frac{1}{K_y}\sum_{k=0}^{K_y-1}f(t,x,y_k,z),\label{eq:spanave}
\end{align}
where $K_y$ is the number of grid points in the spanwise direction.
In the rest of this paper, the spanwise averaging operator $[\bullet]$ is abbreviated for conciseness, but is used together with total and phase averaging operators to increase a number of ensembles. 

\subsubsection{Phase decomposition}
In this subsection, we consider a physical quantity $f$, which is a function of time $t$.
We will conduct a phase decomposition to extract coherent flow structures based on the period that corresponds to the actuation frequency $F^+$.
For this purpose, the instantaneous physical quantity $f(t)$ is decomposed into overall average $\overline{f}$; phase fluctuation $\tilde{f}$; and turbulent fluctuation $f''$ as follows: 
\begin{align}
f(t)
&=\overline{f}+f'(t)
=\overline{f}
+\tilde{f}_\varphi
+f''(t)\\
&=\ang{f}_\varphi+f''(t).\label{phase}
\end{align}
Therefore, the following decomposition holds:
\begin{align}
\overline{f'g'}&=\overline{\tilde{f}_{\varphi}\tilde{g}_{\varphi}}^{\varphi}+\overline{f''g''},\label{eq:overall_double}\\
\ang{f'g'}_\varphi&=\ang{\tilde{f}_{\varphi}\tilde{g}_{\varphi}}_\varphi+\ang{f''g''}_\varphi.\label{eq:phased_double}
\end{align}
These are the definitions of phase decomposition of the Reynolds stress by replacing $f$ and $g$ by $u$ and $w$, respectively.

\subsection{Discrete Fourier transform}
Discrete Fourier transform (DFT) analysis is conducted in a temporal direction to discuss the power spectrum density (PSD) of the wall-normal velocity fields, based on the FFTW developed by Ref.~\onlinecite{Frigo2006}.
All the data is multiplied by the Han window to suppress a sidelobe leakage of the DFT.
In this study, the highest frequency is defined as $f_{\text{highest}}=1/\Delta t$ where $\Delta t$ is the computational time step.
The highest sampling rate corresponds to $St=500$ in this study, which is sufficiently high to capture the turbulent phenomena.
Therefore, the lowest frequency is set as $f_{\text{lowest}}=1/(T_{\text{all}}/10)$, which corresponds to $St=0.1$. 
To increase the number of ensembles, we subdivide the time period $T_{\text{all}}$ into 10 segments: that is, $t_0 +n T_{\text{all}}/10 \leq t < t_0 + (n+1) T_{\text{all}}/10$ where $n=0,1,\ldots,9$.
The DFT is conducted for each time segment, which are then simply averaged.
Furthermore, for this DFT procedure, instantaneous physical quantities are sampled on five $y$-normal planes that are distributed across the spanwise direction, which are then simply averaged to increase the number of ensembles for the DFT results.

\subsection{Linear stability analysis}
In this study, we perform a linear stability analysis to identify the most unstable frequency that strongly amplifies a wall-normal velocity fluctuation in the streamwise direction (based on the KH instability), in the separated shear layer near the leading edge.
\textcolor{\hglght}{
It is assumed that the mean flow is locally parallel in the streamwise direction, and the instability mainly arises from an inviscid mechanism with a minor effect of viscosity. Accordingly, the Rayleigh equation is solved as an eigen value problem for a spatial growth rate of the wall-normal velocity fluctuation.
}
In the right side of Fig.~\ref{fig:grid}, the wall-parallel and wall-normal directions are defined as $x_{\text{wp}}$ and $x_{\text{wn}}$, respectively.
The velocity components in the wall-parallel and wall-normal directions are similarly defined as $u_{\text{wp}}$ and $u_{\text{wn}}$, respectively.
The time- and spanwise-averaged flow is considered as a mean flow profile, where the wall-parallel velocity field, $u_{\text{wp}}(x_{\text{wp}},x_{\text{wn}})$, is extracted from the straight line that is normal to the airfoil surface at the chordwise location of $x_{\text{wp}}$.
Then, the wall-normal velocity fluctuation is assumed to be $u_{\text{wn}}'(x_{\text{wp}},x_{\text{wn}},t)=\hat{u}_{\text{wn}}(x_{\text{wn}})\exp (-\alpha x_{\text{wp}} + \omega t)$, where $\alpha$ and $\omega$ are complex wave numbers in the spatial and temporal direction.
The governing equation of $\hat{u}_{\text{wn}}$ is given as follows:
\begin{align}
&\frac{ {\rm d}^2 \hat{u}_{\text{wn}}}{ {\rm d}x_{\text{wn}}^2}
-
\left(
\frac{1}{\overline{u}_{\text{wp}}-\omega/\alpha}\frac{ {\rm d}^2 \overline{u}_{\text{wp}}}{ {\rm d} x_{\text{wn}}^2} +\alpha^2
\right)\hat{u}_{\text{wn}}=0\quad (0\leq x_{\text{wn}}\leq x_{\text{wn};\text{max}}),\label{eq:LSA0}\\
&\qquad \hat{u}_{\text{wn}}(0)=\hat{u}_{\text{wn}}(x_{\text{wn};\text{max}})=0,\label{eq:LSA}
\end{align}
where $x_{\text{wn};\text{max}}$ corresponds to the 99\% thickness of the boundary layer at each chordwise location.
Note that the time- and spanwise-averaged wall-parallel velocity $\overline{u}_{\text{wp}}$ is modelled by a modified hyperbolic tangent function, which is adopted as a base flow in Eq.~\eqref{eq:LSA0}.
Moreover, the quantities are normalized by the reference velocity $U_{\text{ref}}$ and length $\delta_{\text{ref}}$, which are the maximum value of $\overline{u}_{\text{wp}}(x_{\text{wn}})$ on the wall-normal line and the distance of an inflection point from the airfoil surface, respectively.
The governing equation given by Eq.~\eqref{eq:LSA0} formulates an eigen value problem of $\alpha$ and $\hat{u}_{\text{wn}}$ for the specific $\omega$ (that is, for the specific frequency $St$).
Accordingly, the most unstable eigen mode of $\hat{u}_{\text{wn}}$ is identified so that a real part of the spatial growth rate becomes the largest in the negative direction: $-\alpha_r=-\Re(\alpha)$ is maximized.
The shooting method is adopted since the present problem is formulated as a two-point boundary value problem.
The differential equation is numerically solved by fixing the upper-boundary value, i.e., the values of $\hat{u}_{\text{wn}}(x_{\text{wn};\text{max}})$ and $ {\rm d} \hat{u}_{\text{wn}}/ {\rm d}x_{\text{wn}}|_{x_{\text{wn}}=x_{\text{wn};\text{max}}}$ are given for each estimated $\alpha$, then the numerical integration (four-stage fourth-order Runge-Kutta method in this study) is conducted, and the lower boundary value $\hat{u}_{\text{wn}}(0)$ is determined.
Then, $\alpha$ is iteratively modified using a Newton-Raphson method so that $\hat{u}_{\text{wn}}(0)$ becomes $0$ as given in Eq.~\eqref{eq:LSA}.
Note that the first derivative of $\hat{u}_{\text{wn}}$ at the upper boundary is assumed to be the asymptotic solution:
\begin{align}
\lim_{x_{\text{wn}} \to \infty}\frac{ {\rm d}\hat{u}_{\text{wn}}}{ {\rm d} x_{\text{wn}}}=-C\alpha \exp[-\alpha x_{\text{wn}}].\quad (C\text{: constant})
\end{align}
%
%

\section{Basic characteristics of separation controlled flows}\label{sec:basicflow}
\subsection{Aerodynamic performance of controlled flows with different $\Fp$}\label{subsec:aerocoef}
\begin{figure}[htbp]
\centering
\renewcommand{\size}{0.475}
\centering
\ifCONDITION
   \subfloat[][$C_{\rm{L}}$]{\includegraphics[width=\size\textwidth]{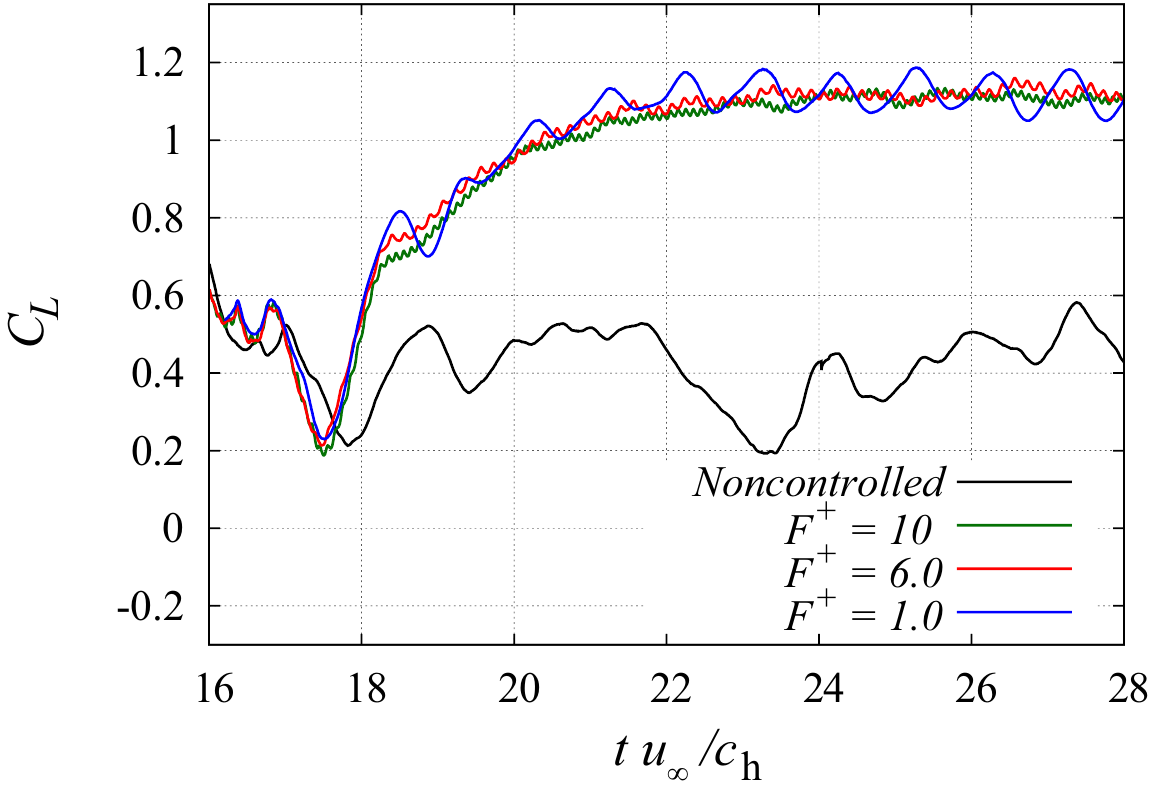}}
   \subfloat[][$C_{\rm D}$]{\includegraphics[width=\size\textwidth]{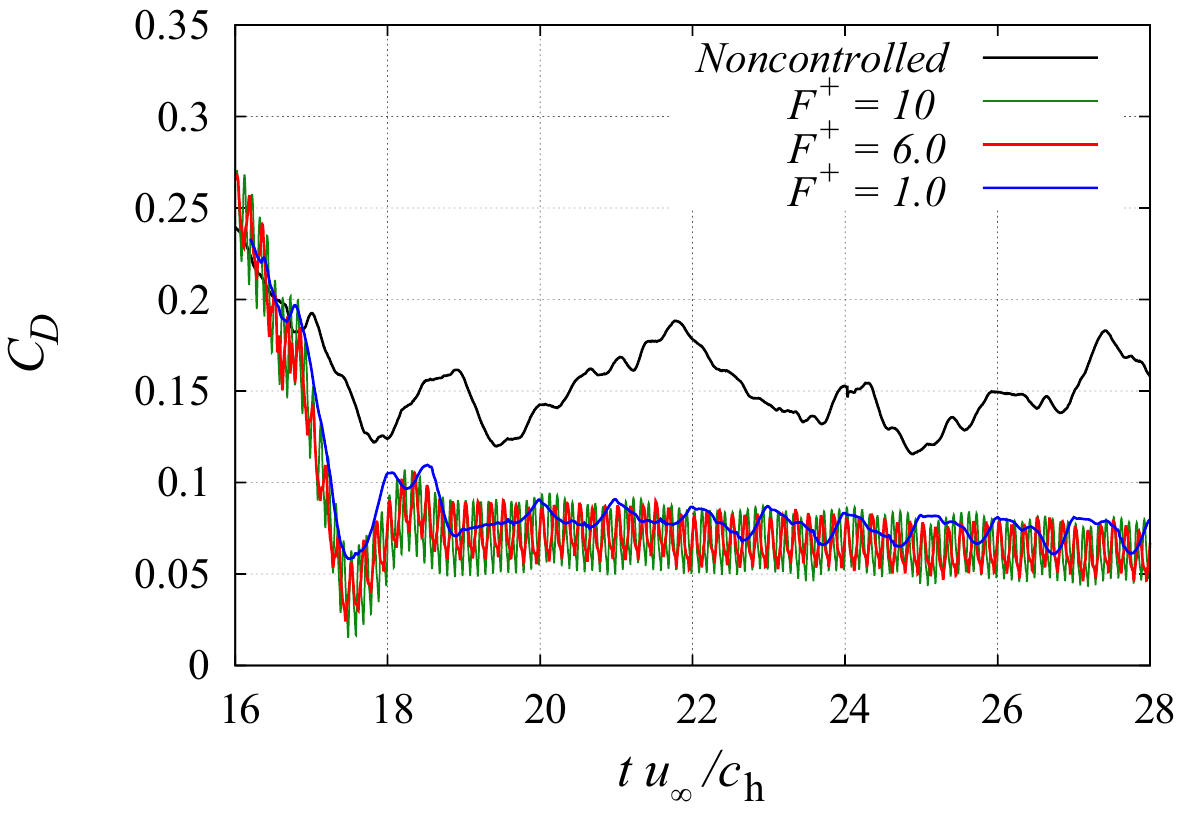}}
\else
\fi
\caption{
Time history of the aerodynamic coefficients in the cases with $\Cm=2\dd{-3}$: (a) lift coefficient $C_{\rm{L}}$; (b) drag coefficient $C_{\rm D}$.
}\label{fig:CLCDhistory}
\end{figure}

Figure~\ref{fig:CLCDhistory} shows time histories of the lift and drag coefficients, $C_{\rm{L}}$ and $C_{\rm D}$, in the noncontrolled and controlled cases.
The horizontal axis shows the nondimensional time, $t=t^* u_\infty/c_{\text{h}}$, and the flow control is performed for $16\leq t \leq 28$.
The black lines, in Figs.~\ref{fig:CLCDhistory}(a) and (b), show $C_{\rm{L}}$ and $C_{\rm D}$ histories of the noncontrolled case.
The blue, red, and green lines show the controlled cases using a strong input momentum ($\Cm=2.0\dd{-3}$) with $\Fp=1.0$, $6.0$, and $10$, respectively.
In both of the controlled cases, $C_{\rm{L}}$ and $C_{\rm D}$ rapidly improved in $16\leq t\leq 20$, where the lift increases and the drag reduces.
This indicates a stall recovery, where both $C_{\rm{L}}$ and $C_{\rm D}$ are significantly improved from the noncontrolled case.
At $t\geq 20$, $C_{\rm{L}}$ and $C_{\rm D}$ reach quasi-steady states, where both $C_{\rm{L}}$ and $C_{\rm D}$ oscillate in specific frequencies that are different for each controlled case.
It is found that these frequencies correspond to the actuation frequencies ($\Fp$), thereby suggesting the existence of a periodic flow structure that is associated with $\Fp$.
The more precise analysis for these periodic flow structures will be shown in Sec.~\ref{subsec:coherent}.

\begin{figure}[htbp!]
\renewcommand{\size}{0.425}
\ifCONDITION
\centering
    \subfloat[][$C_{\rm{L}}$     ]{\includegraphics[width=\size\textwidth]{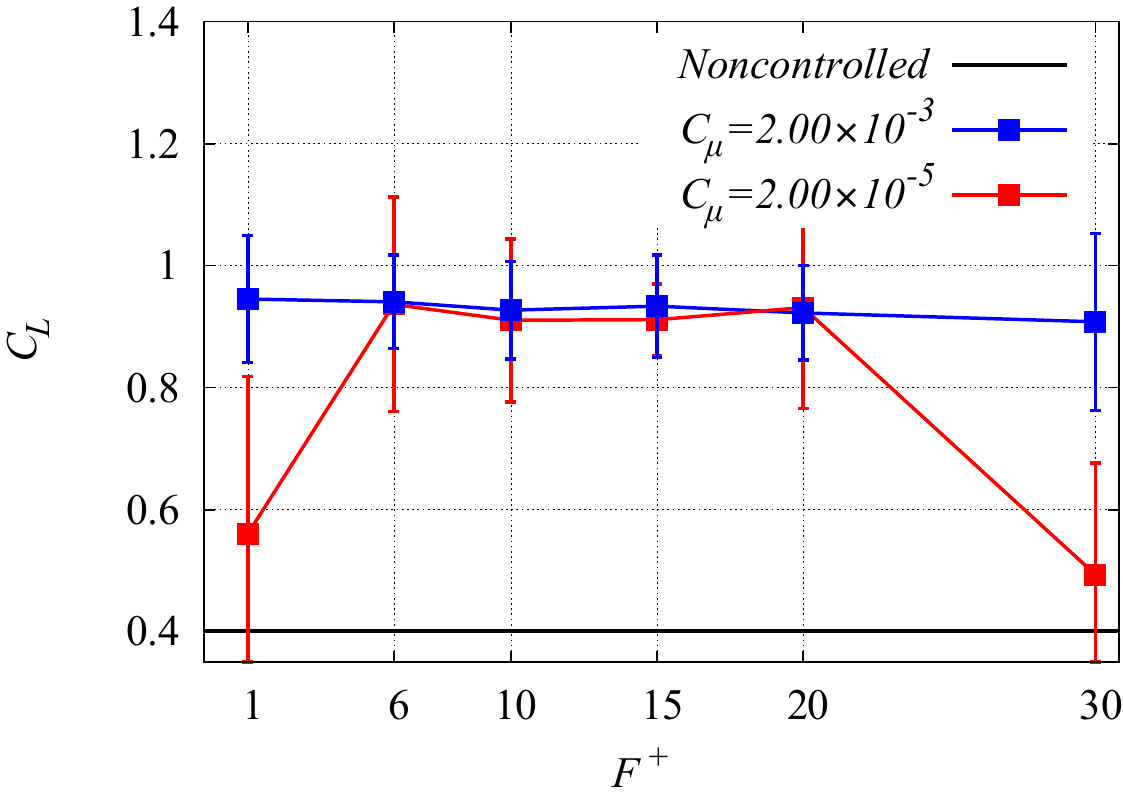}}
    \subfloat[][$C_{\rm D}$     ]{\includegraphics[width=\size\textwidth]{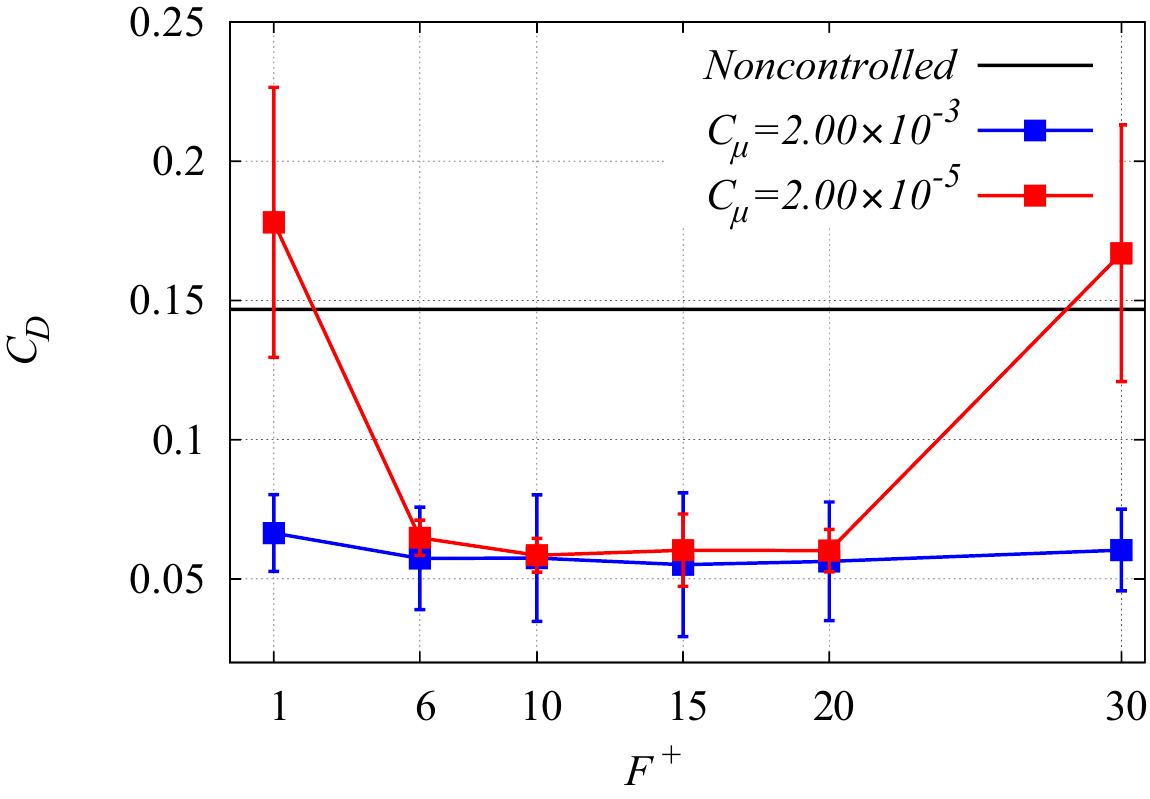}}\\
    \subfloat[][$C_{\rm{L}}/C_{\rm D}$ ]{\includegraphics[width=\size\textwidth]{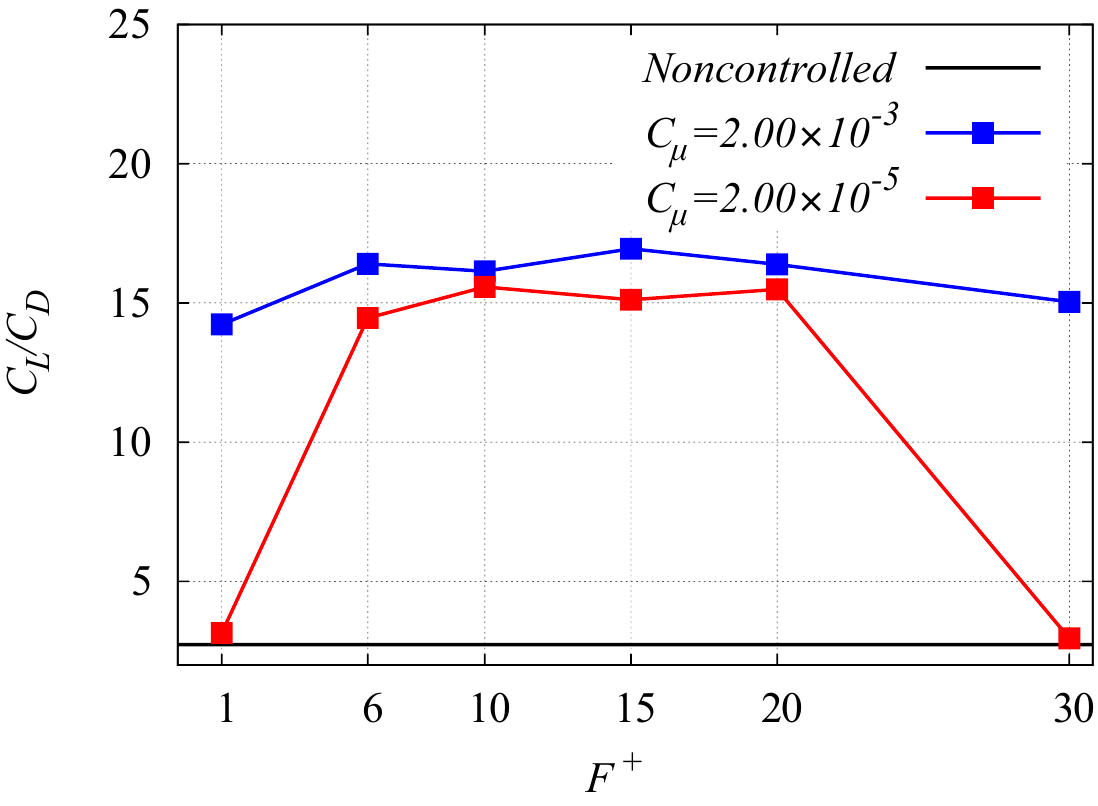}}
\else
\fi
\caption{
Aerodynamic coefficients of $C_{\rm{L}}$, $C_{\rm D}$, and $C_{\rm{L}}/C_{\rm D}$ of time- and spanwise-averaged flows:
black straight lines show the noncontrolled case, blue lines show the strong input case ($\Cm=2.0\dd{-3}$), and red lines show the weak input case ($\Cm=2.0\dd{-5}$).
The standard deviations of $C_{\rm{L}}$ and $C_{\rm D}$ in the time direction are visualized by error bars, whose length is $75$ times larger than the actual scale for ease of visualization.
}\label{fig:freq-clcd}
\end{figure}

Next, we show the summary of time-averaged $C_{\rm{L}}$, $C_{\rm D}$, and $C_{\rm{L}}/C_{\rm D}$ (lift-to-drag ratio) values in Fig.~\ref{fig:freq-clcd}.
Here again, the time period $20\leq t\leq 28$ is regarded as the quasi-steady state and taken for the time averaging procedure as explained in Sec.~\ref{subsec:numerical-tools}.
In Fig.~\ref{fig:freq-clcd}, the effects of $\Fp$ on $C_{\rm{L}}$, $C_{\rm D}$, and $C_{\rm{L}}/C_{\rm D}$ are shown, where black straight line shows the noncontrolled case, the blue line shows the strong input ($\Cm=2.0\dd{-3}$) case, and the red line shows the weak input ($\Cm=2.0\dd{-5}$) case.
The standard deviation of $C_{\rm{L}}$ and $C_{\rm D}$ history (Fig.~\ref{fig:CLCDhistory}) is visualized by error bars in Figs.~\ref{fig:freq-clcd}(a) and (b), the length of which is set to be $75$ times larger than the actual scale for ease of visualization.

The strong input cases ($\Cm=2.0\dd{-3}$: blue lines) increase $C_{\rm{L}}$ and decrease $C_{\rm D}$ significantly, which indicates a stall recovery.
In these cases, the controlled cases with $6 \leq \Fp\leq 20$ improve $C_{\rm{L}}/C_{\rm D}$ better than $\Fp=1.0$ and $30.0$ (Fig.~\ref{fig:freq-clcd}(c)).
Interestingly, such an $\Fp$ effect is more clearly observed in $C_{\rm D}$ of Fig.~\ref{fig:freq-clcd}(b) than $C_{\rm{L}}$ of Fig.~\ref{fig:freq-clcd}(a), where the controlled cases with $\Fp=1.0$ and $30$ represent values larger than the cases with $6 \leq \Fp\leq 20$.
This implies that, in the present controlled cases, the better control is achieved by decreasing the drag, rather than increasing the lift.
Note that hereinafter, the ability of separation control is evaluated by the time-averaged $C_{\rm{L}}/C_{\rm D}$ value.
In the weak input cases ($\Cm=2.0\dd{-5}$: red lines), the controlled cases with $\Fp=1.0$ and $30$ do not improve the lift and drag compared with the noncontrolled case, and therefore the stall recovery is not achieved.
In contrast, the controlled cases with $6\leq \Fp \leq 20$ represent an ability to suppress the separation.
In this way, the observation of strong and weak input cases suggests an optimal regime in the actuation frequency, $6\leq \Fp \leq 20$, to achieve the better control capability.

It should be mentioned that the significance of a drag reduction on the separation control ability compared with a lift increase could largely rely on the flow/actuator conditions.
For example, recent study on separation control over an NACA0012 airfoil using the SJ (Ref.~\onlinecite{Zhang2015}) reported that the time-averaged $C_{\rm{L}}/C_{\rm D}$ with $\Fp=1.0$ is higher than that with $\Fp=4.0$, although the drag reduction by $\Fp=4.0$ is larger than that by $\Fp=1.0$.
This indicates that in their flow condition, increasing the lift is more effective to improve $C_{\rm{L}}/C_{\rm D}$.
The inconsistency with the present result can be explained as follows.
First, the chord Reynolds number of Ref.~\onlinecite{Zhang2015} is $10,000$, which is much lower than our condition ($63,000$).
Second, the SJ is modeled by the two-dimensional (spanwise-uniform) velocity profile on the airfoil surface (so-called ``boundary condition model'') in Ref.~\onlinecite{Zhang2015}.
These two difference can affect the turbulent transition in the separation controlled flow.
Indeed, in our simulation, strong turbulent structures are generated in the SJ cavity, which enhances the turbulent transition in the separation-controlled flow over an airfoil, as will be discussed in Sec.~\ref{subsec:unsteady} and Sec.~\ref{subsubsec:tke}.
The present study performs simulations using the precisely-modeled SJ that includes a cavity deformation; therefore we can properly evaluate the effect of turbulence structure generated by SJ on the flows over an airfoil.

%
%

\subsection{Time-averaged flow fields}\label{subsec:timeave}
\begin{figure}[htbp!]
\centering
    \renewcommand{\func}{u}
    \renewcommand{\size}{\sizepng}
\ifCONDITION
 \subfloat[][Noncontrolled]{\includegraphics[viewport=5 47 400 175,clip,width=\size\textwidth]{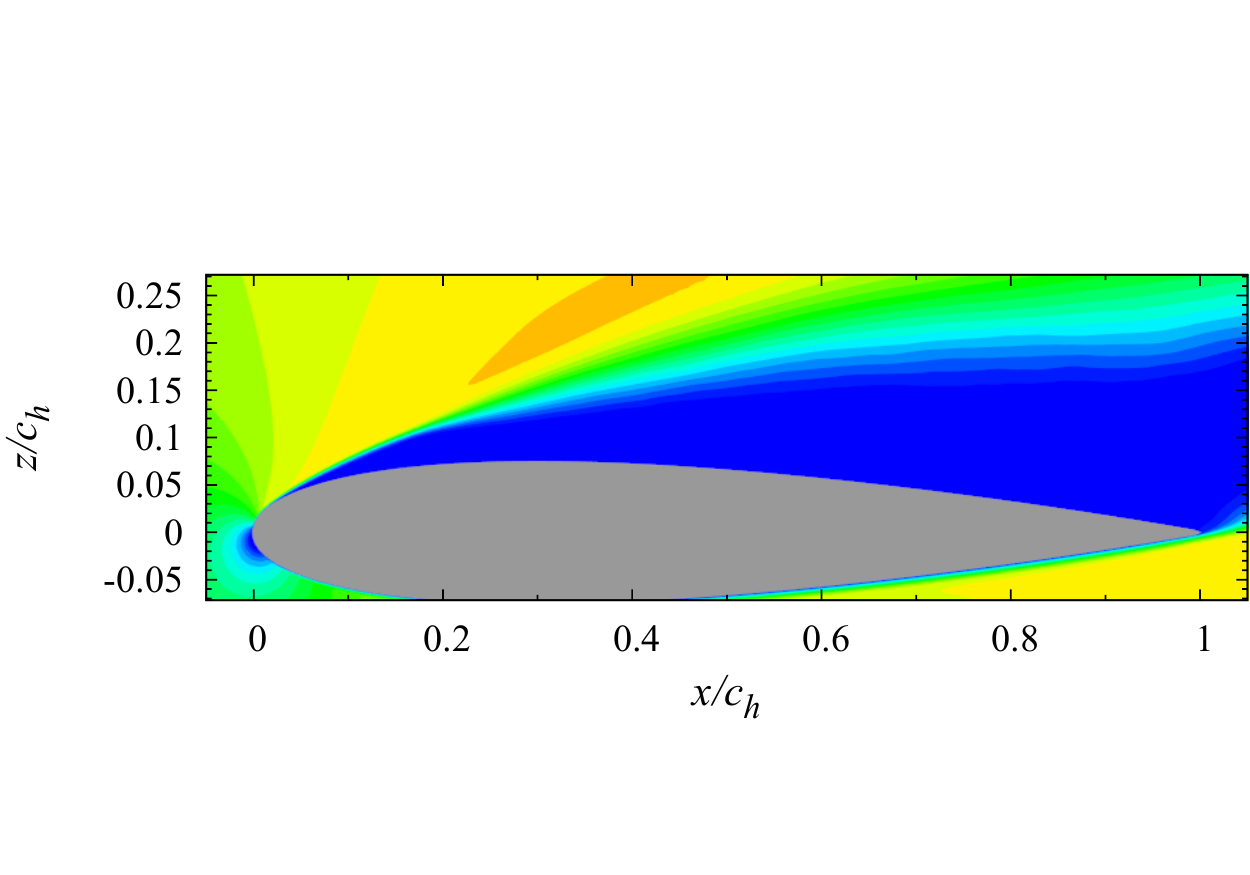}}\\
    \subfloat[][\hcasesjaa]{\includegraphics[viewport=5 47 400 175,clip,width=\size\textwidth]{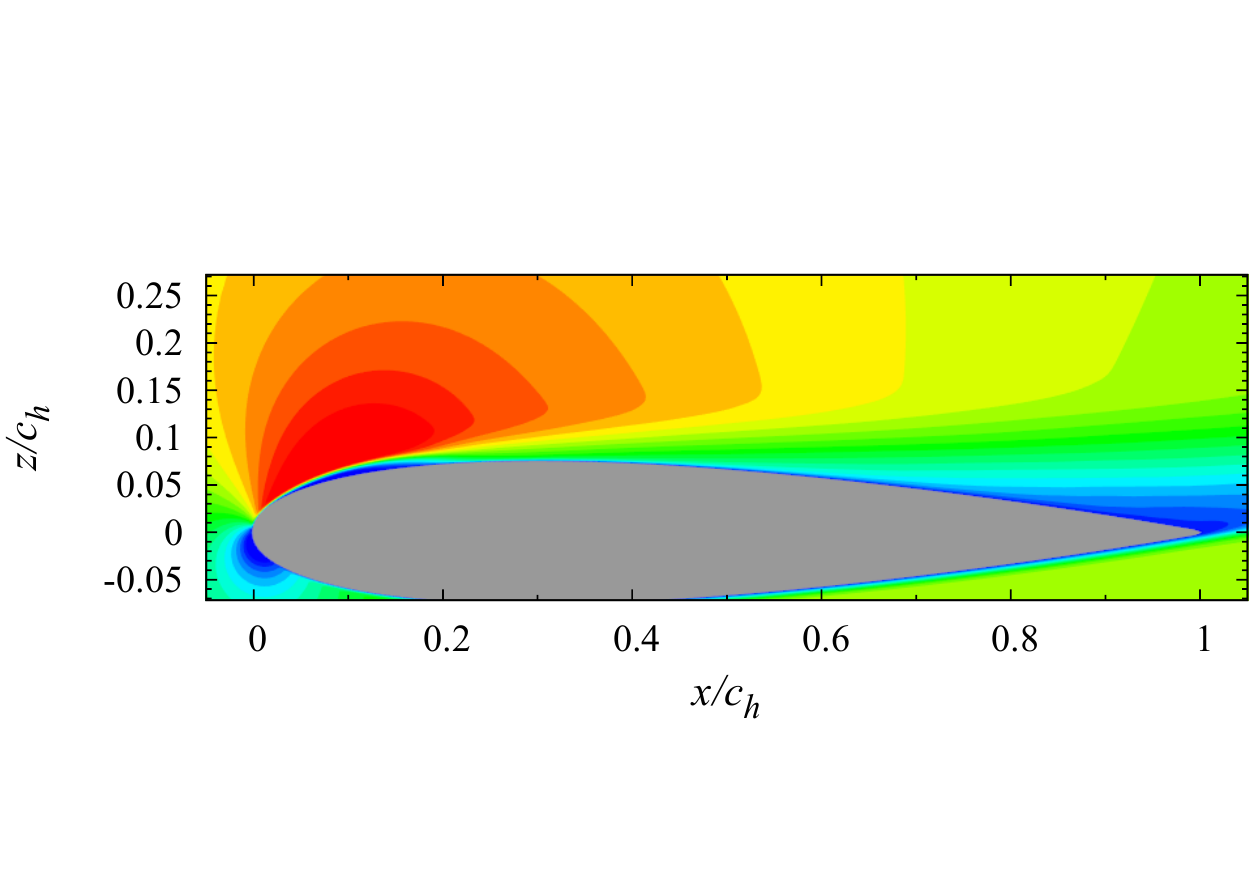}}
    \subfloat[][\hcasesjba]{\includegraphics[viewport=5 47 400 175,clip,width=\size\textwidth]{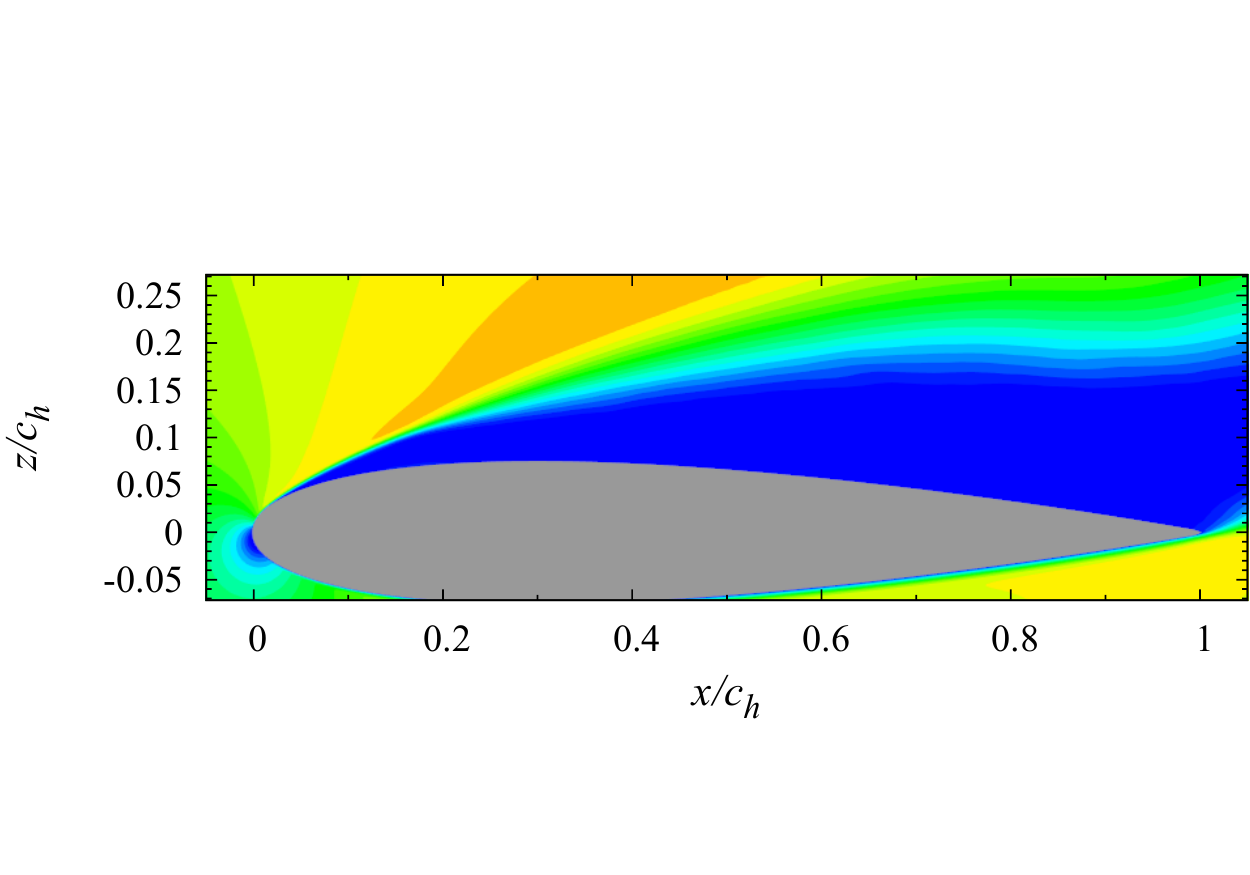}}\\
    \subfloat[][\hcasesjab]{\includegraphics[viewport=5 47 400 175,clip,width=\size\textwidth]{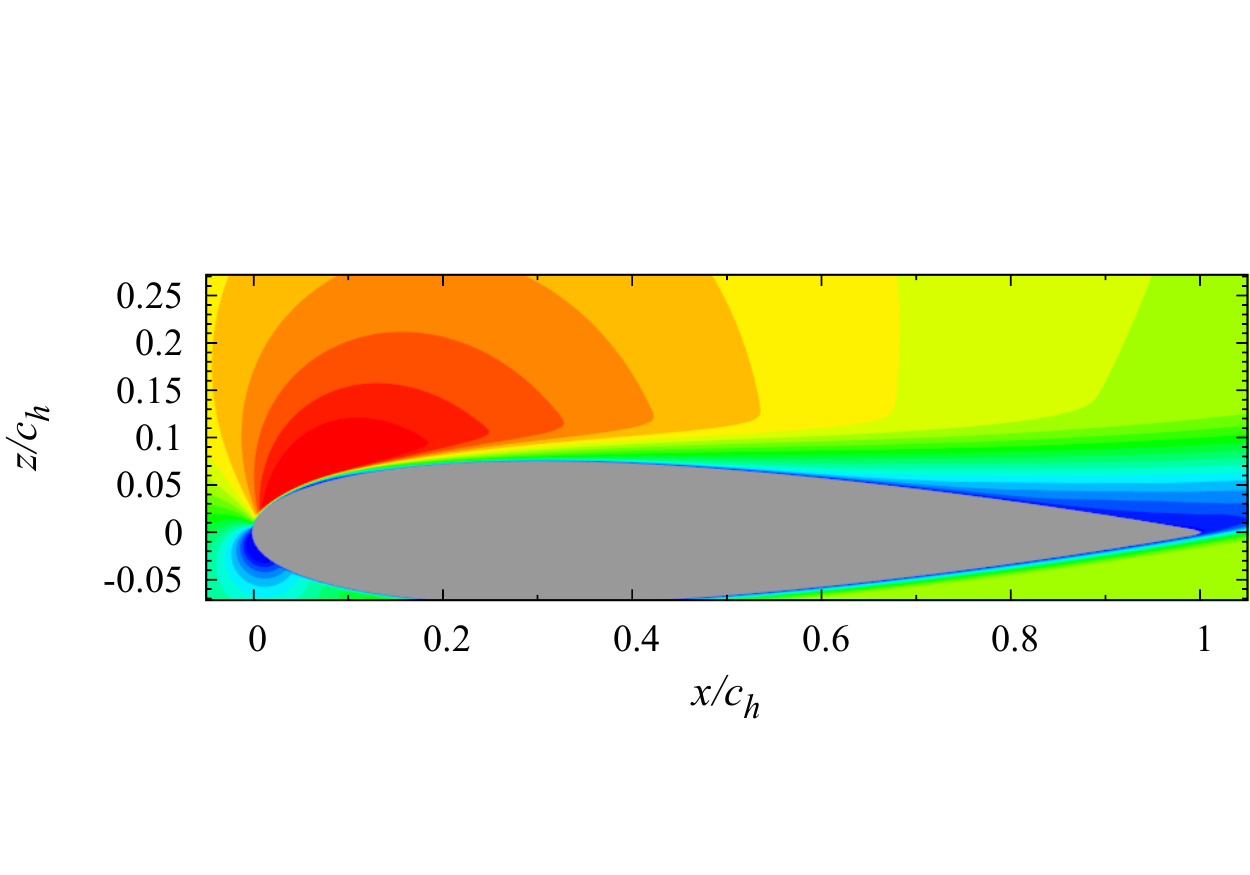}}
    \subfloat[][\hcasesjbb]{\includegraphics[viewport=5 47 400 175,clip,width=\size\textwidth]{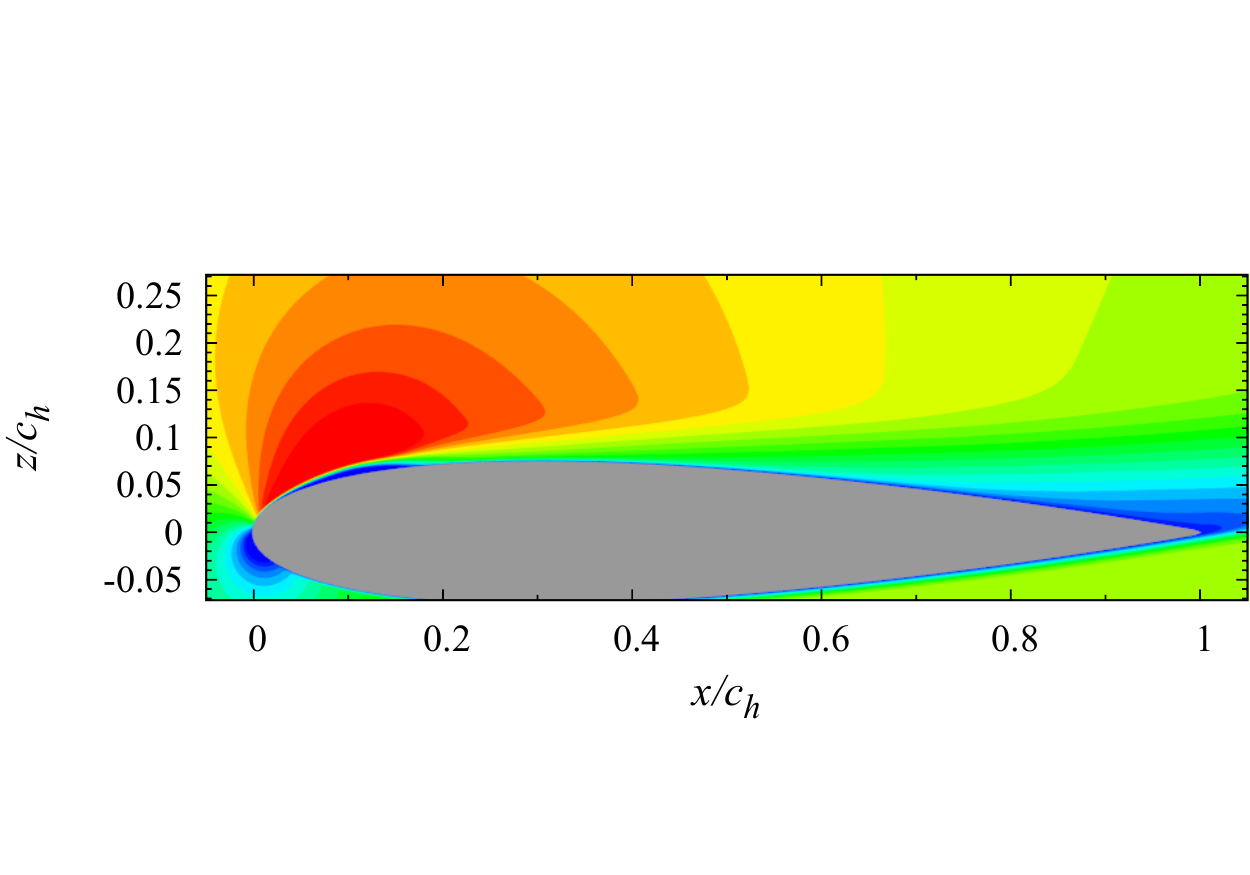}}
\else
\fi
    \caption{
Time- and spanwise-averaged flow fields of the representative cases.
Contour color shows the chordwise velocity component normalized by the freestream velocity: $0.0\leq u/u_{\infty} \leq 1.5$.
    }\label{fig:u}
\end{figure}

Figure~\ref{fig:u} shows time- and spanwise-averaged flow fields of the representative computational cases.
Contour colors visualize the chordwise velocity normalized by the freestream velocity, where blue to red color corresponds to $0.0\leq u/u_{\infty}\leq 1.5$.
Figure~\ref{fig:u}(a) shows the noncontrolled case.
A large blue region appears on the suction side of an airfoil, which corresponds to a reversed flow region.
As such, the flow naturally separates from the vicinity of the leading edge, thereby resulting in the stall condition.
In the controllable cases (Figs.~\ref{fig:u}(b), (d), and (e)), the chordwise velocity becomes positive over the most of the suction side, while the uncontrollable case of Fig.~\ref{fig:u}(c) shows a separated flow similar to the noncontrolled case.
This observation is consistent with the time-averaged aerodynamic coefficients, as discussed in Fig.~\ref{fig:freq-clcd}.
The controllable cases of Figs.~\ref{fig:u}(b), (d), and (e) exhibit reversed flow regions near both of the leading and trailing edges.
As will be seen in Fig.~\ref{fig:ins}, the reversed flow region near the leading edge corresponds to a laminar separation bubble, where the laminar flow separates near the leading edge.
The separated shear layer is then transient to turbulence and reattaches to the airfoil surface, thereby formulating a turbulent boundary layer.
Then, the turbulent boundary layer again separates from the airfoil surface, which corresponds to the reversed flow region near the trailing edge.
This is a fundamental feature of the separation-controlled flow in the present conditions, which exhibits two characteristic separated (reversed flow) regions in the time and spanwise-averaged fields.
The size of these separated flow regions will be discussed in the next paragraph.

%
%
\begin{figure}[htbp!]
\centering
\renewcommand{\size}{0.485}
\ifCONDITION
   \subfloat[][$\Cm=2.0\dd{-3}$]{\includegraphics[width=\size\textwidth]{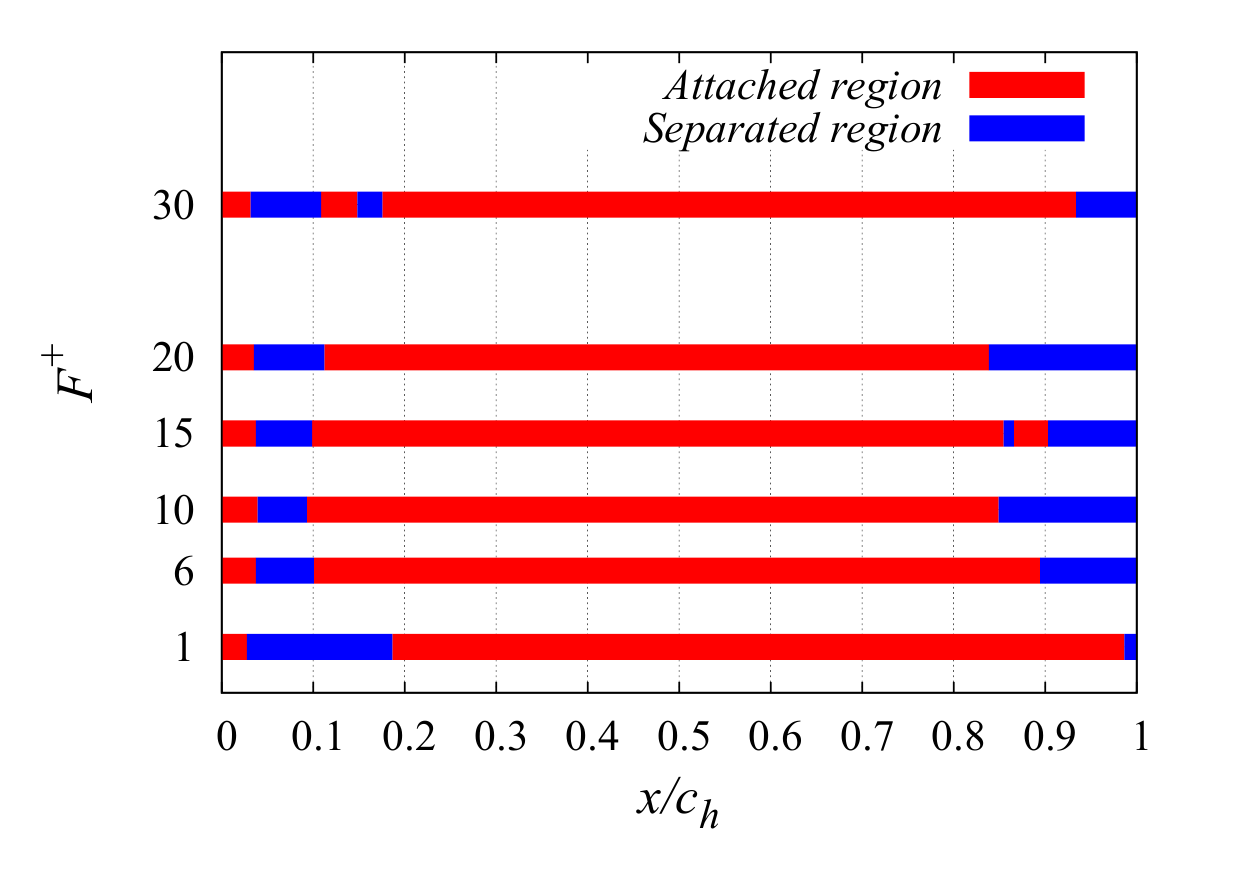}}
   \subfloat[][$\Cm=2.0\dd{-5}$]{\includegraphics[width=\size\textwidth]{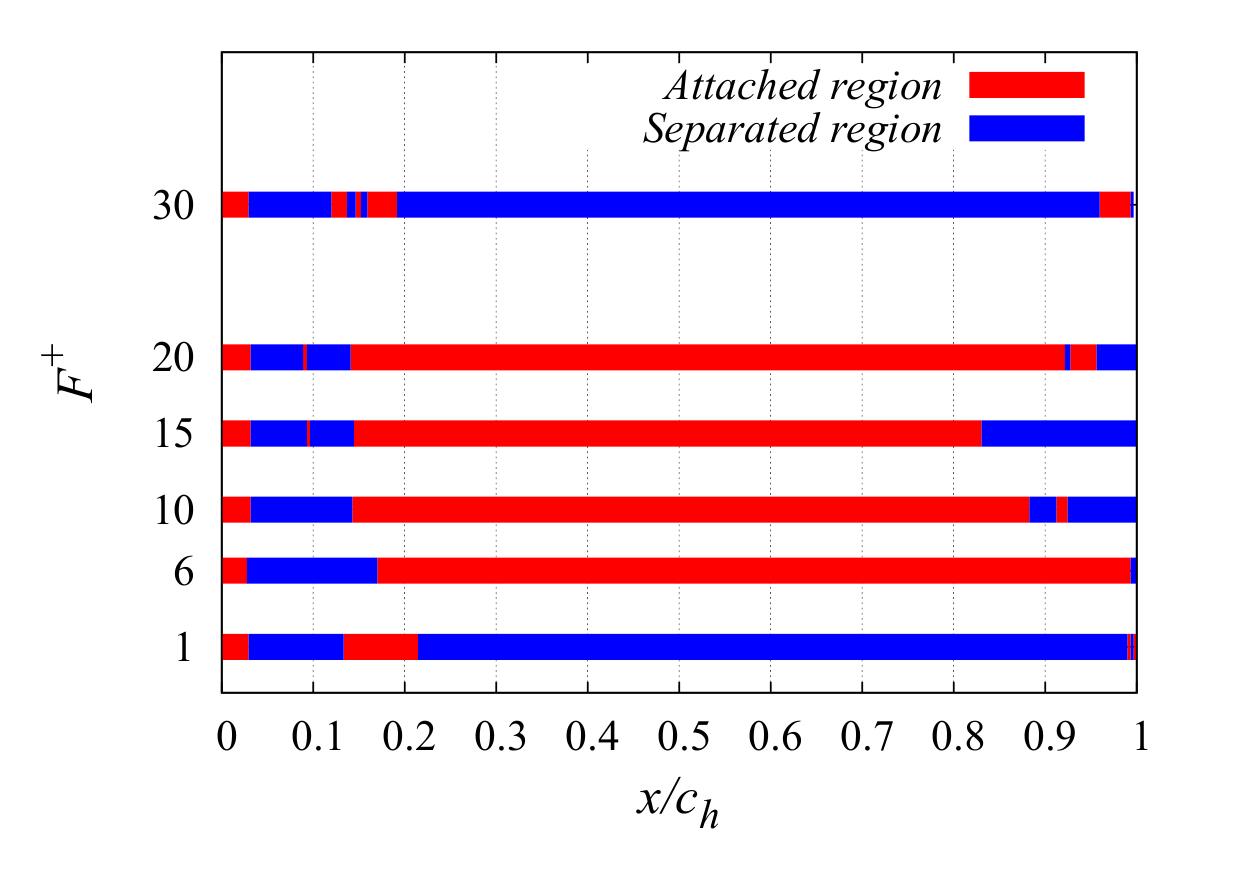}}
\else
\fi
\caption{
Separated and attached flow regions on the suction side of the airfoil is visualized.
Blue and red bars represent the separated and attached regions, respectively, which are evaluated by the sign of the skin friction on the suction side of the airfoil surface.
That is, the red (or blue) region corresponds to the positive (or negative) skin friction, where $\del u_{\text{wp}}/\del x_{\text{wn}}>0$ (or $\leq 0$).
(a) shows strong input cases ($\Cm=2.0\dd{-3}$);
(b) shows weak input cases ($\Cm=2.0\dd{-5}$).
}\label{fig:freq-sep}
\end{figure}

\begin{figure}[htbp!]
\centering
\renewcommand{\size}{0.485}
\ifCONDITION
   \subfloat[][$\Cm=2.0\dd{-3}$]{\includegraphics[width=\size\textwidth]{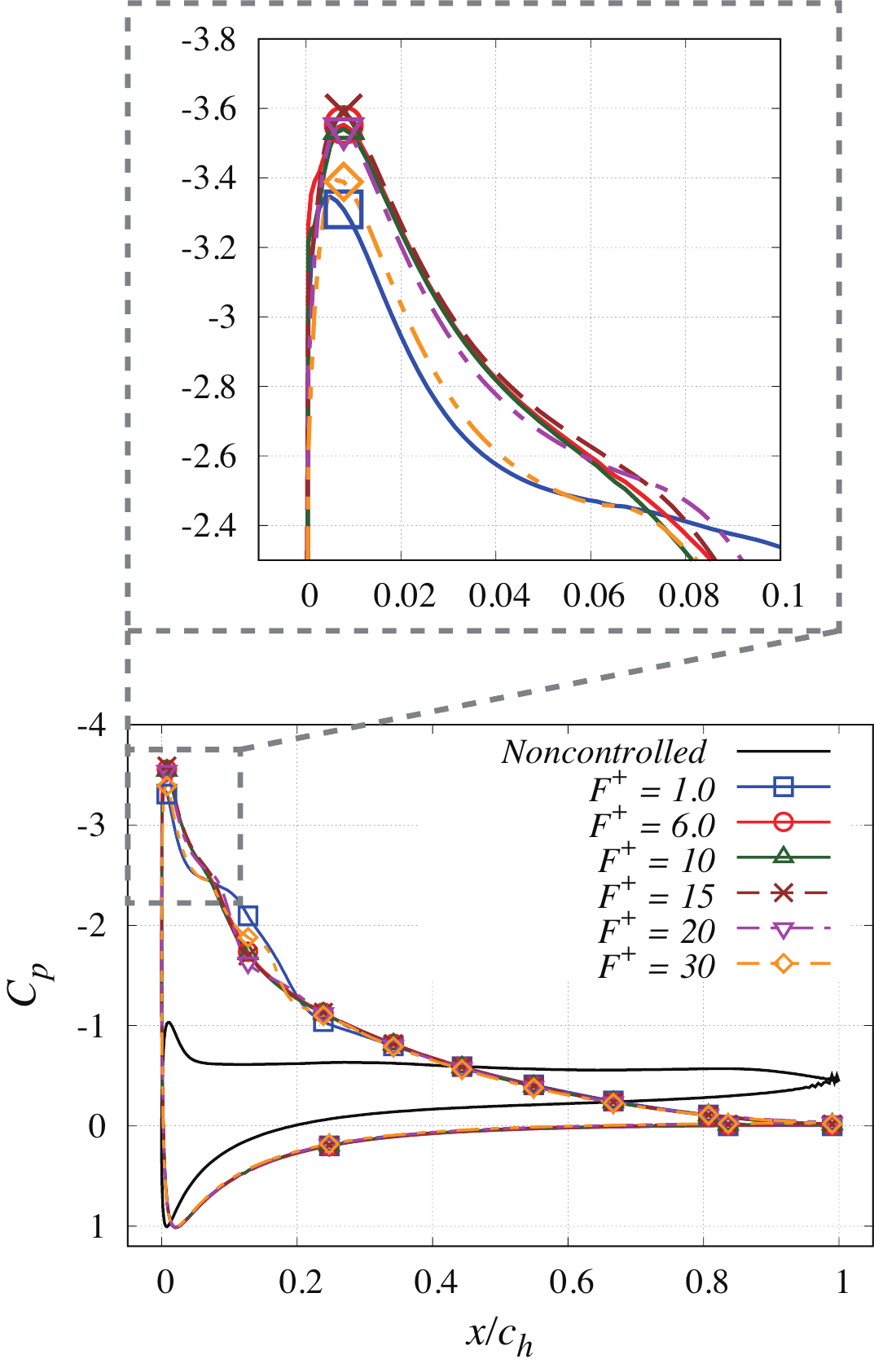}}
   \subfloat[][$\Cm=2.0\dd{-5}$]{\includegraphics[width=\size\textwidth]{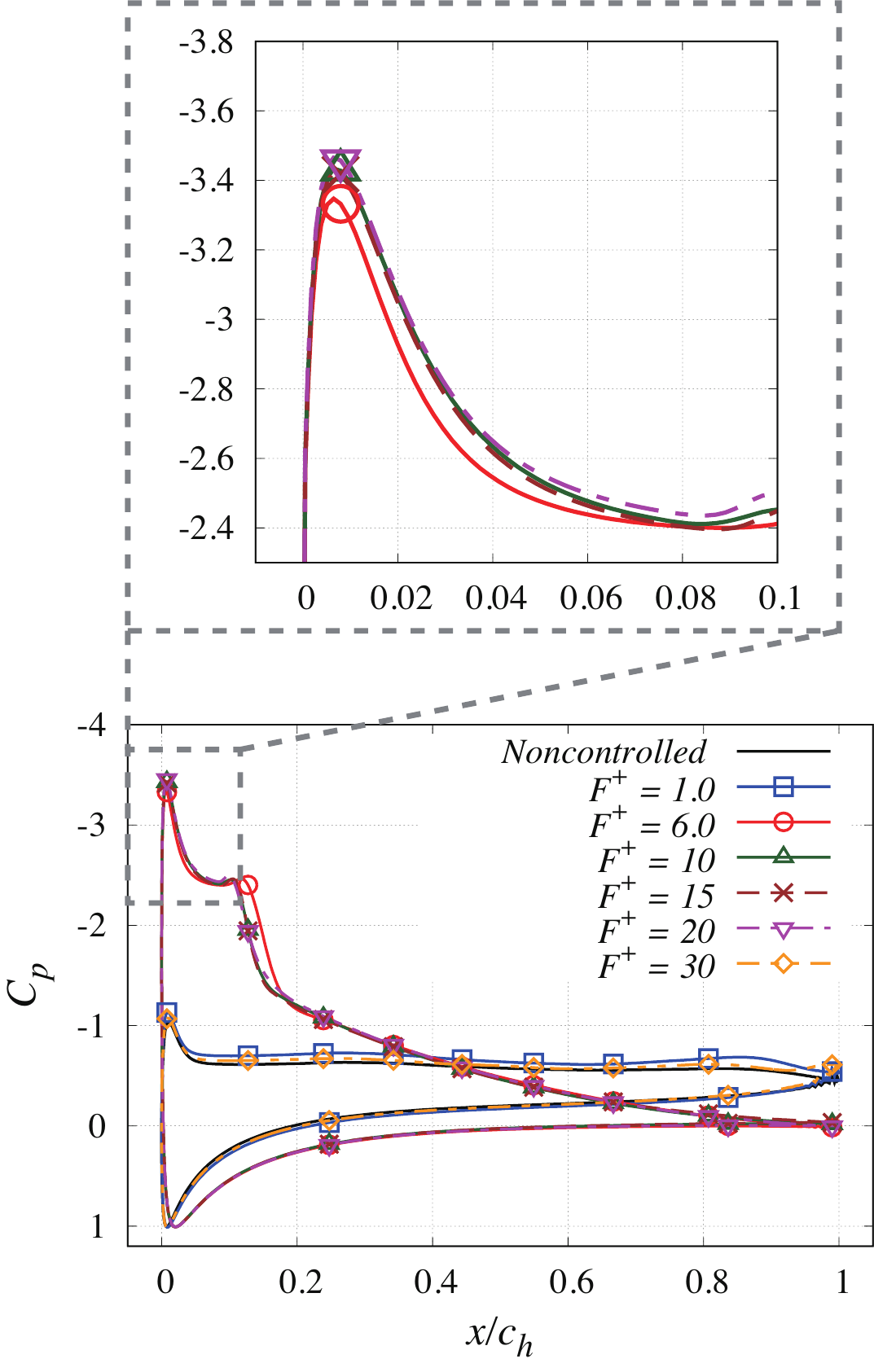}}
\else
\fi
\caption{
Pressure coefficient $C_{\rm p}$ on the airfoil surface, based on the time- and spanwise-averaged flows: (a) $\Cm=2.0\dd{-3}$ and (b) $\Cm=2.0\dd{-5}$.
}\label{fig:freq-cp}
\end{figure}

Figure~\ref{fig:freq-sep} visualizes the separated and attached flow regions with different $\Fp$ and $\Cm$.
The horizontal axis shows a chordwise position on the suction side, $x/c_{\text{h}}$, and the vertical axis shows $\Fp$.
Red and blue color bars represent the attached and separated flow regions, respectively.
These are identical to the regions of positive and negative skin friction in the time- and spanwise-averaged flows.
In other words, the red (or blue) region corresponds the region where $\del u_{\text{wp}}/\del x_{\text{wn}}>0$ (or $\leq 0$) on the suction side of the airfoil surface.
In the strong input cases (Fig.~\ref{fig:freq-sep}(a)), the controlled cases with different $\Fp$ show small separated regions near the leading edge, which correspond to separation bubbles.
The size of the separation bubble becomes smaller at $6.0\leq \Fp\leq 20$ than $\Fp=1.0$ and $30$.
The smaller separation bubble leads to the smaller drag coefficient as the flow over the leading edge is more smoothly accelerated, thereby resulting in the stronger suction peak in the surface pressure at the leading edge, as follows.
Figure~\ref{fig:freq-cp}(a) shows a pressure coefficient on the airfoil surface, $C_{\rm p}$, in the strong input cases with $\Fp=1.0$, $6.0$, and $10$, where the other cases are omitted for the ease of visualization.
The horizontal axis shows a chordwise position normalized by the chord length.
The upper half part corresponds to the suction side;
and the lower part corresponds to the pressure side.
The black line represents the noncontrolled case, where the very weak peak (suction peak), $C_{\rm p}\approx -1.0$, appears near the leading edge of the suction side.
In contrast, the controlled cases show strong suction peaks, $C_{\rm p}\approx -3.5$, near the leading edge.
Specifically, it is observed that the suction peaks of the $\Fp=6.0$ and $10$ cases are stronger (smaller $C_{\rm p}$) than that of the case with $\Fp=1.0$.
As such, in the present controlled cases, the smaller separation bubble leads to the stronger suction peak at the leading edge, thereby resulting in the stronger thrust and smaller drag coefficients.
Another noteworthy feature is that in Fig.~\ref{fig:freq-sep}(a), the separation-controlled flows of $6.0\leq \Fp\leq 20$ separates earlier (at more upstream) than those of $\Fp=1.0$ and $30$, in the vicinity of the trailing edge.
Although the flow attachment near the trailing edge generally contributes to the lift increase, it does not strongly affect the aerodynamic performance in the present condition as seen in Fig.~\ref{fig:freq-clcd}(a).
Nonetheless, it is interesting that the larger separation bubble would result in the larger attached region near the trailing edge, which will be revisited later in Sec.~\ref{subsec:coherent} in the context of coherent spanwise vortex structures on the suction side.
We observe the similar trend in the weak input cases ($\Cm=2.0\dd{-5}$) in Fig.~\ref{fig:freq-sep}(b).
The small separated flow regions (separation bubbles) appear near both leading and trailing edges when the separation is suppressed in $6.0\leq\Fp\leq 20$.
Furthermore, in Fig.~\ref{fig:freq-cp}(b), it is observed that the size of the separation bubble with $\Fp=6.0$ is smaller than that with $\Fp=10$.
Note that in the other cases with $\Fp=1.0$ and $30$, the flow separates on the most of the airfoil surface, which suggests that the flow is in the stall condition.


To summarize, when the aerodynamic coefficient is significantly improved by the control, the separation is effectively suppressed in the time- and spanwise-averaged fields and therefore the separation control is successful.
The controlled flows generally exhibit a laminar separation bubble near the leading edge as well as a small separated region near the trailing edge, which is the essential feature of the attached flows in the present control conditions.
Furthermore, the controlled flows with the optimal frequency range of $6.0\leq\Fp\leq 20$ shows the smaller separation bubble compared with the cases with $\Fp=1.0$ and $30$.
This is explained by the existence of a strong pressure suction peak at the leading edge, which contributes to a drag reduction and therefore improving $C_{\rm{L}}/C_{\rm D}$.
In the next subsection, unsteady characteristics of the controlled flows are investigated.

\subsection{Unsteady characteristics of separation-controlled flows}\label{subsec:unsteady}
\subsubsection{Instantaneous flow fields}\label{subsubsec:ins}
\begin{figure}[htbp!]
\centering
    \renewcommand{\func}{instant-mv1}
    \renewcommand{\size}{0.4}
\ifCONDITION
    \subfloat[][\hcasesjaa]{\includegraphics[width=\size\textwidth]{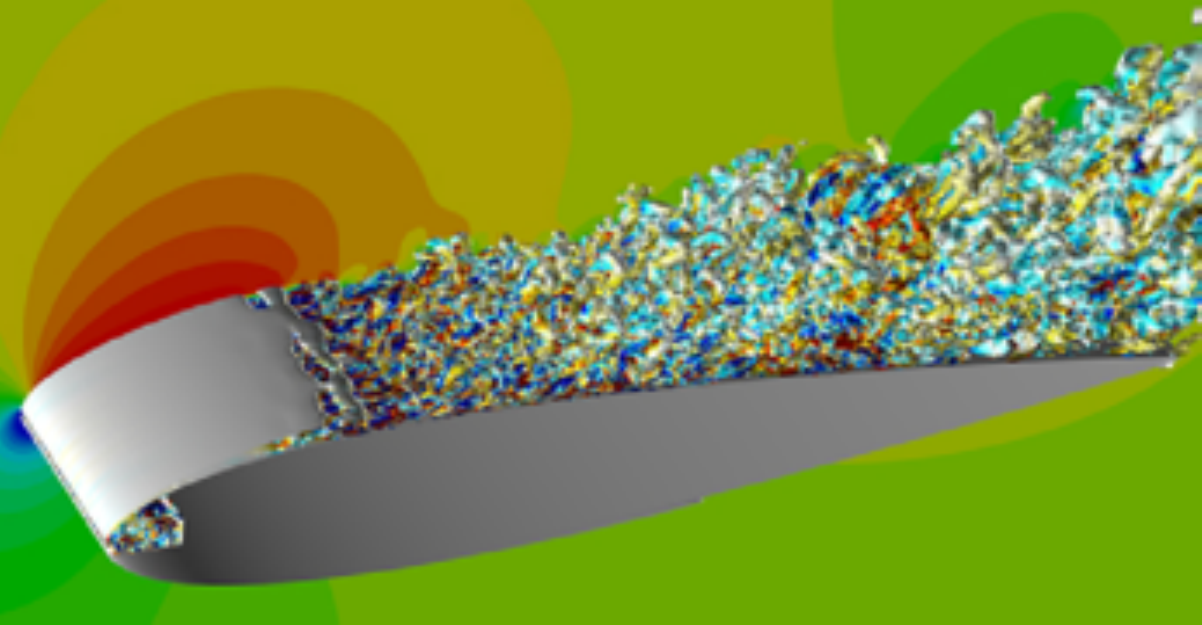}}
    \subfloat[][\hcasesjba]{\includegraphics[width=\size\textwidth]{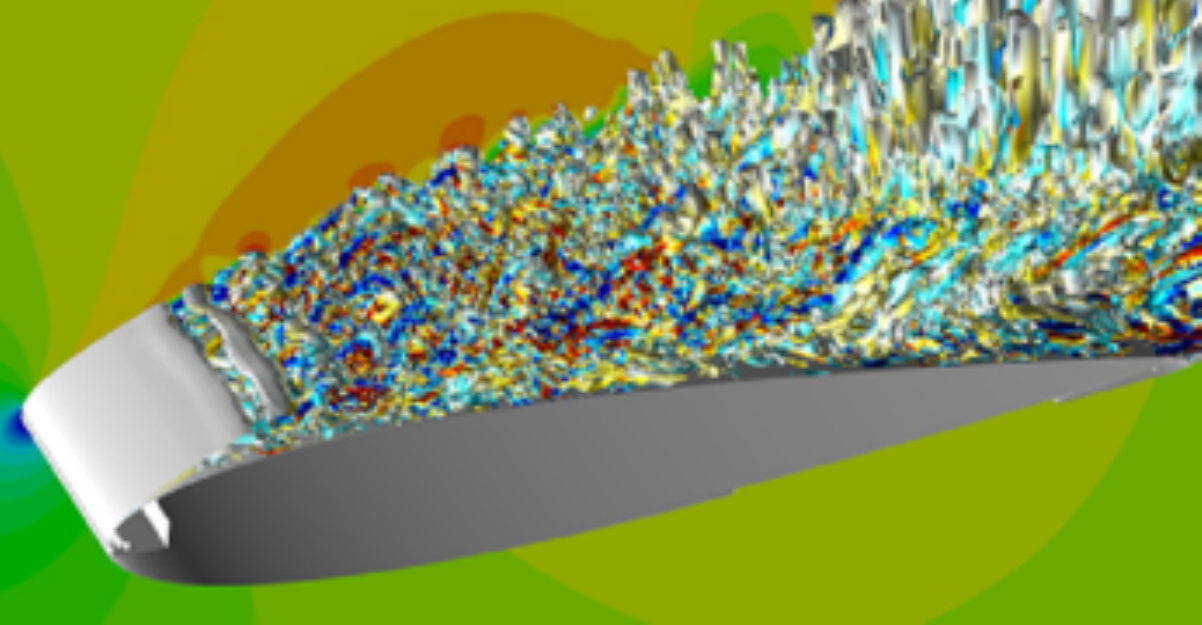}}\\
    \subfloat[][\hcasesjab]{\includegraphics[width=\size\textwidth]{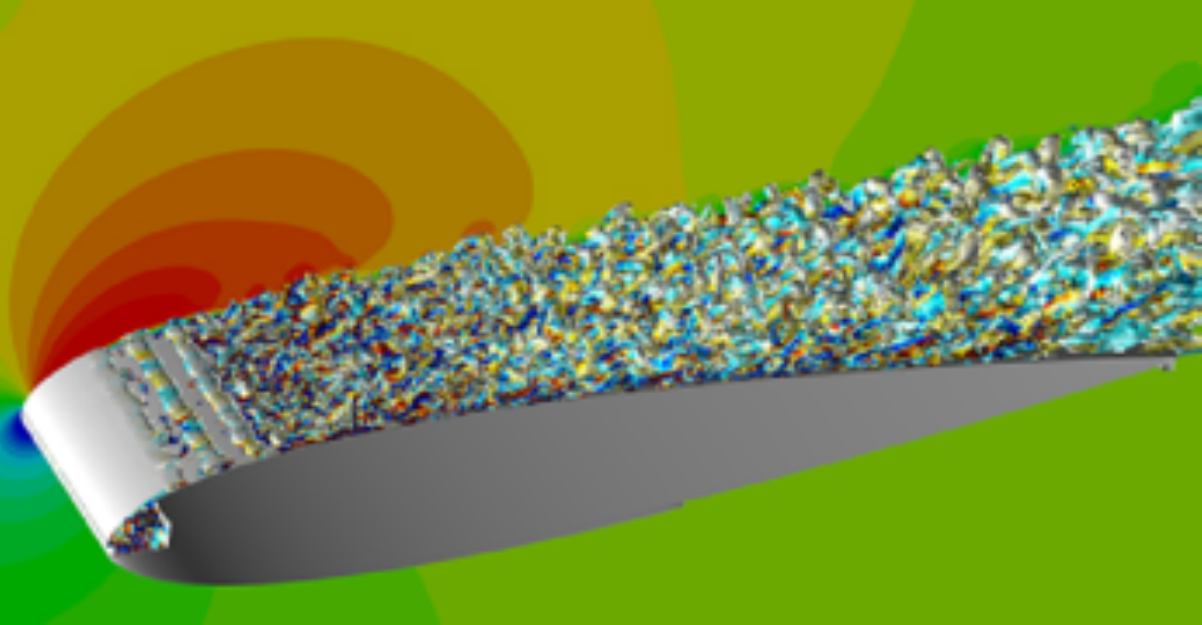}}
    \subfloat[][\hcasesjbb]{\includegraphics[width=\size\textwidth]{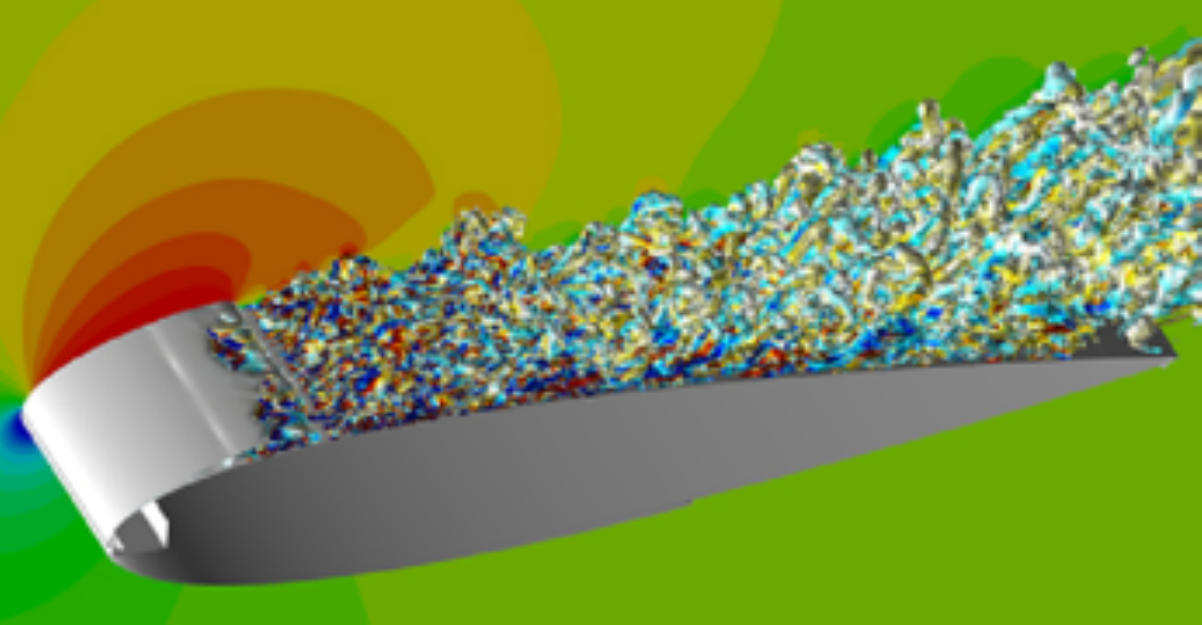}}\\
    \subfloat[][\hcasesjac]{\includegraphics[width=\size\textwidth]{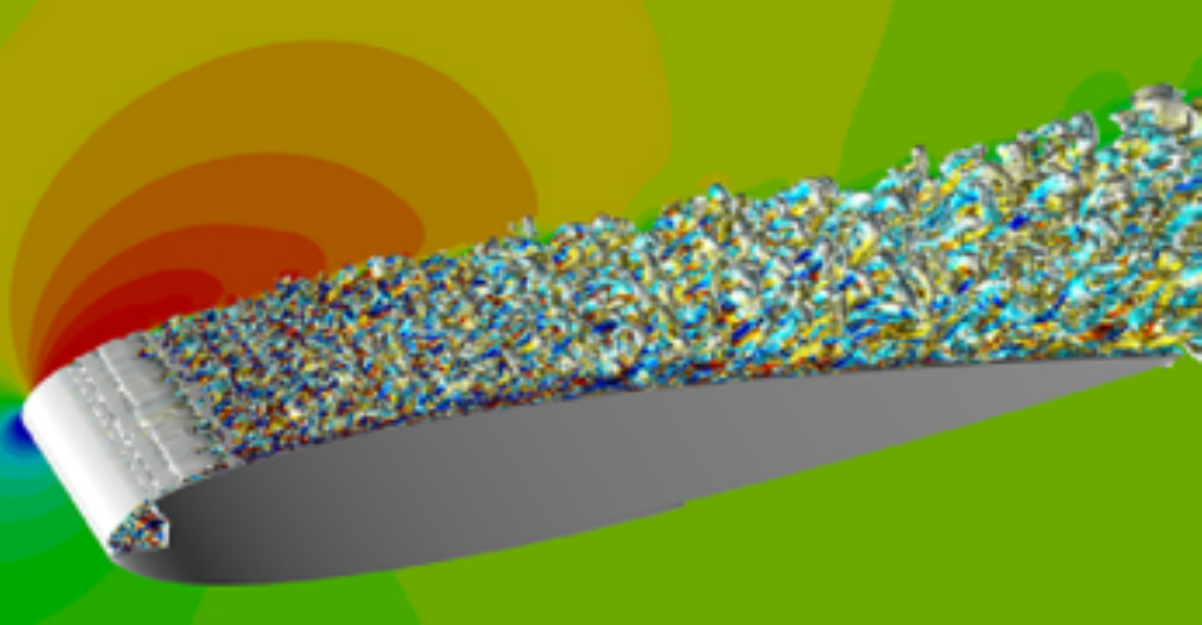}}
    \subfloat[][\hcasesjbc]{\includegraphics[width=\size\textwidth]{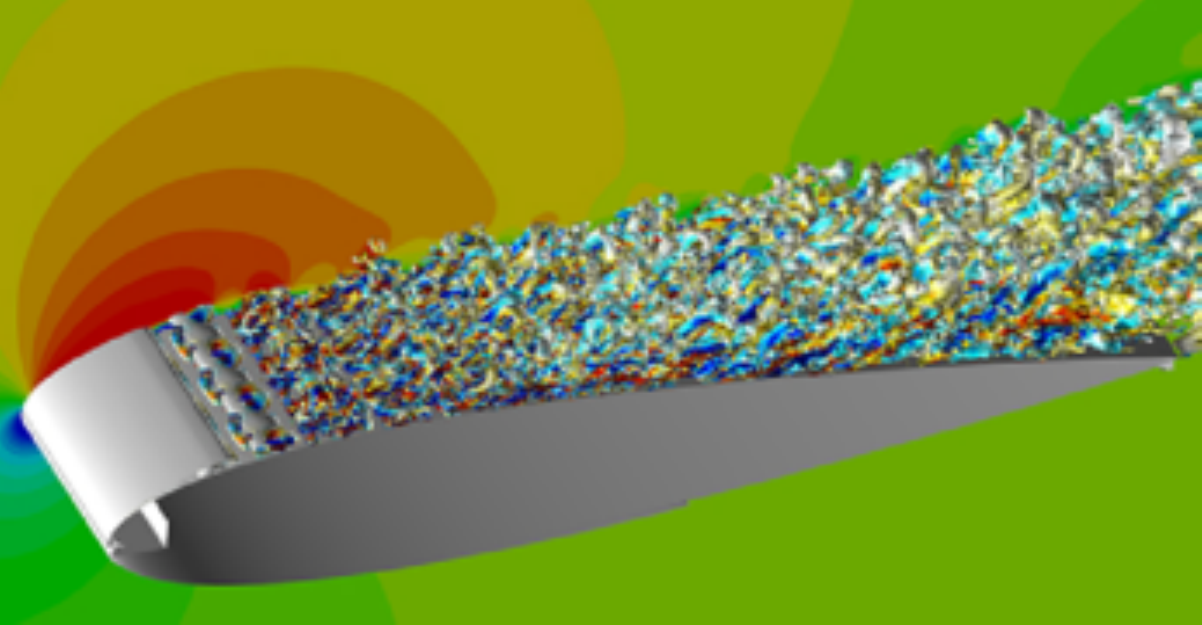}}
\else
\fi
    \caption{
Visualization of the instantaneous flow fields.
Isosurfaces represent a second invariant of the velocity gradient tensor (nondimensional value of $0.3$), which is colored by the vorticity in the chordwise direction (the isosurface is visualized using every two grid points in each direction.) 
The contour plane perpendicular to the spanwise direction is colored by the normalized chordwise velocity component, $0\leq u/u_{\infty}\leq 1.5$.
}\label{fig:ins}
\end{figure}

Figure~\ref{fig:ins} shows instantaneous flow fields of the separation-controlled cases, at $t=19.0$.
Hereinafter, we often compare three different frequencies, i.e., $\Fp=1.0$, $6.0$, and $10$, for each input momentum, where the separation-control ability in the optimal frequency regime, i.e., $6.0\leq\Fp\leq 20$, is effectively characterized.
In Fig.~\ref{fig:ins}, isosurfaces represent a second invariant of the velocity gradient tensor, which is colored by the vorticity in the chordwise direction.
Note that the isosurface is visualized using every two grid points in each direction for saving storage and the burden of post-process.
The contour plane perpendicular to the spanwise direction is colored by the normalized chordwise velocity component, $0\leq u/u_{\infty}\leq 1.5$ (blue to red).
In the strong input cases, all frequency cases (Figs.~\ref{fig:ins}(a), (c), and (e)) achieve the attached flows.
In these controlled cases, a laminar flow separates near the leading edge and becomes turbulent, which afterwards reattaches to the surface, and a turbulent boundary layer develops.
The similar flow picture is observed in the other controllable cases with a weak input (Figs.~\ref{fig:ins}(d) and (f)).
It is again emphasized that three-dimensional vortical structures cover the most of the airfoil surface when the separation is suppressed, which is a typical flow feature of the present separation-controlled cases.
On the other hand, the weak input case with $\Fp=1.0$ (Fig.~\ref{fig:ins}(b)) is not able to suppress the separation, where the flow largely separates in the vicinity of the leading edge.
These observations are consistent with the time-averaged flows discussed in Sec.~\ref{subsec:timeave}.

%
%

\begin{figure}[htbp!]
\centering
    \renewcommand{\size}{0.5}
\ifCONDITION
    \includegraphics[width=\size\textwidth]{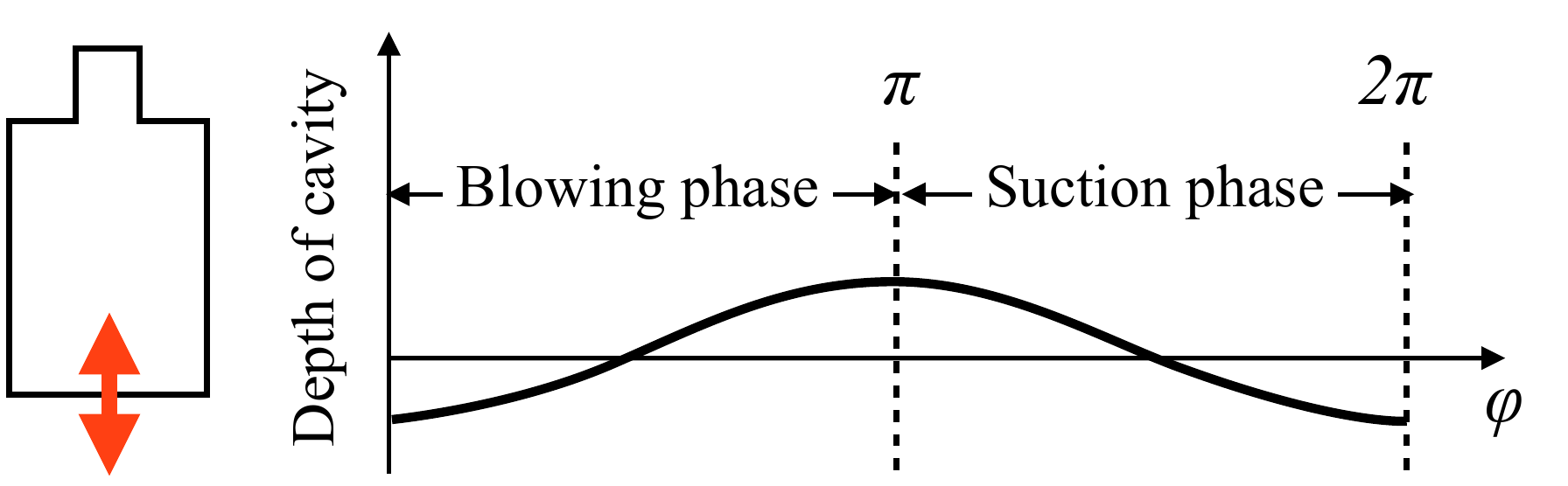}
    \renewcommand{\size}{0.75}
    \subfloat{\includegraphics[width=\size\textwidth]{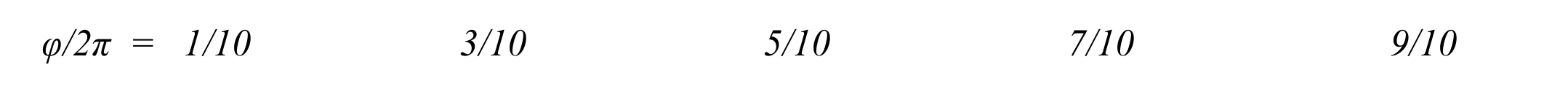}}\vspace{-0.1cm}
    \renewcommand{\func}{mv2}
    \subfloat{\includegraphics[width=\size\textwidth]{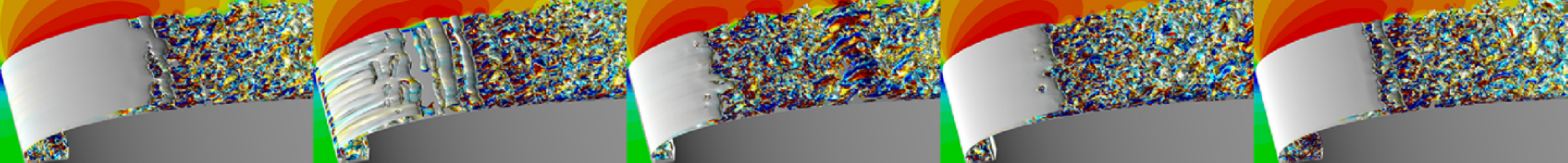}}
    \renewcommand{\func}{mv3}
    \setcounter{subfigure}{0}
    \subfloat[\hcasesjaa]{\includegraphics[width=\size\textwidth]{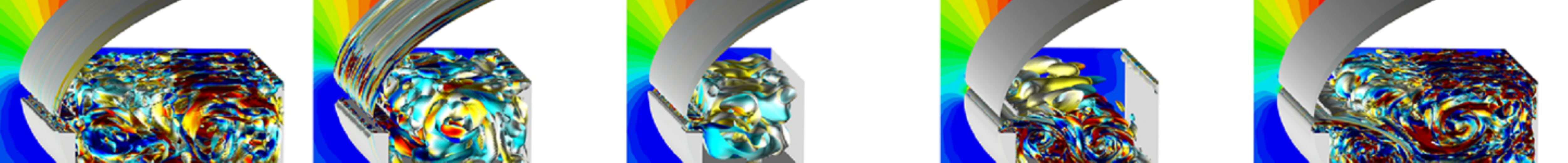}}
    \renewcommand{\func}{mv2}
    \subfloat{\includegraphics[width=\size\textwidth]{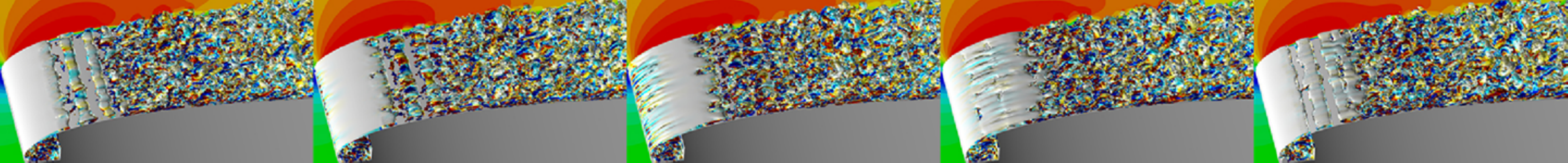}}
    \renewcommand{\func}{mv3}
    \setcounter{subfigure}{1}
    \subfloat[][\hcasesjab]{\includegraphics[width=\size\textwidth]{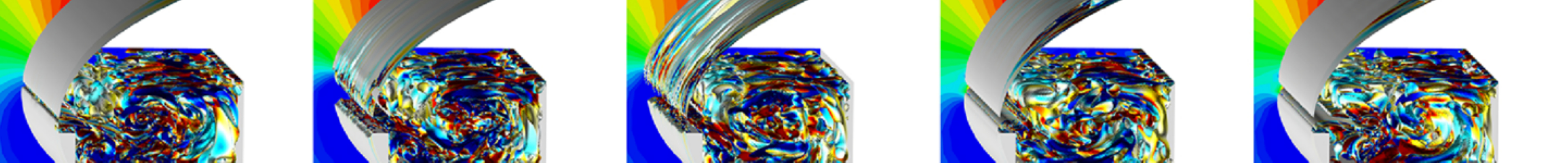}}
    \renewcommand{\func}{mv2}
    \subfloat{\includegraphics[width=\size\textwidth]{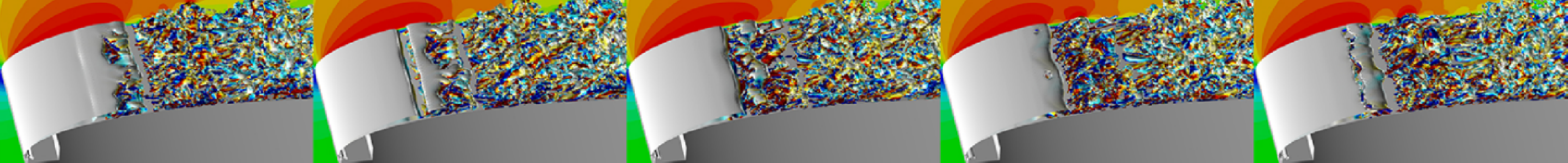}}
    \renewcommand{\func}{mv3}
    \setcounter{subfigure}{3}
    \subfloat[][\hcasesjbb]{\includegraphics[width=\size\textwidth]{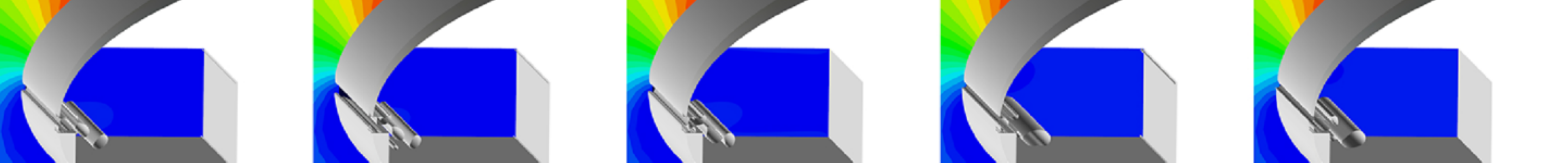}}
\else
\fi
    \caption{
Visualization of the phase-locked view of the instantaneous flow fields, based on $\Fp$.
An isosurface represents a second invariant of the velocity gradient tensor (nondimensional value of $0.3$), which is colored by the vorticity in the chordwise direction (the isosurface is visualized using every two grid points in each direction.) 
The contour plane perpendicular to the spanwise direction is colored by the normalized chordwise velocity component, $0\leq u/u_{\infty}\leq 1.5$.
The top and bottom figures in each case sequentially show the zoom-up views of the phase-locked flow around the leading edge and SJ cavity, respectively. }\label{fig:ins-phase}
\end{figure}

Next, Fig.~\ref{fig:ins-phase} shows instantaneous flows at different phases that are based on the actuation frequency $\Fp$:
$\varphi/2\pi=1/ 10, 3/ 10, 5/ 10, 7/ 10$, and $9/ 10$.
The SJ is blowing during the phases at $0\leq \varphi\leq \pi$;
the SJ is in the suction phase at $\pi< \varphi\leq 2\pi$.
The top and bottom figures show zoom-up views of the phase-locked instantaneous flows around the leading edge and inside the SJ cavity, respectively.
In the strong input cases, the bottom figures of Figs.~\ref{fig:ins-phase}(a) and (b) show turbulent structures inside the SJ cavity.
Those are then injected to the outer flow, which strongly disturbs a laminar separated shear layer near the leading edge.
Accordingly, the separation bubble in the top figures of each case periodically contracts and expands during $0\leq \varphi\leq 2\pi$.
When the separation bubble contracts the most ($\varphi/2\pi=3/10$ in Fig.~\ref{fig:ins-phase}(a) and $\varphi/2\pi=1/10$ in Fig.~\ref{fig:ins-phase}(b)), spanwise coherent structures emerge immediately behind the separated shear layer.
Although those coherent structures contain spanwise disturbances, their shapes are almost uniform in the spanwise direction, and thus, they are called spanwise coherent structures hereafter.
Most importantly, in the case of $\Fp=6.0$ at $\varphi/2\pi=3/10$, the spanwise coherent structures appear at the more upstream location than those in the case of $\Fp=1.0$ at $\varphi/2\pi=1/10$. 
Therefore, a turbulence transition takes place at the more upstream location in the case of $\Fp=6.0$ than $\Fp=1.0$, and thus, the smaller separation bubbles are maintained.

By contrast, the weak input case (Fig.~\ref{fig:ins-phase}(d)) generates spanwise-uniform (two-dimensional) vortical structures inside the cavity. This is also obeserved in the other frequency cases, which are not presented here for brevity.
Therefore, it is suggested that the vortical structures inside the cavity are strongly dependent on the input momentum $\Cm$ provided that $\Cm$ is between $2.0\dd{-3}$ and $2.0\dd{-5}$ as in the present study.
The weak input momentum case results in a spanwise uniform structure inside the cavity as was pointed out by Ref.~\onlinecite{Okada2012a}; nevertheless the outer flows around the airfoil become turbulent similarly to the cases with the strong input momentum.
The spanwise coherent structures emerge behind the laminar separated shear layer, which contain spanwise fluctuation and break down into turbulence afterwards.

The emergence of spanwise coherent structures behind the laminar shear layer is observed in both weak and strong input cases, which suggests a vortex breakdown mechanism triggered by the Kelvin-Helmholtz (KH) instability.
Such a linear instability is well known as a turbulent transition in laminar separation bubbles, c.f., Refs.~\onlinecite{Rist2002,Rist2006,Simoni2014}.
We will revisit the KH-instability-related mechanism later in the context of the optimal $\Fp$ for separation-control ability and contribution to a momentum exchange.

\subsubsection{Turbulent transition in the separation-controlled flows}\label{subsubsec:tke}

\begin{figure}[htbp!]
\centering
   \renewcommand{\func}{TKE}     
   \renewcommand{\size}{1.0}
\ifCONDITION
   \subfloat[][$\Cm=2.0\dd{-3}$]{\includegraphics[width=\size\textwidth]{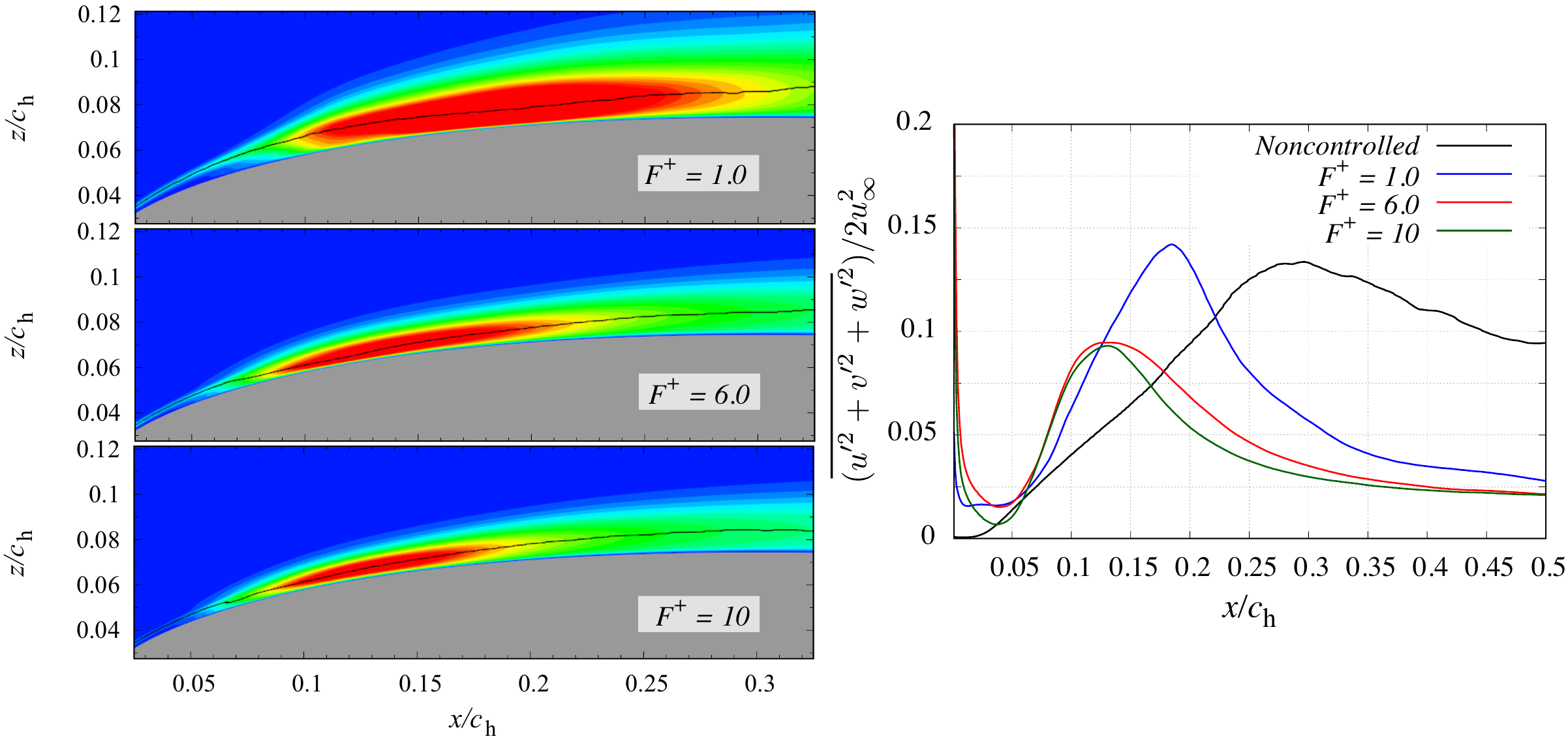}}\\
   \subfloat[][$\Cm=2.0\dd{-5}$]{\includegraphics[width=\size\textwidth]{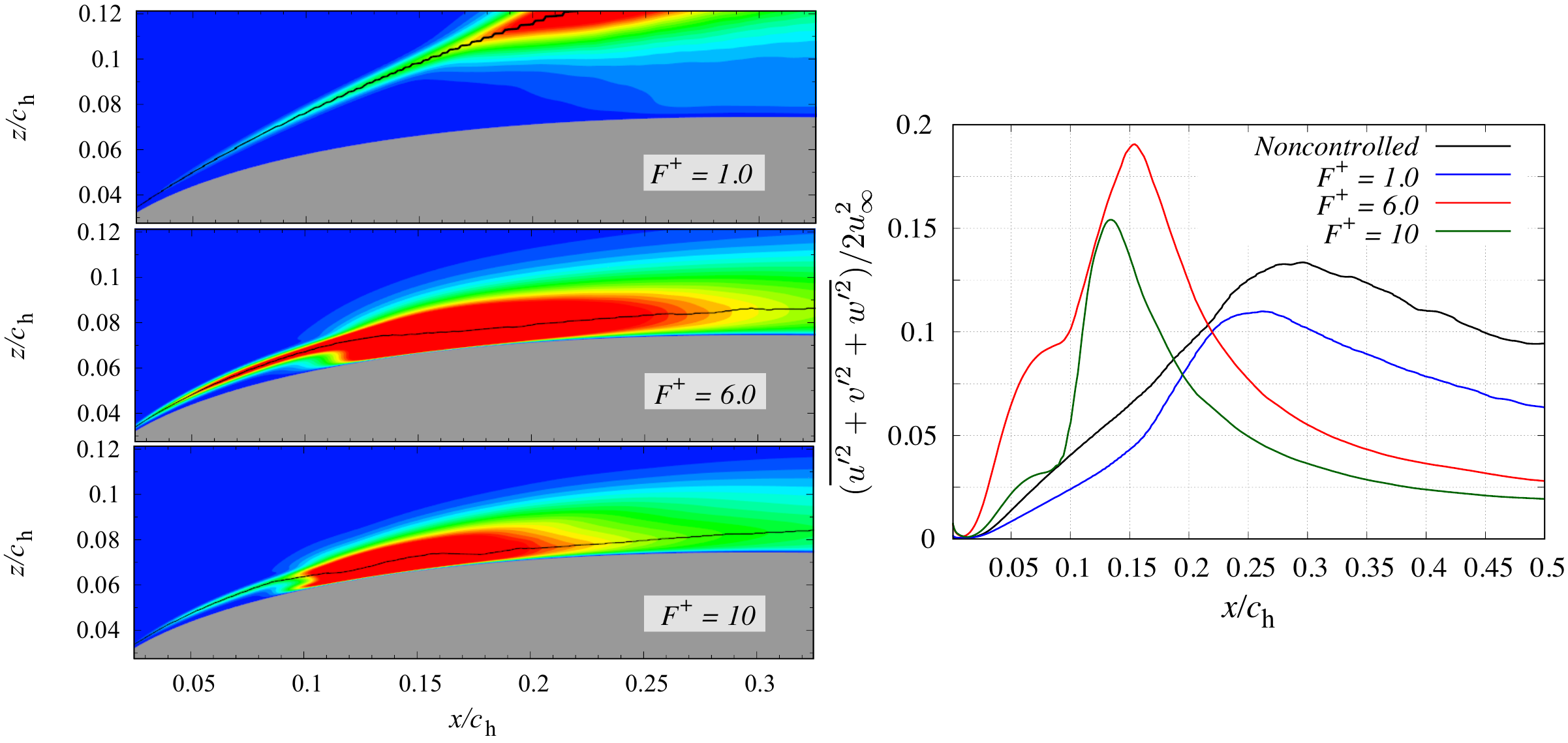}}
\else
\fi
\caption{
Visualization of the TKE in time- and spanwise-averaged fields of (a)$\Cm=2.0\dd{-3}$ and (b)$\Cm=2.0\dd{-5}$.
The left contour plots show the TKE distribution normalized by the freestream velocity (i.e., $\overline{u'^2+v'^2+w'^2}/2u_\infty^2$) between $0$ and $0.075$ with 20 contours, where the black lines indicate the positions where the TKE becomes maximum on each wall-normal grid line (in the $\zeta$ direction) that is approximately perpendicular to the airfoil surface, which is called the TKE-max line.
The line plots in the right side represent the TKE distribution on the black line (TKE-max line) in the left figures, which therefore shows the maximum TKE values on each wall-normal grid line, which is called the TKE-max plot.
}\label{fig:TKEmax}
\end{figure}
Next, we investigate the location of turbulent transition in the separation-controlled cases, based on the turbulent kinetic energy (TKE) distribution.
Figure~\ref{fig:TKEmax} shows the TKE distribution normalized by the freestream velocity in the time- and spanwise-averaged fields, i.e., $\overline{u'^2+v'^2+w'^2}/2u_\infty^2$.
In each input momentum case, the left three figures show zoom-up views of the normalized TKE with 20 contours between $0$ and $0.075$.
The black lines represent the positions where the TKE becomes maximum on each wall-normal grid line (in the $\zeta$ direction) that is approximately perpendicular to the airfoil surface, which is called a TKE-max line hereinafter.
The line plots in the right side represent the TKE distribution on the TKE-max line of left figures (black line), which therefore shows the spatial distribution of the maximum TKE values on each wall-normal grid line.
We herein call this plot as a TKE-max plot.
In each controlled case, the higher TKE region (colored by yellow to red) is observed near the leading edge.
This region approximately corresponds to the location where the spanwise coherent structures are released from the laminar separated shear layer.
It is clear that in the strong input cases (Fig.~\ref{fig:TKEmax}(a)), the size of the higher TKE region becomes smaller as $\Fp$ increases, which corresponds to the reduction of the separation bubble size and the improvement of the separation control ability.
Furthermore, the TKE-max plots of $\Fp=6.0$ and $10$ show smaller peaks in the more upstream position compared with that of $\Fp=1.0$.
These results support the observation that the $\Fp=6.0$ and $10$ cases achieve a smooth and quick turbulent transition compared with the $\Fp=1.0$ case as is also shown in the instantaneous flow fields (Figs.~\ref{fig:ins} and \ref{fig:ins-phase}).
Note that the similar localized peak in the TKE distribution has been observed in the turbulent transition process by Refs.~\onlinecite{Alam2000,Dandois2007}.
The weak input cases (Fig.~\ref{fig:TKEmax}(b)) show the similar trend, where the case with $\Fp=10$ shows smaller and more upstream peak in the TKE-max plots.

As such, the location of turbulent transition in the separated-controlled flows is different for each $\Fp$ and $\Cm$.
The TKE-max plots in Fig.~\ref{fig:TKEmax} support the observation that in the instantaneous flow fields, the quick and smooth turbulent transition is closely related to the formation of the smaller separation bubble, thereby achieving higher control capabilities. 

\subsubsection{Temporal spectral analysis of velocity fluctuation}\label{subsubsec:psd}
\begin{figure}
\centering
    \renewcommand{\func}{fft}
    \renewcommand{\size}{1.0}
\ifCONDITION
    \includegraphics[width=\size\textwidth]{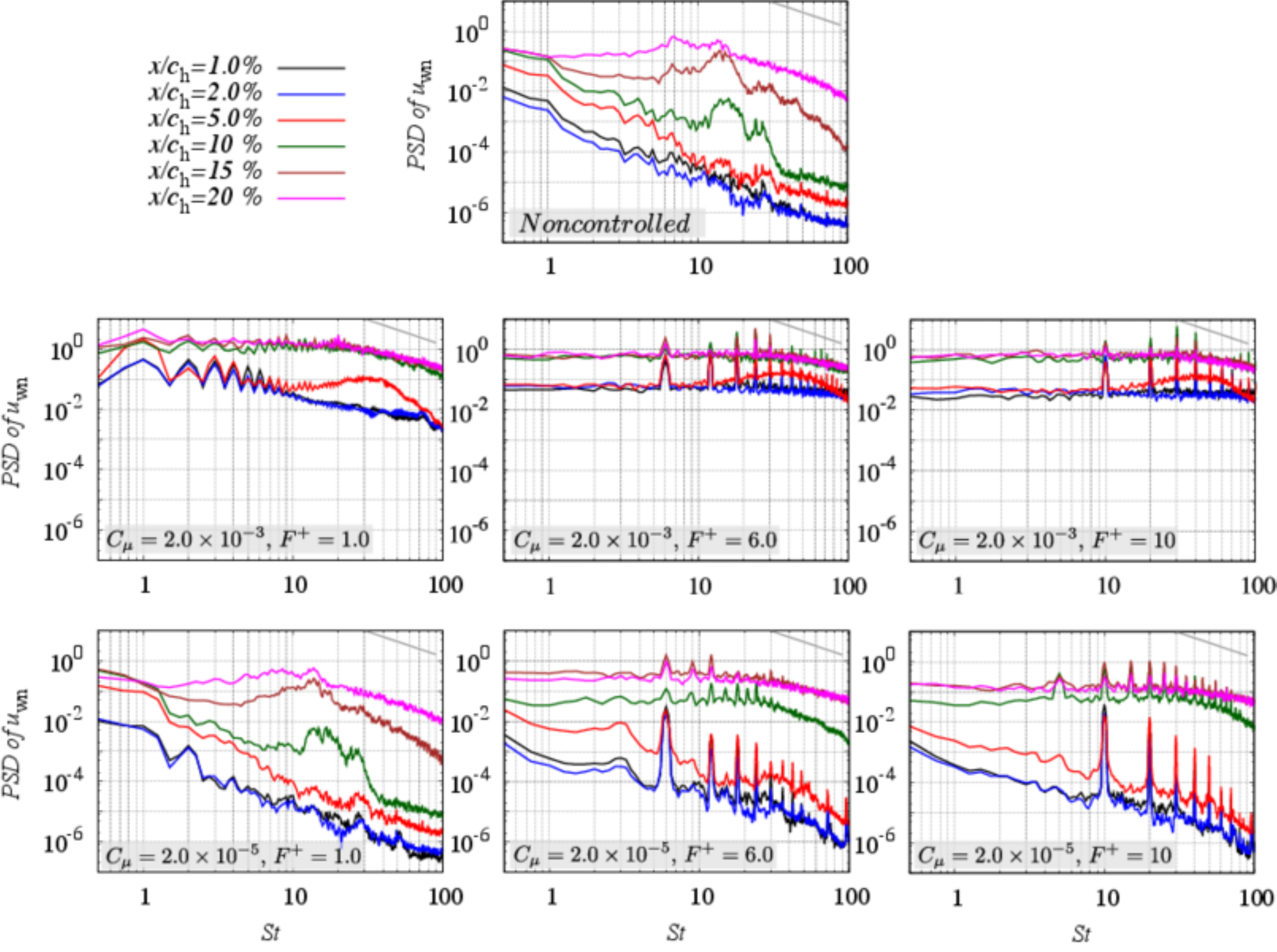}
\else
\fi
    \caption{
The PSD of the wall-normal velocity component is plotted.
The horizontal axis shows a frequency normalized by the freestream velocity and the chord length, $\St$.
The line plots colored in black to magenta show the PSD at $x/c_{\text{h}}=0.01$ to $0.2$.
All sampling points are defined on the TKE-max line in Fig.~\ref{fig:TKEmax}.
The grey straight line in the right top corner of each figure shows Kolmogorov's $5/3$ law.
    }\label{fig:freq-fft}
\end{figure}

Figure~\ref{fig:freq-fft} shows the PSD of the wall-normal velocity component ($u_{\text{wn}}$) at $x/c_{\text{h}}=0.01$, $0.02$, $0.05$, $0.1$, $0.15$, and $0.2$.
The details of the post process are represented in Sec.~\ref{sec:method}.
A straight grey line visualized in the right top corner of each figure indicates Kolmogorov's $-5/3$ law.
The horizontal axis shows the Strouhal number ($St$) that is a nondimensional frequency normalized by the freestream velocity and the chord length.
Note again that the normalization of $St$ is in the same way as the actuation frequency $\Fp$.
The location of the sampling point is defined so that the TKE takes its maximum on the wall-normal grid line at each $x/c_{\text{h}}$.
Therefore, all the sampling points are located on the TKE-max lines (the black lines in the left figures of Fig.~\ref{fig:TKEmax}).

In the strong input cases ($\Cm=2.0\dd{-3}$), PSD profiles show peaks at each actuation frequency, i.e., $\St=\Fp$.
These peaks remain strong further downstream, for example, at $x/c_{\text{h}}=0.2$.
This indicates that the flow is periodically disturbed even in the turbulent boundary layer, and thus the existence of a characteristic flow structure with the period of $\St=\Fp$ is expected.
In addition, the PSD profiles show strong peaks at harmonic frequencies of  $St = n\Fp$ ($n=1,2,\ldots$).  For example, $St=6.0$, $12.0$, $18.0$, $\cdots$, $6n$, $\cdots$ are observed in the case of $\Fp=6.0$.  
Such harmonic frequencies are generally associated with a nonsinusoidal periodic motion in the flow structures, which will be discussed in more detail in the next section based on the phase-averaging procedure with the time period of $\Fp$.
The other significant feature is that at $x/c_{\text{h}}=0.05$ (red lines), the PSD of $\St\approx 40$ is widely increased, irrespective of the actuation frequency of $\Fp$.
The location of $x/c_{\text{h}}=0.05$ approximately corresponds to the first half of the laminar separation bubble.
Therefore, it is expected that a flow disturbance involving a high-frequency motion is selectively amplified in the vicinity of the leading edge, which would be related to the KH instability as was mentioned in the previous section.

In the weak input cases ($\Cm=2.0\dd{-5}$), the cases with $\Fp=6.0$ and $10$ show the similar feature as in the strong input cases.
The PSD profiles show peaks at $\St=\Fp$ and its harmonics, even in the turbulent boundary layer that develops in the further downstream of separation bubbles.
In the vicinity of the leading edge ($x/c_{\text{h}}=0.01,0.02$, and $0.05$), the magnitude of the PSD of all frequencies is smaller than the strong input cases, which would be simply due to the weaker input momentum.
The smaller PSD in the higher frequency modes ($40\leq\St$) supports the fact that the weak input cases do not generate strong turbulent structures inside the SJ cavity, as is qualitatively discussed in the instantaneous flow fields (Fig.~\ref{fig:ins-phase}).
Finally, the weak input case with $\Fp=1.0$ is not able to suppress the separation, whose PSD profile is similar to that of the noncontrolled case.
In these uncontrollable cases, the magnitude of PSD at $\St\simeq 15$ is selectively amplified between $x/c_{\text{h}}=0.05$ and $0.2$, which is related to the KH instability in the separated shear layer.

As such, in the separation-controlled cases, the PSD profile shows peaks at the actuation frequency and its harmonics, i.e., $\St=n\Fp (n=1,2,\ldots)$, even within the turbulent boundary layer.
This implies the existence of a coherent periodic flow structure that emerges in the turbulent boundary layer.
We will further investigate such a coherent flow structure in the next section, which is strongly related to the mechanism of the chordwise momentum exchange.
The other remarkable finding is that in the strong input case, the PSD of $\St\simeq 40$ is selectively amplified, irrespective of the actuation frequency $\Fp$.
This is related to the KH instability in the separation bubble, which will be discussed later in detail with regards to the optimal range of $\Fp$  (Sec.~\ref{sec:chap2-control}).

%
%
\section{Chordwise momentum exchange in controlled flows}\label{sec:momentum}
An effective separation control needs a constant injection of momentum in the chordwise direction (chordwise momentum) in the vicinity of the airfoil surface.
In the present study, most of the chordwise momentum is not directly injected from the SJ as the orifice of SJ is normal to the airfoil surface, and the momentum coefficient is relatively small compared with the freestream value ($\mathcal{O}(10^{-3})$ and $\mathcal{O}(10^{-5})$ of the freestream momentum).
Therefore, as was mentioned in Sec.~\ref{sec:intro}, it is expected that the chordwise momentum is injected (entrained) from the freestream away from the airfoil surface, primarily using unsteady flow structures.
Such unsteady momentum exchange between the freestream and flows near the airfoil surface can be quantified by the Reynolds stress distribution.
Specifically, we focus on $-\overline{u'w'}$ that represents a correlation of the chordwise velocity fluctuation to the vertical velocity fluctuation.

\subsection{Overall phase decomposition of the Reynolds stress $-\overline{u'w'}$}\label{subsec:phase}
In this subsection, the phase averaging procedure is conducted for the unsteady flow fields to extract the coherent structures in the turbulent boundary layer.
The Reynolds stress of the overall fluctuation, $-\overline{u'w'}$, is decomposed into a periodic ($-\overline{\tilde{u}_{\varphi}\tilde{w}_{\varphi}}^{\varphi}$) and nonperiodic ($-\overline{u''w''}$; that is also called as a ``turbulent'' component) components based on $\Fp$, where  and  correspond to the contribution from the periodic and turbulent fluctuations, respectively.
The detail of the phase decomposition was described in Sec.~\ref{subsec:numerical-tools}.
These components are related as follows:
\begin{align}
\overline{u'w'}&=\overline{\tilde{u}_{\varphi}\tilde{w}_{\varphi}}^{\varphi}+\overline{u''w''}\label{eq:reyoverall}.
\end{align}
Note that all quantities represent spanwise-averaged values and the symbol $[\bullet]$ of them is omitted for the brevity.

\begin{figure}
\centering
    \renewcommand{\func}{decomp}
    \renewcommand{\size}{1.0}
\ifCONDITION
    \subfloat[][\hcasesjaa]{\includegraphics[width=\size\textwidth]{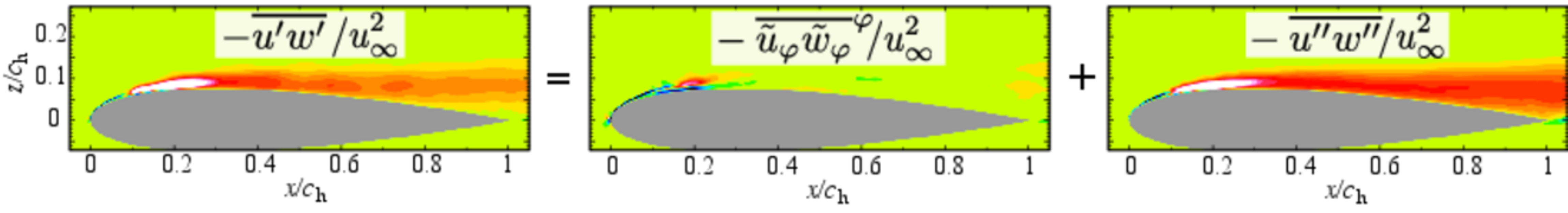}}\\
    \subfloat[][\hcasesjab]{\includegraphics[width=\size\textwidth]{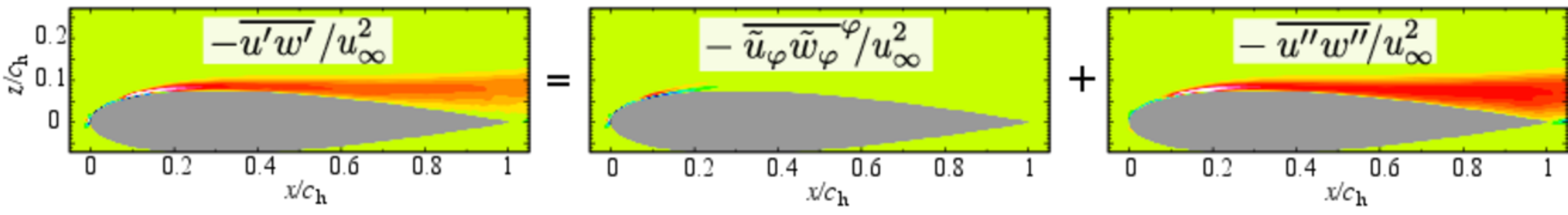}}
\else
\fi
    \caption{
Decomposition of the Reynolds stress is visualized: left column shows the overall component of $-\overline{u'w'}/u_{\infty}^2$, middle and right column show the periodic and turbulent components of $-\overline{\tilde{u}_{\varphi}\tilde{w}_{\varphi}}^{\varphi}/u_{\infty}^2$ and $-\overline{u''w''}/u_{\infty}^2$, where the contour range is fixed to be from $-0.0125$ to $0.0125$.
    }\label{fig:Rey}
\end{figure}

Figure~\ref{fig:Rey} shows the phase decomposition of the Reynolds stress in the $\Fp=1.0$ and $6.0$ cases with the strong and weak input ($\Cm=2.0\dd{-3}$).
The other controlled cases including the weak input cases show the similar trend, and thus they are omitted here.
The left column shows the overall component normalized by the freestream velocity ($-\overline{u'w'}/u_\infty^2$), middle and right columns show the periodic and turbulent components normalized by the freestream velocity ($-\overline{\tilde{u}_{\varphi}\tilde{w}_{\varphi}}^{\varphi}/u_\infty^2$ and $-\overline{u''w''}/u_\infty^2$). 
The contour range is fixed to be from $-0.0125$ to $0.125$.
Figure~\ref{fig:Rey} indicates that the turbulent component is dominant in the Reynolds stress so that an exchange of the chordwise momentum is mainly caused by the turbulent (nonperiodic) fluctuation.
This is a common feature among the separation-controlled flows, regardless of $\Fp$ and $\Cm$.
It is interesting that the periodic component $-\overline{\tilde{u}_{\varphi}\tilde{w}_{\varphi}}^{\varphi}$ is not dominant although in the PSD profile of Fig.~\ref{fig:freq-fft}, there is a clear indication that periodic coherent flow structures exist in the turbulent boundary layer.
Therefore, we will proceed to the further analysis on the relationship between the turbulent Reynolds stress and the periodic coherent flow structures in the next subsection.

%
%

\subsection{Coherent vortex structures and momentum exchange in phase-averaged fields}\label{subsec:coherent}
\begin{figure}
\centering
\renewcommand{\size}{0.3}
\ifCONDITION
   \subfloat{\includegraphics[viewport=0 0 28 350,clip,width=0.037\textwidth]{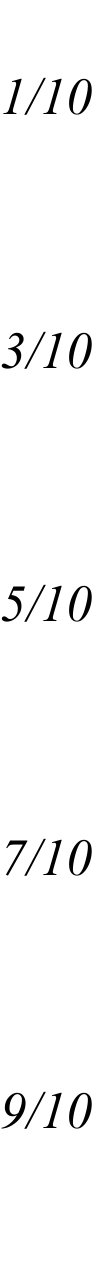}}
   \setcounter{subfigure}{0}
   \subfloat[][{\scriptsize $\Fp=1.0$,  $-\ang{\tilde{u}_{\varphi}\tilde{w}_{\varphi}}_{\varphi}/u_{\infty}^2$}]{\includegraphics[width=\size\textwidth]{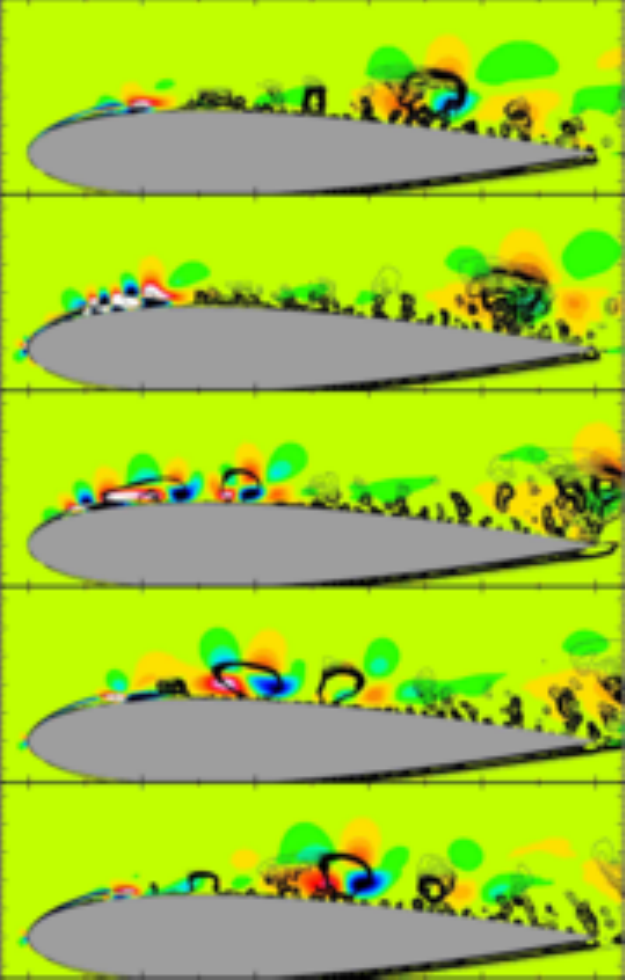}}
   \subfloat[][{\scriptsize $\Fp=1.0$,  $-\ang{u''w''}_{\varphi}/u_{\infty}^2$}]{\includegraphics[width=\size\textwidth]{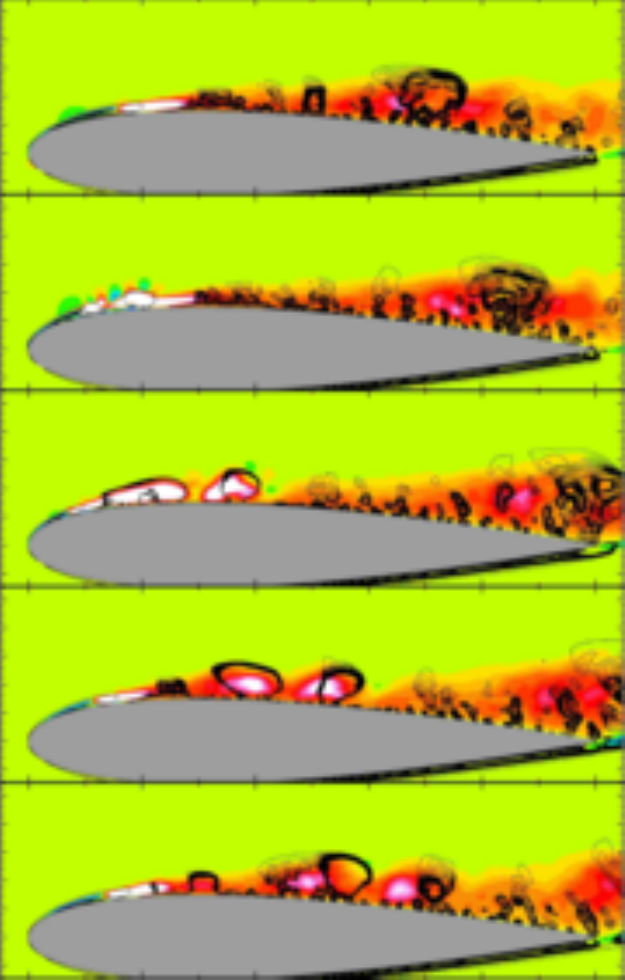}}\\
   \subfloat{\includegraphics[viewport=0 0 28 350,clip,width=0.037\textwidth]{phasel.pdf}}
   \setcounter{subfigure}{2}
   \subfloat[][{\scriptsize $\Fp=6.0$,  $-\ang{\tilde{u}_{\varphi}\tilde{w}_{\varphi}}_{\varphi}/u_{\infty}^2$}]{\includegraphics[width=\size\textwidth]{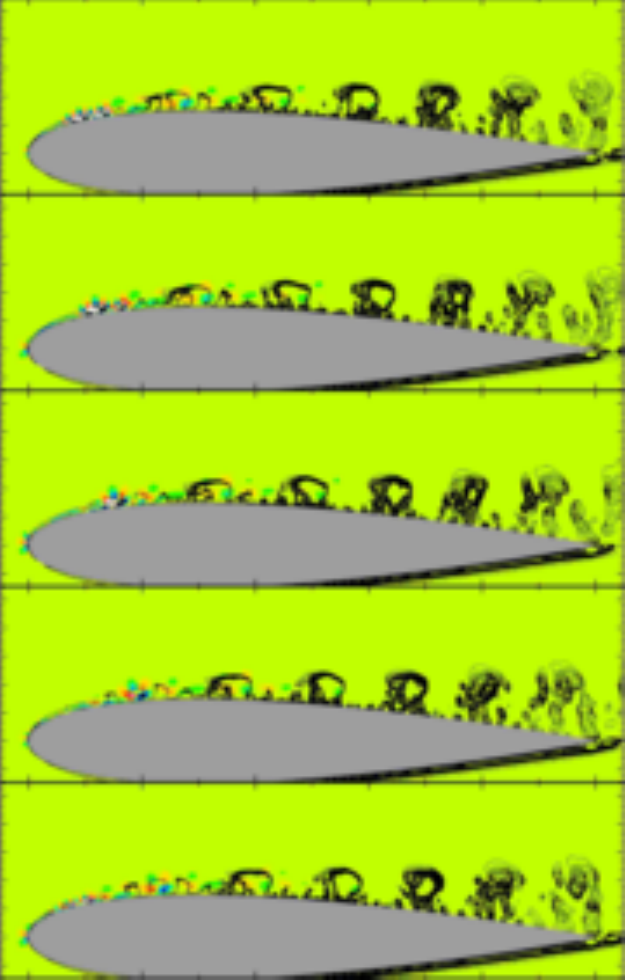}}
   \subfloat[][{\scriptsize $\Fp=6.0$,  $-\ang{u''w''}_{\varphi}/u_{\infty}^2$}]{\includegraphics[width=\size\textwidth]{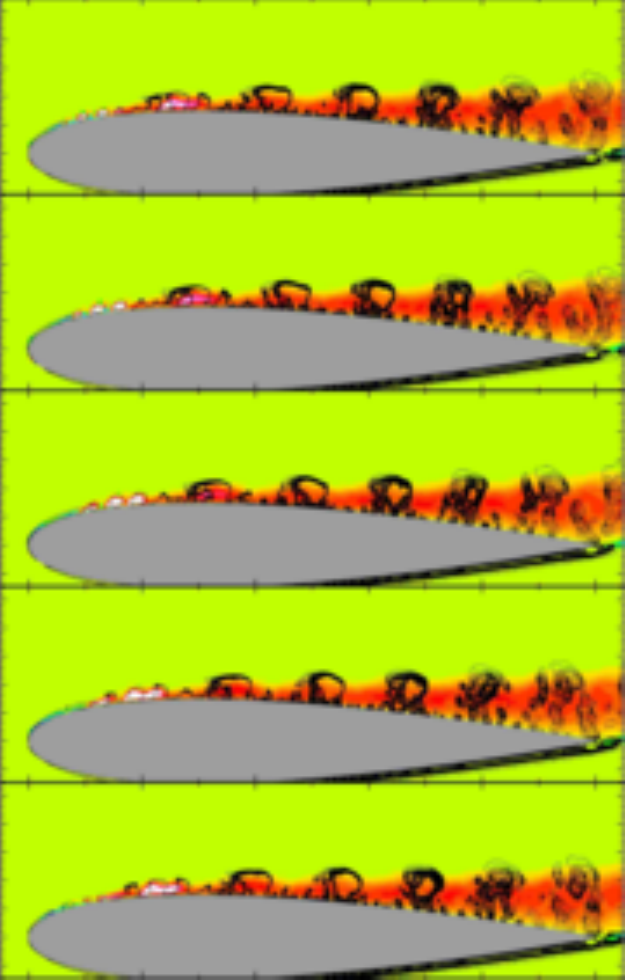}}
\else
\fi
\caption{
Visualization of the periodic and turbulent components of the Reynolds stress at each phase, i.e., $-\ang{\tilde{u}_{\varphi}\tilde{w}_{\varphi}}_{\varphi}/u_{\infty}^2$ and $-\ang{u''w''}_{\varphi}/u_{\infty}^2$, in the strong input cases of $\Cm=2.0\dd{-3}$ with $\Fp=1.0$ and $6.0$.
The left column shows the number of phases ($\varphi/(2\pi)=1/10$, $3/10$, $5/10$, $7/10$, and $9/10$).
Contour color represents each Reynolds stress between $-0.0125$ to $0.0125$.
Black contour lines show a second invariant of the velocity gradient tensor normalized by the freestream velocity (20 lines between $0.005$ and $3.5$), which is based on the phase- and spanwise-averaged velocity field at each phase, i.e., $\ang{u}_\varphi$, $\ang{v}_\varphi$, and $\ang{w}_\varphi$, which are the same between the left and right columns.
}\label{fig:Rey-phase-cmu2d-3}
\end{figure}

In this subsection, we investigate the relationship between the periodic coherent flow structure and phase-decomposition of the Reynolds stress.
First, the phase- and spanwise-averaged velocity fields, $\ang{u}_\varphi$, $\ang{v}_\varphi$, and $\ang{w}_\varphi$, are computed from Eqs.~\eqref{eq:phaseave} (we omit the symbol for the spanwise-averaging operator for brevity).
Figure~\ref{fig:Rey-phase-cmu2d-3} shows the strong input cases ($\Cm=2.0\dd{-3}$), where the black contour lines represent a second invariant of the velocity gradient tensor (between $0.02$ and $0.1$) based on $\ang{u}_\varphi$, $\ang{v}_\varphi$, and $\ang{w}_\varphi$.
These contour lines successfully extract the periodic coherent flow structures associated with $\Fp$, which are shown in the phases of $\varphi/(2\pi)=1/10$, $3/10$, $5/10$, $7/10$, and $9/10$.
Furthermore, the contour color (blue to red) represents the periodic and turbulent components of the Reynolds stress at each phase, $-\ang{\tilde{u}_{\varphi}\tilde{w}_{\varphi}}_{\varphi}$ and $-\ang{u''w''}_{\varphi}$, as is defined by Eq.~\eqref{eq:phased_double}.
Note that in Figs.~\ref{fig:Rey-phase-cmu2d-3}, (a) and (b) (or (c) and (d)) show each component of the Reynolds stress in the case of $\Fp=1.0$ (or $\Fp=6.0$), where the black contour lines are the same between Figs.~\ref{fig:Rey-phase-cmu2d-3}(a) and (b) (or Figs.~\ref{fig:Rey-phase-cmu2d-3}(c) and (d)).

In the present decomposition, a periodic component of the Reynolds stress, $-\ang{\tilde{u}_{\varphi}\tilde{w}_{\varphi}}_{\varphi}$, becomes strong in which the periodic flow motion with the period of $\Fp$ appears.
In Figs.~\ref{fig:Rey-phase-cmu2d-3}(a) and (c), periodic coherent flow structures are identified by black contour lines.
A separated shear layer is visualized near the leading edge, which expands and contracts periodically.
During $\varphi/(2\pi)=1/10$ to $3/10$, the separated shear layer becomes the shortest, emitting multiple small vortex structures.
These vortex structures convect in the downstream direction, and one of them remains strong further downstream as a coherent vortex structure. 
Such a coherent vortex is generated via the merging and diffusion process of small vortices that are released from the separation bubble.
The periodic component of the Reynolds stress, $-\ang{\tilde{u}_{\varphi}\tilde{w}_{\varphi}}_{\varphi}$, exhibits a quadrupole distribution surrounding each coherent vortex structure.
Such a quadrupole distribution of the Reynolds stress generally appears around the strong two-dimensional (spanwise uniform in this study) vortex structures. 
As such, the phase-averaged visualizations illustrate that the separation bubble periodically oscillates in the period of $\Fp$, and the small vortex structures are emitted, and eventually the one of which remains strong further downstream.
The resultant coherent vortex consecutively convects along the airfoil surface and is released from the trailing edge periodically (in the same period as $\Fp$).
The periodic fluctuation of the aerodynamic coefficients in Fig.~\ref{fig:CLCDhistory} is related to the generation of the coherent vortex structure.
Indeed, the number of coherent vortices approximately corresponds to $\Fp$. 
Furthermore, in the vicinity of the trailing edge, the $\Fp=1.0$ case shows strong coherent vortex structure more clearly than the $\Fp=6.0$ case.
Because the $\Fp=1.0$ case attains smaller separated region near the trailing edge in Fig.~\ref{fig:freq-sep}, it is expected that the strong coherent vortex in the lower actuation frequency ($\Fp=1.0$) can suppress the separation near the trailing edge more effectively.
The similar trend has been reported in the previous study for the difference in between ``{\it MF }'' and ``{\it HF }'' (middle and high frequencies) cases in Ref.~\onlinecite{Zhang2015}.
Finally again, the significant finding so far is that the periodic component of the Reynolds stress is localized only around the coherent vortices, which, however, is not dominant on the most of the suction side of the airfoil.

On the other hand, the turbulent component of the Reynolds stress becomes strong where the unsteady flow motion is nonperiodic (not related to the period of $\Fp$).
Such unsteady flow motion frequently appears together with the three-dimensional turbulent vortex structures.
In Figs.~\ref{fig:Rey-phase-cmu2d-3}(b) and (d), the strong turbulent component covers almost all the airfoil surface in each phase, which supports the discussion on the decomposition of the total Reynolds stress in Fig.~\ref{fig:Rey}. 
In particular, the turbulent component is locally enhanced inside the small vortex structures that are emitted from the separation bubble in between $\varphi/(2\pi)=1/10$ and $3/10$, where the strong three-dimensional fluctuation emerges due to turbulent transition.
The important finding is that such a locally-enhanced turbulent component convects downstream together with the coherent vortex.
In addition, the turbulent component is dominant and further stronger than the periodic component.
This indicates that the strong turbulent component is entrained by the coherent vortex that convects in the downstream direction, thereby contributing to the chordwise momentum exchange by the three-dimensional turbulent vortex structure.
To this end, the formation of the periodic coherent vortex structures is significant in terms of the contribution to the chordwise momentum exchange via the entrained turbulent component of the Reynolds stress all over the airfoil surface, rather than the direct contribution via the periodic component of the Reynolds stress.

As such, the coherent vortex structure is generated with its period of $\Fp$ through the merging and diffusion process of small vortex structures that are emitted from the separated shear layer.
The periodic component of the Reynolds stress exhibits a quadruple distribution around the coherent vortex, whose strength is not as strong as that of the turbulent component.
To summarize, the most important result is that the chordwise momentum exchange is achieved mainly by the turbulent component of the Reynolds stress, which is entrained by the coherent vortex structure and distributed all over the airfoil surface.
On the other hand, the coherent vortex itself does not strongly contribute to the chordwise momentum exchange via the periodic component of the Reynolds stress.
The formation of the coherent vortex will be discussed in more detail in the next subsection.

%
%

\subsection{Formation of the coherent vortex}
\begin{figure}
\centering
\ifCONDITION
   \subfloat[][Vertical view]{\includegraphics[viewport=10 10 300 255,clip,width=0.35\textwidth]{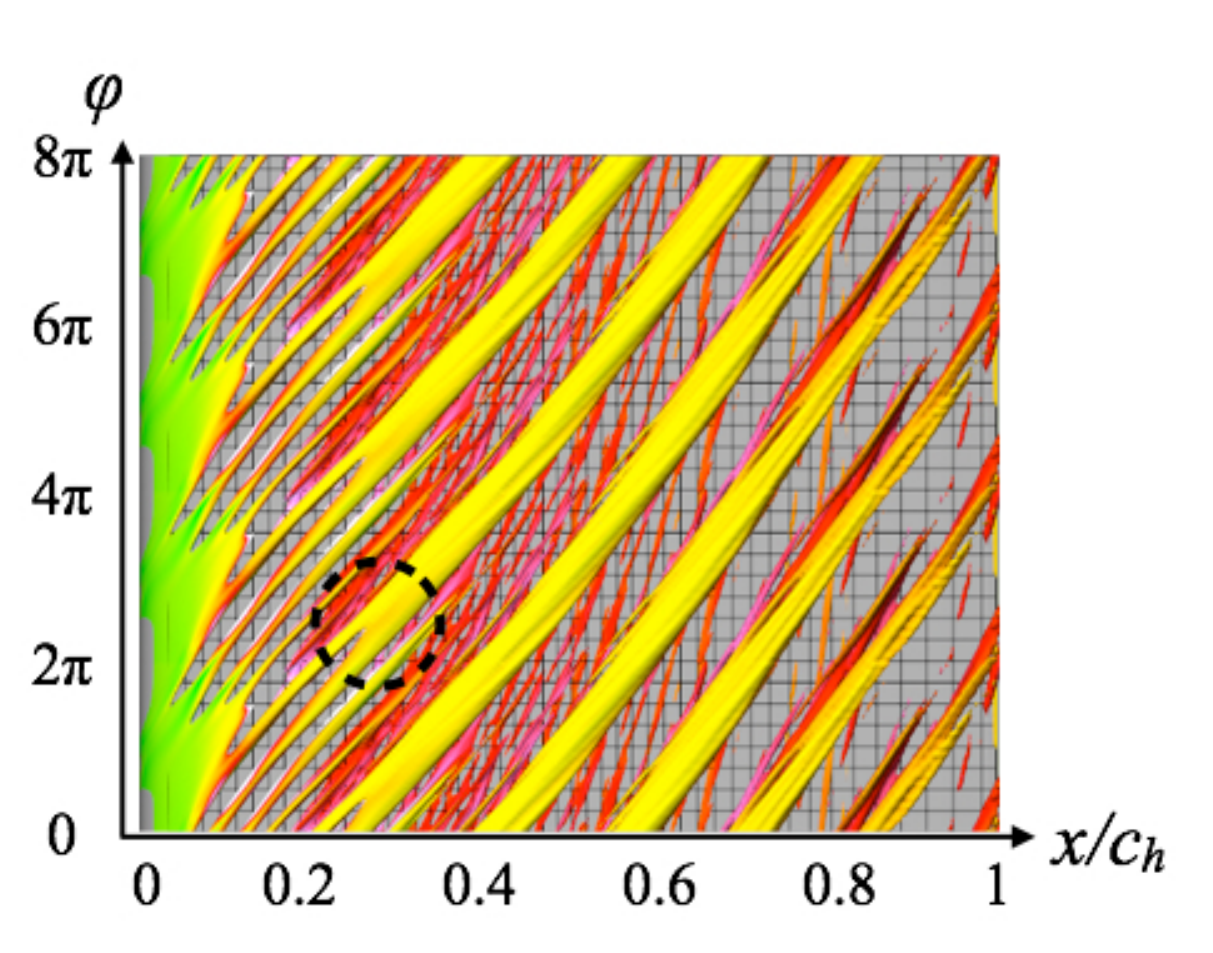}}
   \subfloat[][Perspective view]{\includegraphics[width=0.6\textwidth]{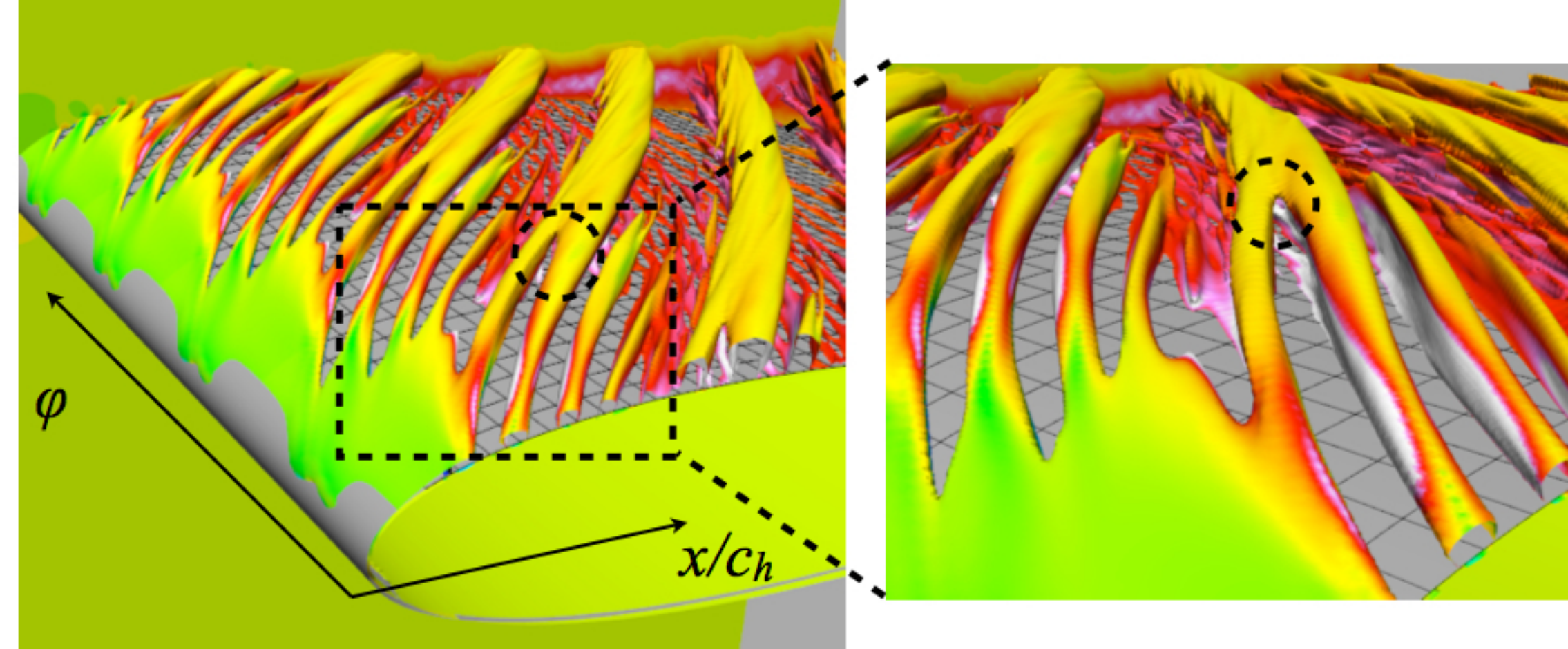}}
\else
\fi
\caption{
Time-space visualization of the convecting coherent vortex structures in the strong input case ($\Cm=2.0\dd{-3}$) with $\Fp=6.0$.
$\varphi$ indicates the phase between $0$ and $8\pi$ (four periods), and $x/c_{\text{h}}$ indicates the chordwise coordinate.
An isosurface represents a second invariant of the velocity gradient tensor of the phase- and spanwise-averaged flows (of which value is $1.0$ and normalized by the freestream velocity), i.e., $\ang{u}_\varphi$, $\ang{v}_\varphi$, and $\ang{w}_\varphi$, which is colored by the turbulent component of the Reynolds stress at each phase, $-0.0125\leq -\ang{u''w''}_\varphi/u_{\infty}^2\leq 0.0125$.
}\label{fig:Vortex-cmu2d-3-F06}
\end{figure}

In this subsection, the formation of the coherent vortex is discussed for the strong input case with $\Fp=6.0$.
Figure \ref{fig:Vortex-cmu2d-3-F06} shows a space-time combined visualization of the coherent vortex motion and the turbulent component of the Reynolds stress.
The $\varphi$-axis indicates the phase angle based on $\Fp=6.0$ for $0 \leq \varphi \leq 8\pi$ (four periods), and the $x/c_{\text{h}}$-axis indicates the spatial chordwise direction.
At each phase, the phase- and span-averaged flow fields are visualized, where the isosurface shows a second invariant of the velocity gradient tensor.
Therefore, the isosurface in Fig.~\ref{fig:Vortex-cmu2d-3-F06} shows the convection of the coherent vortex structures which are illustrated by black contours in Figs.~\ref{fig:Rey-phase-cmu2d-3}(c) and (d).
The isosurface is colored by the turbulent component of the Reynolds stress at each phase: $-\ang{u''w''}_\varphi$.

In Fig.~\ref{fig:Vortex-cmu2d-3-F06}(a), the separation bubble near the leading edge periodically expands and contracts, from which the small vortex structures are emitted. 
Those small vortex structures convect in the downstream direction, and two of them are merging at $x/c_{\text{h}}\simeq 0.25$ and $\varphi\simeq 4\pi$ while the other structures are gradually dissipated, which eventually form single coherent vortex (see the black-dotted circle).
The resultant coherent vortex remains strong at further downstream, which is then periodically released from the trailing edge.
As such, there are clear merging and diffusion processes of small vortex structures that are originally generated from the separated shear layer, which form a single coherent vortex in the period of $\Fp$.
%
%

\section{Spatial growth of wall-normal velocity fluctuations}\label{sec:spatialgrowth}
In this section, a spatial growth of the wall-normal velocity fluctuations is discussed to identify the mechanism behind the optimal actuation frequency.
Thus far, it is shown that the separation controlled flows typically exhibit separation bubbles near the leading edge, from which small vortex structures are emitted periodically.
Those small vortex structures are merged and dissipated in the downstream of the separation bubble, and a single coherent vortex structure in the phase- and spanwise-averaged field is formed.
Such a coherent vortex structure is generated consecutively and convected in the period of $\Fp$, and also entrains turbulent vortices.
As a consequence, the chordwise momentum near the airfoil surface, which is required to suppress the separation, is mainly entrained from the freestream via the turbulent component of the Reynolds stress, $-\ang{u''w''}$.
Therefore, the key mechanism of separation control under the present conditions is considered to be a formation of the coherent vortex structure  as well as strong turbulent fluctuation, which both contribute to the chordwise momentum exchange.
%

However, it is not yet clear how the optimal actuation frequency (that is identified as a frequency band between $\Fp=6.0$ and $20$ in this study) is associated with the mechanism behind the effective momentum exchange as described above.
To elucidate this point, we perform an inviscid linear stability analysis on the wall-normal velocity fluctuations, which is expected to help us identify the relationship between the mechanism of an effective momentum exchange and the most unstable frequency that excites the KH instability in the vicinity of the LSB, thereby providing a strategy to select an optimal $\Fp$.
The similar linear stability analyses have been attempted in a considerable number of studies (as is reviewed in Sec.~\ref{sec:intro});
however, the prior studies have mostly focused on the agreement between the unstable frequency predicted by the LST and that directly calculated from the DNS / LES (or experimental) results, in the region where the fluctuation obeys a linear approximation (linear instability regime).
For example, in the context of separation control, Ref.~\onlinecite{Dandois2007} presented a streamwise evolution of the natural frequency $f_{kh}$ (that is calculated by the most unstable frequency in the LST) showing a good agreement with that calculated from the LES data.
Accordingly, they also presented a comparison of a spatial growth of the most unstable fluctuation in between the controlled and uncontrolled cases (Figs.~33 and 34 in Ref.~\onlinecite{Dandois2007});
however, it was only based on the LST using the time-averaged mean flow, and the spatial growth of the fluctuations was not rigorously investigated for the LES data that should not necessarily fit the LST results.
Indeed, the results presented in Sec.~\ref{sec:momentum} suggest that the significant momentum exchange is maintained by the consecutive coherent vortices that entrain turbulent fluctuation, of which mechanism is beyond the LST as long as the time-averaged flow field is adopted as a base flow.
Therefore, it is worth revisiting the inviscid LST to precisely identify the relationship between the KH instability and the separation-control mechanisms behind the effective momentum exchange that is associated with the optimal $\Fp$.
This section investigates the spatial growth of the wall-normal velocity fluctuations in both linear and nonlinear growth regimes, based on the spatial growth rate that is predicted from the LST for a time-averaged flow in comparison with the spatial growth rate that is directly calculated from the LES data.

The present discussion starts with the validation of the post-process tool that is introduced in Sec.~\ref{sec:method}, based on the noncontrolled case.
Next, the controlled cases are focused, where the spatial growth rate that is estimated by the LST is precisely compared with that directly calculated from the LES data, and the emergence of both linear and nonlinear growth regimes are associated with the mechanisms of the effective momentum exchange in separation control.
Finally, spatial growth of the PSD that is extracted from the LES data is visualized for each frequency, which provides us with the strategies to identify the optimal $\Fp$ for effective separation control.
Such an attempt to rigorously investigate a spatial growth rate of the wall-normal velocity fluctuations and its relation to the mechanism of a momentum exchange would be the first time in controlling a massively separated flow, to the best of our knowledge.

\subsection{Noncontrolled case}\label{sec:chap2-noncontrol}
In the context of turbulent transition, it is well known that the KH instability plays a significant role provided that a laminar separated shear layer is developed and an inflection point exists in the base flow~\cite{Alam2000,Rist2006,Dandois2007,Marxen2011}.
The wall-normal velocity fluctuation that is \textcolor{\hglght}{exponentially} amplified via the KH instability can be estimated as the most unstable eigenmode of the Rayleigh equation given by Eq.~\eqref{eq:LSA}.
Accordingly, the spatial growth rate of the wall-normal fluctuation, $-\alpha_i$, is predicted at each chordwise location for each frequency (normalized as $St$), which is called the spatial growth rate based on the inviscid linear stability theory ($-\alpha_i$ by the LST) hereinafter.
On the other hand, the spatial growth rate can be directly computed from the LES data as well, where the PSD of the wall-normal velocity fluctuation is extracted from the TKE-max line (see Fig.~\ref{fig:TKEmax}).

\begin{figure}
\centering
\renewcommand{\size}{1.0}
\ifCONDITION
\includegraphics[width=\size\textwidth]{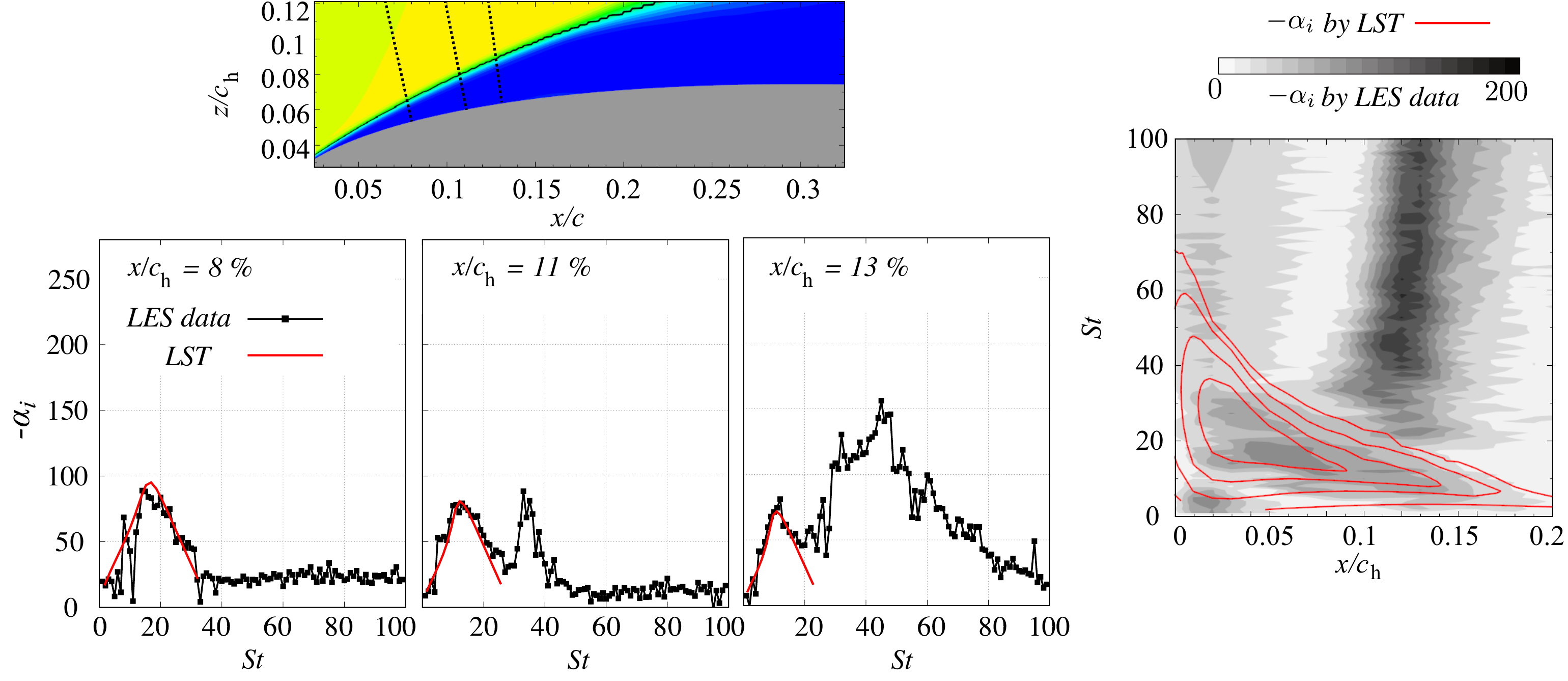}
\else
\fi
\vspace{-0.2cm}
\caption{
Spatial growth rate $-\alpha_i$ in the noncontrolled case is visualized through the analysis of LST and FFT of the LES data.
In the left top figure, contour colors represent the time- and spanwise-averaged chordwise velocity, i.e., $u/u_\infty$;
three black dashed lines indicate the path on which the eigen mode analysis of LST is performed based on Eq.~\eqref{eq:LSA};
the black solid line represents the TKE-max line on which the FFT is performed to compute $-\alpha_i$ from the LES data.
Three line plots in the left bottom corner represent the spatial growth rate $-\alpha_i$ at different chordwise positions, which are estimated by LST (red lines) and FFT of the LES data (black lines).
The right side plot shows $-\alpha_i$ in the $St$-$x/c_{\text{h}}$ plane, where the red contour lines and black-to-white contours represent $-\alpha_i$ based on the LST and LES data, respectively (contour range is $0\leq -\alpha_i\leq 200$). 
}\label{fig:LSA1-off}
\end{figure}

First, the spatial growth rate based on the LES data is compared with that estimated by the LST.
Figure \ref{fig:LSA1-off} shows the comparison of spatial growth rate $\alpha_i$ by the LST and LES data.
In each case, left top figure shows the time- and spanwise-averaged chordwise velocity field, $u/u_\infty$, where the black line shows the TKE-max line and black-dotted lines show the location of taking a wall-tangential velocity component for the base flow profile in the LST analysis.
Left bottom figures show the spatial growth rate $-\alpha_i$ at different chordwise positions, which are estimated by LST (red lines) and FFT of LES data (black lines with points).
The spatial growth rate estimated by the LST and LES data shows reasonably good agreement at $x/c_{\text{h}}=0.08$, which means that the wall-normal fluctuation is amplified by the KH instability and \textcolor{\hglght}{exponentially} grows in the downstream direction.
In this linear growth regime, the linear instability frequency is identified as $St\simeq 20$.
At $x/c_{\text{h}}=0.11$, the linear instability frequency is estimated as $St\simeq 15$ (red solid line).
Meanwhile, the most unstable frequency is $St\simeq 30$ that is directly calculated from the LES data (black dots).
The frequency $St\simeq 30$ corresponds to the harmonic mode of the linear instability frequency of $St\simeq 15$.
In this region, the fluctuation sufficiently grows in the upstream of $x/c_{\text{h}}=11\%$ such that the linear approximation cannot be applied to predict the most unstable frequency, which is then transient to the nonlinear-growth regime.
The fluctuation at $x/c_{\text{h}}=0.13$ is also in the nonlinear growth regime, where the linear instability frequency $St\simeq 10$ is not the most unstable frequency, but the higher frequency modes ($St\simeq 50$) exhibit larger spatial growth rate.
Such a larger growth rate in the higher frequency modes indicates that the nonlinear interaction between fluctuations of multiple frequencies occurs rapidly so that these higher frequency modes are amplified faster than the linear instability mode.
This region is also characterized by a turbulent transition that is promoted by small vortex structures convecting and developing in the downstream direction.
The right figure in Fig.~\ref{fig:LSA1-off} shows the distribution of the spatial growth rate in the $St$-$x/c_{\text{h}}$ plane, where the red contour lines and black-to-white contours show $-\alpha_i$ identified by the LST and LES data, respectively (the contour range is $0\leq -\alpha_i\leq 200$). 
In this contour plot, a transient of the most unstable frequency in the downstream direction is more clearly observed:
the spatial growth rate that is computed from the LES data agrees better with the LST result near the leading edge.
In the downstream location ($x/c_{\text{h}}\geq 0.1$), the higher frequency modes represent a larger spatial growth rate, $-\alpha_i$, where the fine vortex structures are rapidly amplified via harmonic modes of the linear instability mode.

In this way, the spatial growth of the wall-normal velocity fluctuation in the separated shear layer can be explained by the following sequential regimes:
the first regime is a linear growth regime, where the spatial growth of the fluctuation can be approximated by the inviscid linearized equation, and thus the most unstable mode is identified by the LST as a linear instability mode;
the second regime is a nonlinear growth regime, where the amplitude of the fluctuation is already saturated such that the unstable mode no longer follows the linearized approximation, and higher frequency modes show the larger growth rate compared to that of the linear instability mode.
This is a typical description of turbulent transition triggered by the KH instability~\cite{Gaster1966,Nishioka1989,Dandois2007,Simoni2014}.

%
%

\subsection{Separation controlled cases}\label{sec:chap2-control}
\subsubsection{Spatial growth rate}
In this part, spatial growth rate of the wall-normal fluctuation is discussed for the separation-controlled cases of $\Cm=2.0\dd{-3}$ with $\Fp=1.0$ and $6.0$, and $\Cm=2.0\dd{-5}$ with $\Fp=6.0$.
\textcolor{\hglght}{
The applicability of the LST to controlled flows has to be carefully addressed since the strength of the wall-normal fluctuation in the controlled flow is largely dependent on $\Cm$ of the SJ, which can be so large that the linearization is not legitimated in the governing equation.
Based on the discussion in Fig.~\ref{fig:freq-fft}, the PSDs of the wall-normal fluctuation in the non-controlled flow are mostly smaller than $O(10^{-1})$ up to 5\% of the chord length ($x/c_{\text{h}}\leq 5.0\%$). 
Considering the fact that the noncontrolled flow clearly exhibits the linear growth mode at $x/c_{\text{h}}=8\%$ as in Fig.~\ref{fig:LSA1-off}, the cases of $\Cm=2.0\dd{-5}$ can be also the scope of LST because the PSDs of wall-normal fluctuation are approximately smaller than $O(10^{-1})$ at $x/c_{\text{h}}\leq 5.0\%$ in Fig.~\ref{fig:freq-fft}.
On the other hand, the cases with $\Cm=2.0\dd{-3}$ represent strong three-dimensional flow structures inside the cavity (Figs.~\ref{fig:ins-phase}(a)-(c)) thereby resulting in large PSDs in Fig.~\ref{fig:freq-fft}.
Therefore, the applicability of the LST to the cases of $\Cm=2.0\dd{-3}$ is not clear and needs to be carefully compared with the LES results. 
We will discuss the above-mentioned applicability of the LST to strong input cases for controlled flows as well.
}

\begin{figure}
\centering
\renewcommand{\size}{0.825}
\ifCONDITION
\subfloat[][$\Cm=2.0\dd{-5},\Fp=6.0$]{\includegraphics[width=\size\textwidth]{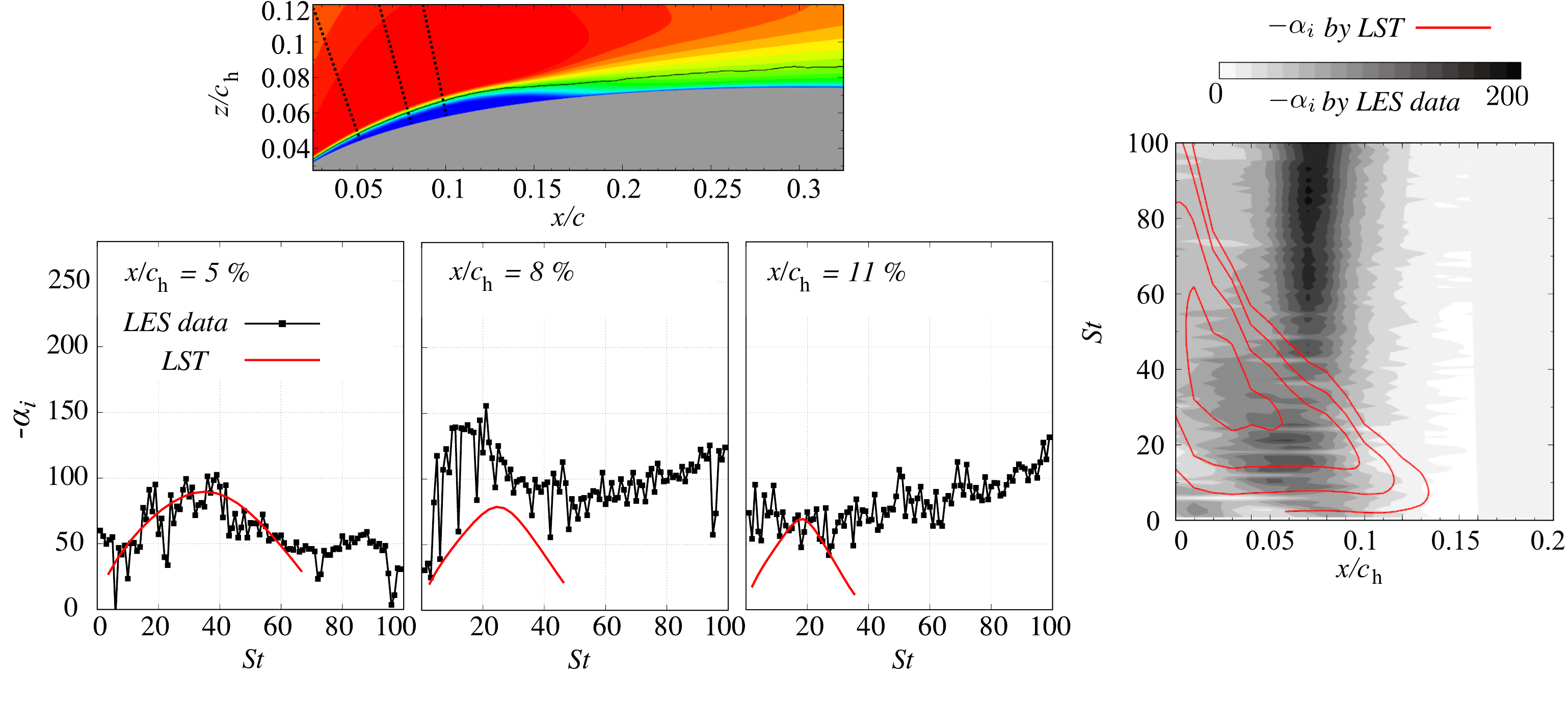}}\\
\subfloat[][$\Cm=2.0\dd{-3},\Fp=1.0$]{\includegraphics[width=\size\textwidth]{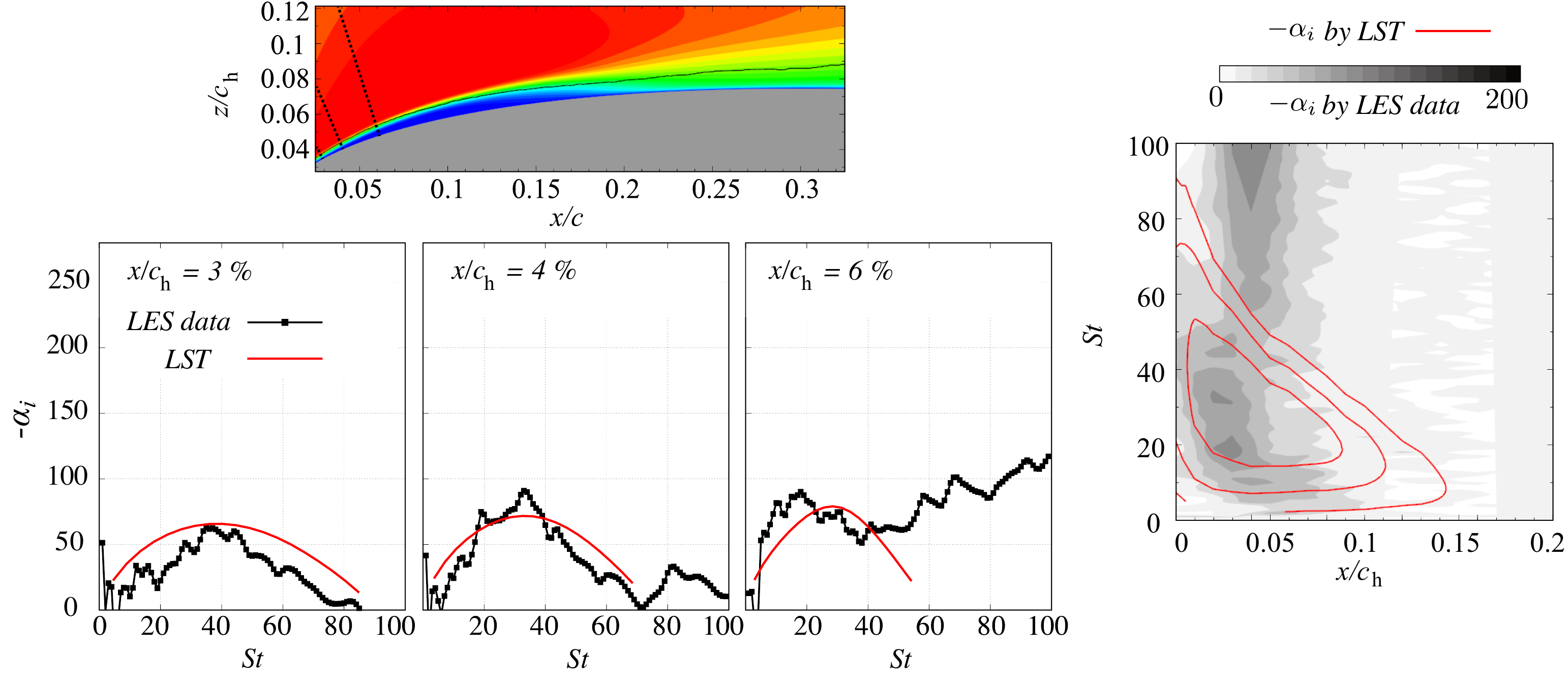}}\\
\subfloat[][$\Cm=2.0\dd{-3},\Fp=6.0$]{\includegraphics[width=\size\textwidth]{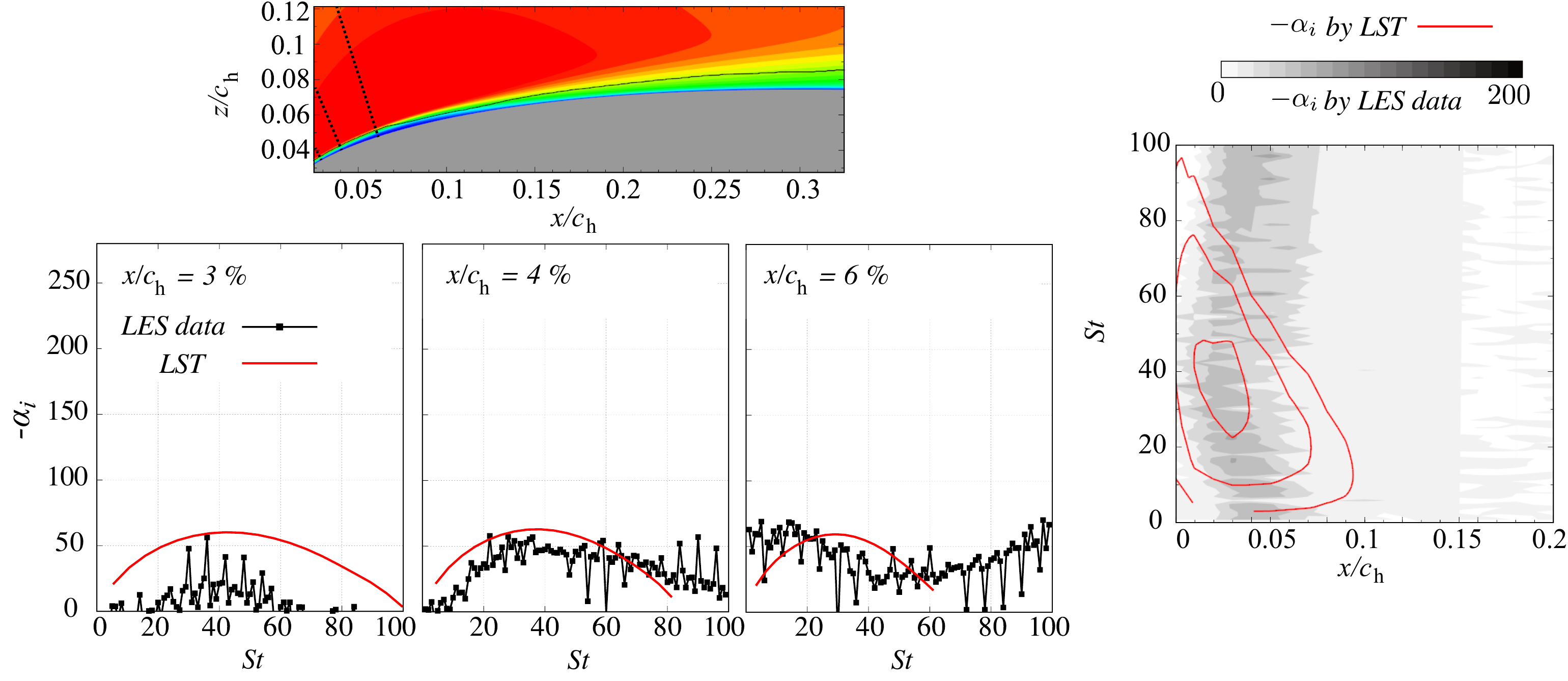}}
\else
\fi
\caption{
Spatial growth rate $-\alpha_i$ in the representative controlled cases is visualized through the analysis of LST and FFT of the LES data.
The plots are similar to Fig.~\ref{fig:LSA1-off}, of which description is omitted here.
}\label{fig:LSA1-controlled}
\end{figure}

Figure~\ref{fig:LSA1-controlled}(a) shows a comparison between the LST and LES results in the weak input case of $\Cm=2.0\dd{-5}$ with $\Fp=6.0$. 
In the left three figures, the spatial growth rate calculated from the LES data (black lines with squares) is more oscillatory than those of the noncontrolled case in Fig.~\ref{fig:LSA1-off}, which is due to wall-normal fluctuation introduced by the SJ. 
\textcolor{\hglght}{
As discussed in Fig.~\ref{fig:freq-fft}, the PSD of wall-normal fluctuation with $\Cm=2.0\dd{-5}$ is approximately the same level as that of the noncontrolled case in $x/c_{\text{h}}\leq 0.05$, and thus, it is expected that the instability can be predicted by the LST similarly to the noncontrolled case.
Indeed, at $x/c_{\text{h}}=0.05$ in Fig.~\ref{fig:LSA1-controlled}(a), both of the LES and LST results broadly increase at $St\simeq 40$, and the most unstable frequency is identified as $St\simeq 40$.
Therefore, the LST is applicable to the weak input case of $\Cm=2.0\dd{-5}$ for predicting the unstable frequency in the vicinity of the leading edge.
The wall-normal fluctuation is exponentially amplified according to the LST in this region, which is identified as the linear instability regime.}
At $x/c_{\text{h}}=0.08$ in Fig.~\ref{fig:LSA1-controlled}(a), the linear instability frequency is identified as $St\simeq 25$, which gives the maximum spatial growth ratio in the LST curve;
on the other hand, the profile of the LES data deviates from the LST curve, and the higher ($St> 25$) and lower ($St<25$) frequencies show larger spatial growth rate than that of the linear instability frequency.
Such a large spatial growth rate in the higher frequency modes is observed similarly in the noncontrolled case (see Sec.~\ref{sec:chap2-noncontrol}), where the fluctuation is largely amplified so that the nonlinear effects cannot be neglected.
Furthermore, the spatial growth rate of the lower frequencies becomes larger than that of the linear instability mode.
This is caused by a merging process of small vortex structures, which are emitted from the separated shear layer and formulating periodic coherent vortices in the period of $\Fp$.
It is supported by the fact that in Fig.~\ref{fig:freq-fft}, the PSD of the wall-normal velocity fluctuation exhibits a clear peak of $\St=\Fp$ at $x/c_{\text{h}}=0.05$, where the fluctuation from the SJ directly contributes to the formation of periodic coherent vortices in the period of $\Fp$.
Such a larger growth rate in the lower frequency mode is one of the distinctive characteristics in the present controlled flows, and the similar merging process has been reported in \cite{Yarusevych2019,Lambert2019} for the LSB with and without forcing.
At $x/c_{\text{h}}=0.11$ in Fig.~\ref{fig:LSA1-controlled}(a), the LST curve no longer agrees with the LES data, where the higher frequency modes show the larger spatial growth ratio compared to the linear instability mode, i.e., $St=20$, and the vortex breakdown and turbulent transition are considered dominant in this nonlinear regime.

Figure~\ref{fig:LSA1-controlled}(b) shows the case of $\Cm=2.0\dd{-3}$ with $\Fp=1.0$.
As was discussed in the first paragraph of this section, the PSD of wall-normal fluctuation in $\Cm=2.0\dd{-3}$ is approximately 100 times stronger than that of the noncontrolled or weak control cases in Fig.~\ref{fig:freq-fft}. 
The LST is not generally applicable to such a strong fluctuation case as the nonlinear effect may not be neglected in the dominant instability unlike the noncontrolled and weak-input cases.
Nevertheless, Fig.~\ref{fig:LSA1-controlled}(b) shows a coincidence of the most unstable frequencies between the LST and LES near the leading edge, i.e., $x/c_{\text{h}}=0.003$ and $0.004$, where the spatial growth rate is amplified at around $St\simeq 40$ and $35$, respectively.
It is noteworthy that a linear instability mode is dominant at the fore portion of the LSB, and thus, the primary instability can be predicted from the LST even in such a strong input case.
Then, the difference between the LST and LES becomes more significant at $x/c_{\text{h}}=0.06$, where the LST identifies $St\simeq 30$ as the most unstable frequency while the higher and lower frequency modes show larger spatial growth ratio in the LES results.
In this region, the nonlinear effect cannot be neglected as was discussed in the noncontrolled and weak-input cases. 
The higher frequency mode ($St > 30$) is amplified due to a vortex breakdown and turbulent transition, while lower frequency mode ($St < 30$) represents a vortex merging and formulation of the coherent vortex structure in the period of $\Fp$.

In the case of $\Cm=2.0\dd{-3}$ with $\Fp=6.0$ (Fig.~\ref{fig:LSA1-controlled}(c)), the profile based on the LES data is very oscillatory at $x/c_{\text{h}}=0.03$ and $0.04$ unlike the weak input case of $\Fp=6.0$ (Fig.~\ref{fig:LSA1-controlled}(b)) and the strong input case of $\Fp=1.0$ (Fig.~\ref{fig:LSA1-controlled}(b)).
This is because the present input momentum is stronger than the case with $\Cm=2.0\dd{-5}$, and the input frequency is higher than the case of $\Fp=1$, and thus the harmonic modes of input frequency are more sharply excited as was observed in Fig.~\ref{fig:freq-fft}.
However, the spatial growth rates based on the LST and LES are broadly in good agreement at $x/c_{\text{h}}=0.03$ and $0.04$, where the most unstable frequency can be predicted as $\St\simeq 40$ using the LST.
At $x/c_{\text{h}}=0.11$, the unstable frequency is different from the linear instability frequency, where the nonlinear effects cannot be negligible as was observed in the other controlled cases.

To summarize, the spatial growth rate of the wall-normal velocity fluctuation in the controlled flows can be characterized by the linear and nonlinear growth regimes, which is similar to the noncontrolled case.
In the linear growth regime, the spatial growth of the fluctuation follows the LST, where the KH instability is dominant.
In the nonlinear growth regime, the most unstable frequency does not correspond to the linear instability frequency.
Specifically, it is expected that the lower frequency modes correspond to the emergence of periodic coherent vortices in the period of $\Fp$, which is excited by the disturbances directly introduced from the SJ.
Furthermore, it is noteworthy that the controlled cases with $\Cm=2.0\dd{-3}$ exhibit the linear stability regime in the vicinity of the leading edge although the PSD of the wall-normal fluctuation is approximately 100 times stronger than that of the noncontrolled and weak-input cases. 
This finding indicates that even with a strong disturbance from the SJ, a primary instability can be explained by the simple KH instability, which is associated with the LST, as long as the LSB is maintained in the separation-controlled flow. Therefore, in the present controlled cases including both of $\Cm=2.0\dd{-3}$ and $2.0\dd{-3}$, it is worth investigating the spatial growth of each instability mode more in detail, i.e., linear and higher/lower-nonlinear instability modes, which will be shown in Sec.~\ref{subsubsec:SpatialGrowth}.

%
%

\subsubsection{Linear growth mode in the controlled flows}
\begin{figure}
\centering
\renewcommand{\size}{1.0}
\ifCONDITION
    \subfloat[][$C_\mu=2.0\dd{-5}$, $\Fp=6.0$ ($St_{\text{target}}=36.92$)]{\includegraphics[width=\size\textwidth]{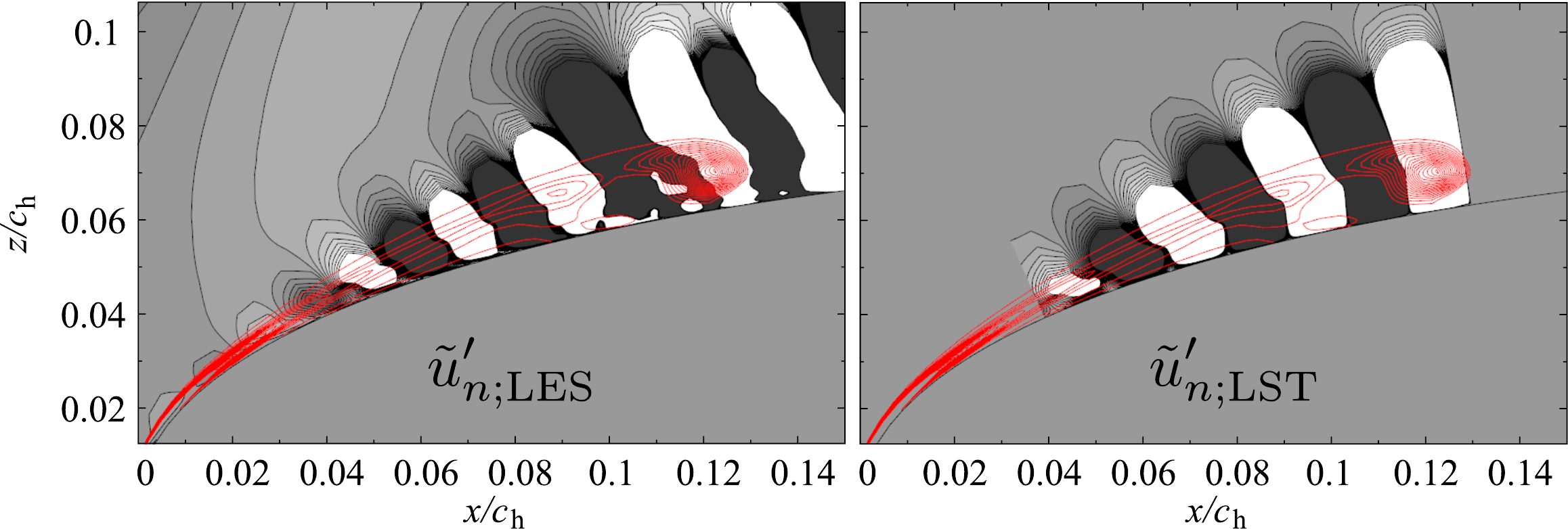}}\\
    \subfloat[][$C_\mu=2.0\dd{-3}$, $\Fp=1.0$ ($St_{\text{target}}=40.00$)]{\includegraphics[width=\size\textwidth]{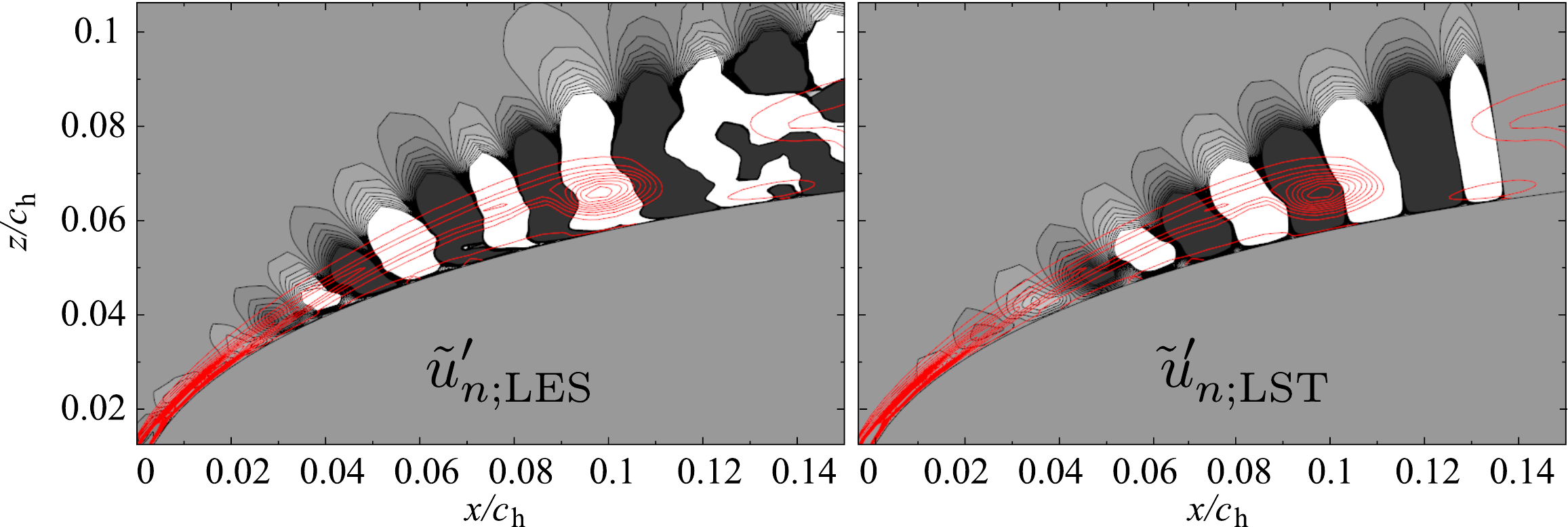}}\\
    \subfloat[][$C_\mu=2.0\dd{-3}$, $\Fp=6.0$ ($St_{\text{target}}=40.00$)]{\includegraphics[width=\size\textwidth]{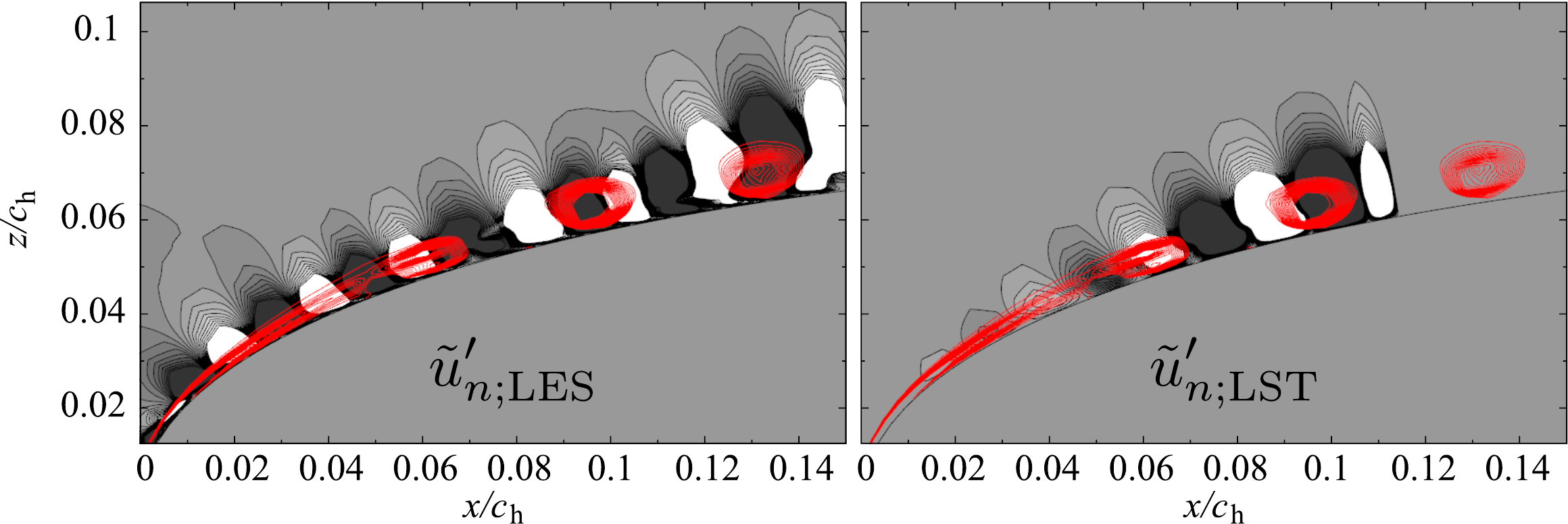}}
\else
\fi
\caption{
Visualization of the linear growth mode (most unstable eigenmode) given by $\hat{u}_{\text{wn};\text{LST}}$ and $\hat{u}_{\text{wn};\text{LES}}$ in Eq.~\eqref{LSTmode} in the cases of Fig.~\ref{fig:LSA1-controlled}.
White-to-black contours show $\hat{u}_{\text{wn};\text{LST}}$ and $\hat{u}_{\text{wn};\text{LES}}$, and the red contour lines show a second invariant of the velocity gradient tensor of instantaneous flows.
}\label{fig:LSA2-controlled}
\end{figure}

In this part, the most unstable eigenmode based on the LST (called as the linear growth mode, hereinafter) is visualized and compared with that directly calculated from the LES data. 
A spatial distribution of the linear growth mode, $\hat{u}_{\text{wn};\text{LST}}(x_{\text{wp}},x_{\text{wn}})$, is computed from the spatial growth rate $\alpha$ and the corresponding eigenmode $\hat{u}_{\text{wn}}$ at the specific frequency, $St=St_{\text{target}}$, as follows~\cite{Rowley2002}:
\begin{align}
\hat{u}_{\text{wn};\text{LST}}(x_{\text{wp}},x_{\text{wn}})=\hat{u}_{\text{wn}}(x_{\text{wp}},x_{\text{wn}})\exp\left[i\int_0^{x_{\text{wp}}}\alpha(X_{\text{wp}}) dX_{\text{wp}}\right].\label{LSTmode}
\end{align}
The unstable mode $\hat{u}_{\text{wn}}$ is normalized by its maximum value, and $St_{\text{target}}$ is set as the most unstable frequency in each case from the LST results shown in Fig.~\ref{fig:LSA1-controlled}.
The line integral in the $x_{\text{wp}}$ direction is conducted along the airfoil surface.
For comparison, the DFT is also conducted for the wall-normal fluctuation to extract the spatial growth mode directly from the LES data.
For the DFT analysis, four periods of $\Fp$ is taken from the beginning of each phase, assuming that the flow is periodic.
The phase average of a real part is then obtained as $\hat{u}_{\text{wn};\text{LES}}$.
Figure~\ref{fig:LSA2-controlled} shows the linear growth mode in the controlled cases that are discussed in the previous subsection.
The white-to-black contours represent $\hat{u}_{\text{wn};\text{LST}}$ and $\hat{u}_{\text{wn};\text{LES}}$, respectively, and the red contour lines show a second invariant of the velocity gradient tensor of instantaneous flows, which visualizes a release of spanwise vortex structures from the separated shear layer near the leading edge.

In Fig.~\ref{fig:LSA1-controlled}, $St_{\text{target}}=36.92$ is chosen in the weak input case of $\Cm=2.0\dd{-5}$ with $\Fp=6.0$, while $St_{\text{target}}=40$ is set in the strong input cases ($\Cm=2.0\dd{-3}$ and $\Fp=1.0$ and $6.0$).
These are close to the linear instability frequencies as shown in Fig.~\ref{fig:LSA1-controlled}.
In each case, $\hat{u}_{\text{wn};\text{LST}}$ and $\hat{u}_{\text{wn};\text{LES}}$ are in qualitatively good agreement, which supports the existence of the linear growth regime in these controlled cases.
Note that the spatial wave number of $\hat{u}_{\text{wn};\text{LST}}$ is slightly higher than that of $\hat{u}_{\text{wn};\text{LES}}$ (that is, a wavelength is smaller), which could be caused by assuming a parallel flow and neglecting the curvature and pressure gradient in the streamwise direction in the LST adopted in this study.

%
%

\subsubsection{Spatial growth of the PSD of the wall-normal velocity fluctuation}\label{subsubsec:SpatialGrowth}
Based on the discussion for the spatial growth rate of the wall-normal velocity fluctuation, it is shown that the separation controlled flows contain a linear growth regime near the leading edge.
On the other hand, in the downstream of the linear growth regime, both higher and lower frequency modes result in a spatial growth rate larger than that of the linear instability mode, which is considered as the nonlinear growth regime.
In this part, an emergence of the nonlinear growth regime is precisely identified based on the spatial growth of the PSD for the wall-normal fluctuation.

\begin{figure}
\centering
\renewcommand{\size}{0.41}
\ifCONDITION
    \subfloat[][$\Cm=2.0\dd{-3},\Fp=1.0$]{\includegraphics[width=\size\textwidth]{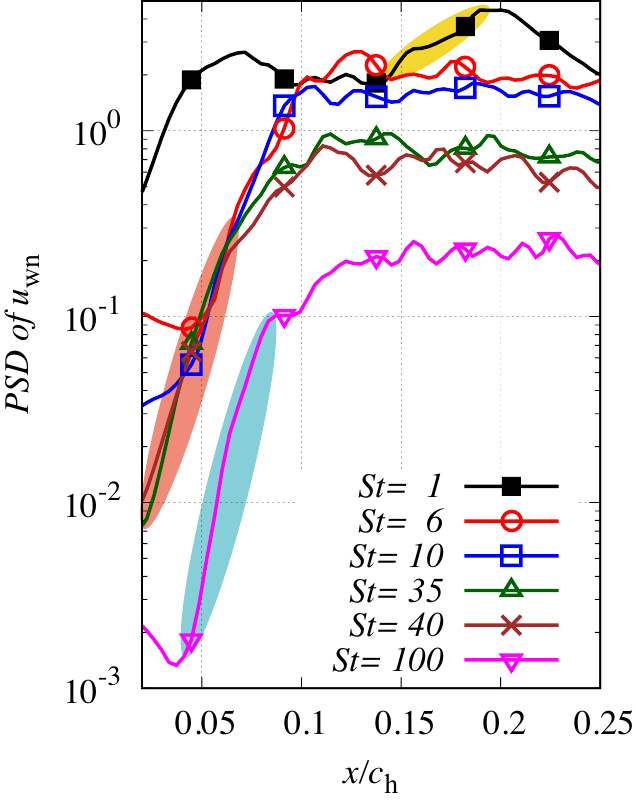}}
    \subfloat[][$\Cm=2.0\dd{-3},\Fp=6.0$]{\includegraphics[width=\size\textwidth]{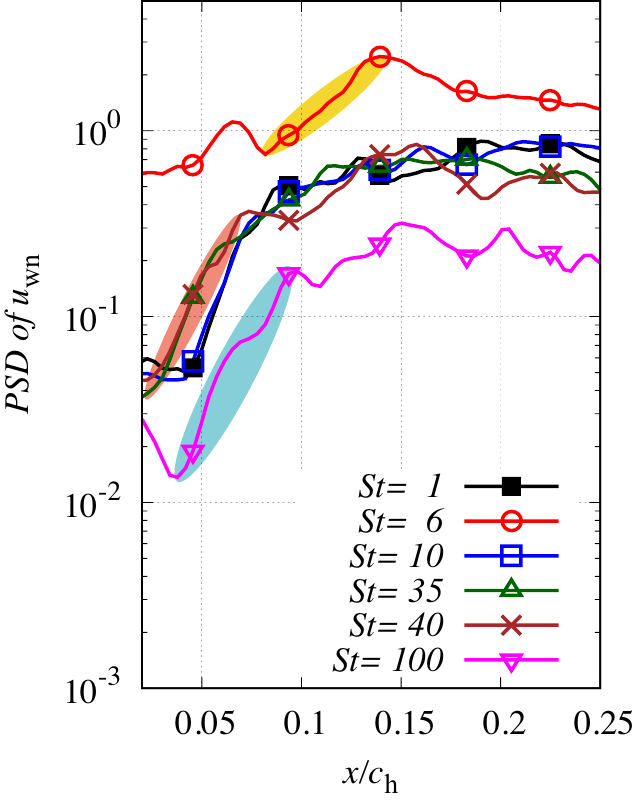}}\\
    \subfloat[][$\Cm=2.0\dd{-5},\Fp=6.0$]{\includegraphics[width=\size\textwidth]{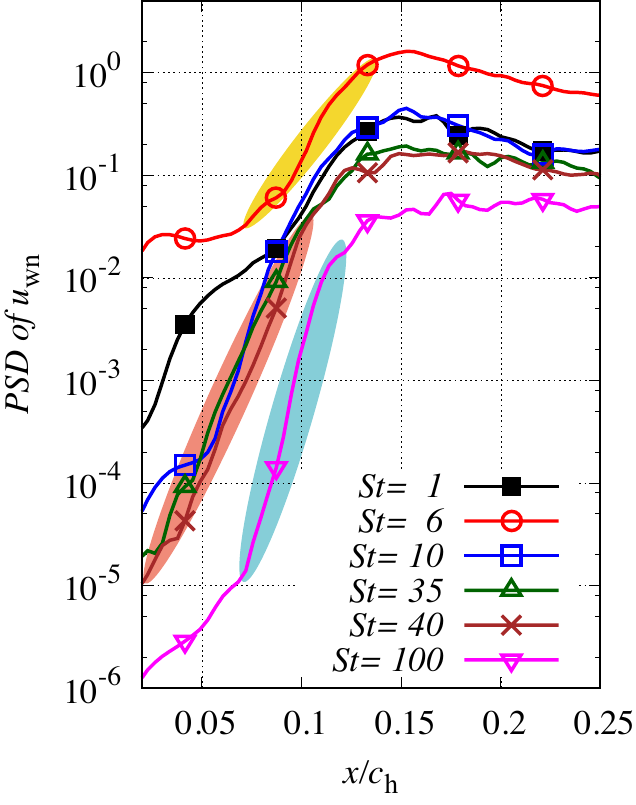}}
    \subfloat[][Schematic of the PSD growth]{\includegraphics[width=\size\textwidth]{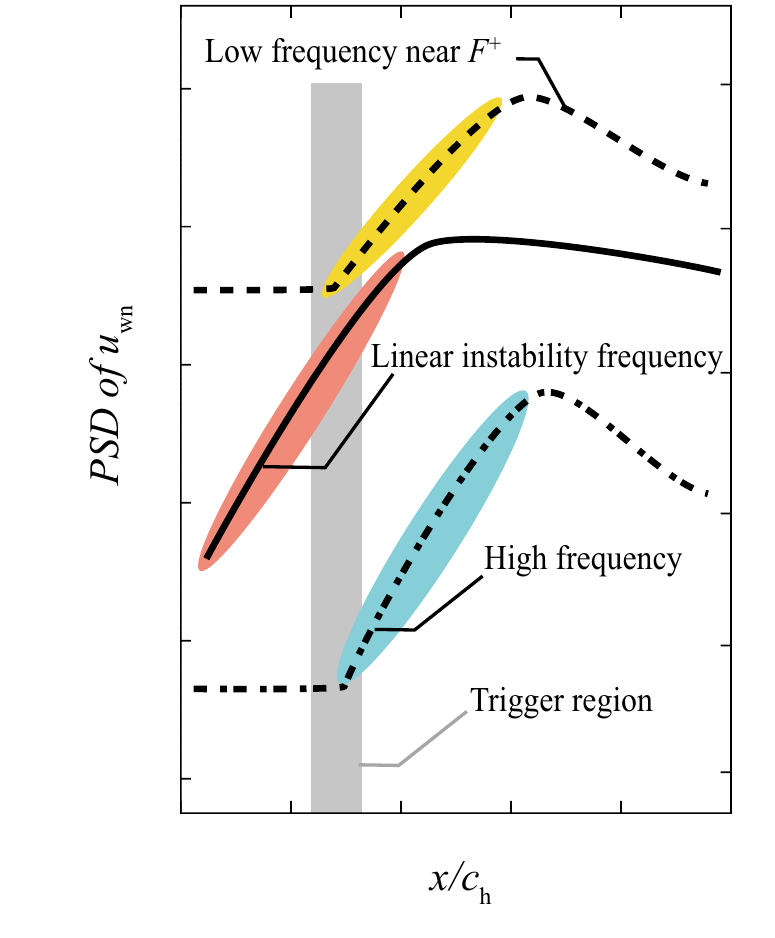}}
\else
\fi
\vspace{-0.2cm}
\caption{Spatial growth of the PSD of wall-normal velocity fluctuation on the TKE-max line (Fig.~\ref{fig:TKEmax}) in the controlled cases.
In (a), (b), and (c), each line represents the PSD of different frequencies ($St=1.0$, 6.0, 10, 35, 40, and 100).
Red, yellow, and blue ellipse regions indicate spatial growth regimes of linear instability mode, low-frequency (near $\Fp$) mode, and high-frequency mode, respectively.
The schematic of a spatial growth of representative modes are shown in (d).
}\label{fig:sgkline-cmu2d-3}
\end{figure}

Figure~\ref{fig:sgkline-cmu2d-3}(a), (b), and (c) shows the spatial growth of the PSD of wall-normal fluctuation in the controlled cases.
The PSD is sampled on the TKE-max line (see Fig.~\ref{fig:TKEmax} and discussion in Sec.~\ref{subsubsec:tke}), which is then spatially filtered  to remove noise based on \cite{Gaitonde2000}.
The black to magenta lines show profiles of $St=1$, 6, 10, 35, 40, and 100, respectively.
In Fig.~\ref{fig:sgkline-cmu2d-3}(a), spatial growth of the PSD in the $\Cm=2.0\dd{-3}$ case with $\Fp=1.0$ is plotted.
The PSD of the linear instability frequency, $St=35$ (and $40$), is small near the leading edge ($\approx 0.01$), which then rapidly grows until $x/c_{\text{h}}\simeq 0.06$.
This corresponds to the linear growth regime, which is highlighted by a red ellipse region.
The PSD of the input frequency ($St=1.0$, black line) is large ($\approx 0.5$) near the leading edge, which immediately grows in $x/c_{\text{h}}\leq 0.05$.
After $x/c_{\text{h}}\simeq 0.05$, the PSD of $St=1.0$ remains almost constant until $x/c_{\text{h}}\simeq 0.15$.
Then it starts to grow again, resulting in the maximum value at $x/c_{\text{h}}\simeq 0.2$, which is highlighted by a yellow ellipse region.
This region corresponds to emergence of the periodic coherent vortex as discussed in Sec.~\ref{subsec:coherent}, which plays an important role in the chordwise momentum exchange.
The other low-frequency modes of $St=6.0$ and $10$ remain almost constant in $x/c_{\text{h}}\leq 0.05$, which start growing afterwards.
It is noteworthy that these low-frequency modes, except for the rapid growth of $\Fp=1.0$ near the leading edge, remain approximately constant until the linear instability mode (e.g., $St=35$ and $40$) sufficiently grows.
Similarly, the high-frequency mode ($St=100$ in Fig.~\ref{fig:sgkline-cmu2d-3}(a)) represents a very small PSD at the leading edge ($\approx 10^{-3}$), which abruptly starts growing at $x/c_{\text{h}}\simeq 0.05$.
This is highlighted by a blue ellipse region in Fig.~\ref{fig:sgkline-cmu2d-3}(a), which corresponds to the nonlinear growth regime involving higher frequency modes as explained in Sec.~\ref{sec:chap2-noncontrol}.
Such rapid growth of high frequency modes is associated with a turbulent transition, generating fine vortex structures.
Hence, there is a specific region where the PSD of lower (and close to $\Fp$) and higher frequency modes are triggered to start growing in the downstream direction, which takes place after the growth of the linear instability mode.
A schematic of the present description is illustrated in Fig.~\ref{fig:sgkline-cmu2d-3}(c).

The similar trend is observed in the case with $\Fp=6.0$ (Fig.~\ref{fig:sgkline-cmu2d-3}(b)), where the linear instability modes ($St=35$ and $40$) initially grow (red ellipse region), then the input frequency mode of $St=6.0$ is triggered to grow in the downstream direction (yellow ellipse region) as well as the other lower frequency modes like $St=1.0$ and $10$.
The higher frequency mode of $St=100$ also rapidly grows after the linear instability modes are sufficiently amplified (blue ellipse region).
On the other hand, the PSD of the linear instability frequencies ($St=35$ and $40$) at the leading edge is larger than that in the case with $\Fp=1.0$.
This is because the input frequency $\Fp=6.0$ is closer to the linear instability frequencies ($St=35$ and $40$) compared with $\Fp=1.0$, and thus the fluctuation around the linear instability frequencies are strongly introduced as a harmonics of the input frequency mode.
The PSD of the higher frequency mode ($St=100$) becomes stronger than that of the $\Fp=1.0$ case, due to the same reason.
Eventually, the spatial growth of the linear instability mode more quickly saturates, and the nonlinear growth regime (including the spatial growth of higher and lower frequency modes) emerges in the more upstream position.
This corresponds to the smooth and quick turbulent transition in the $\Fp=6.0$ case, compared with the $\Fp=1.0$ case as discussed in Sec.~\ref{subsubsec:tke}, which leads to the smaller separation bubble and thus the better control ability.
The similar observation was reported in Refs.~\onlinecite{Yarusevych2017JFM,Yarusevych2017}, where the emergence of roll-up vortices in the more upstream location results in smaller LSB formation.
It should be also noted that in laminar separation control using pulsed vortex generator jets by Ref.~\onlinecite{Postl2011}, the linear instability frequency can be amplified as one of the harmonic frequencies of actuation frequency only when the actuation frequency is equal to or lower than the unstable frequency. For this reason, they concluded that the lower frequency control is more effective than higher frequency controls.

In the case of $\Cm=2.0\dd{-5}$ with $\Fp=6.0$ (Fig.~\ref{fig:sgkline-cmu2d-3}(c)), the PSD of each mode is much smaller than the strong input case in Fig.~\ref{fig:sgkline-cmu2d-3}(b) due to the small input momentum.
Meanwhile, it is clear that the linear instability mode ($St=35$ and $40$) grows first, then the low frequency mode of $\St=6.0$ grows afterwards.
The high frequency mode of $St=100$ also slowly grows from a very small PSD value until $x/c_{\text{h}}\simeq 0.075$, then it starts to grow extensively.
These observations are similar to that in the strong input cases (Figs.~\ref{fig:sgkline-cmu2d-3}(a) and (b)).
On the other hand, the PSD of the fluctuation introduced by the SJ is much smaller than that in the strong input cases.
Therefore, the end of the linear growth regime is located at the more downstream position, where the long distance is required for the linear instability mode to sufficiently grow.
This indicates that turbulent transition takes place at the more downstream location and the size of a separation bubble becomes larger than the strong input cases, thereby resulting in the inferior performance for separation control.

To this end, the significant finding is that even in the controlled cases, the fluctuation with the linear instability frequency initially grows, and then the other higher and lower frequency modes start growing, which are related to the promotion of turbulent transition and generation of periodic coherent vortices, respectively.
Specifically, the lower frequency modes are initially introduced by the input frequency (which is generally much lower than the linear instability frequency), i.e., $St=1.0$ in this study.
The higer separation-control performance would be maintained by a smooth and quick transition from the linear to higher- and lower-nonlinear growth regime, which causes the difference of separation-control capabilities with $\Fp=1.0$ and $6.0$ or higher frequencies.

Therefore, the significant points for the effective separation control are the smooth and quick growth of both
(i) lower frequency modes for the generation of periodic coherent vortices, and
(ii) higher frequency modes for the quick and smooth turbulent transition.
The linear instability of the higher frequency modes should be utilized first for the quick and smooth turbulent transition which leads to the earlier nonlinear spatial growth.   
This is achieved by introducing the fluctuations with the linear instability frequency (typically, $St=30$ to $40$) by the actuation frequency or its harmonics. 
However, the weak input case ($\Cm=2.0\dd{-5}$) with $\Fp=30$, in which the linear instability is excited by the actuation frequency itself, does not show the ability to suppress the separation (Fig.~\ref{fig:freq-clcd}). 
This is because exciting only the higher frequency modes to promote turbulent transition is not sufficient for the effective flow control. 
To compensate for this, the lower frequency mode should be excited together with (ii) the higher frequency modes, where the lower frequency mode leads to the generation of periodic coherent vortex that plays an important role in the chordwise momentum exchange (see Sec.~\ref{subsec:coherent}). 
Therefore, both lower and higher frequency modes should be introduced, for example in the present cases, by the direct actuation frequency $\Fp$ and the harmonics of itself ($n\Fp$), where the high-frequency modes are more strongly excited by adopting $\Fp$ such that it is three to five times lower than the linear instability frequency.
For example, $\Fp=6.0$ shows better performance than $\Fp=1.0$, because in the case of $\Fp=6.0$, the PSD of $St=30$ is more strongly introduced as a harmonic mode of $\Fp$ and $St=30$ is closer to $\Fp=6.0$ than $\Fp=1.0$.
More specifically, in the case of $\Fp=6.0$, the ratio of the linear instability frequency (from $St=30$ to $40$) to $\Fp$ is $\mathcal{O}(1)$ (more strictly, $5$ to $7$), which shows the better control ability than the $\Fp=1.0$ case where the ratio of the linear instability frequency to $\Fp$ is $\mathcal{O}(10)$ (more concretely, $30$ to $40$).
Therefore, the better control ability is achieved by introducing both lower and higher frequency modes to generate periodic coherent vortices and to promote a turbulence transition, respectively.
Consequently, a superior actuation frequency $\Fp$ is generally close to but lower than the linear instability frequency ($St\simeq 30$) so that the linear instability mode can be excited as a harmonic mode of $\Fp$ itself.

%
%

\clearpage
\section{Conclusions}\label{sec:conclusion}
This study investigated the mechanism of separated flow control using the SJ.
Effects of the actuation frequency $\Fp$ on the separation control ability was focused regarding the relationship between the optimal $\Fp$ and the linear instability frequency in the separated shear layer, and the mechanism of exchanging a chordwise momentum between the near-wall surface and freestream was also precisely discussed. 
We consider a separated flow around a NACA0015 airfoil at the 12 $^\circ$ angle of attack and the chord Reynolds number of $Re=63,000$, which completely separates at $x/c_{\text{h}}\approx 2.5\%$.
The SJ consists of a deforming cavity that has a simple two-dimensional (spanwise-uniform) shape and is embedded in the airfoil surface at the leading edge.
The present study performed the LES of separated flow control for different input momentum $\Cm$ ($=2.0\dd{-3}$ and $2.0\dd{-5}$) and actuation frequency $\Fp$ ($1.0$ to $30$), which resolves turbulent structures inside the deforming cavity of the SJ.

In the present flow and actuation conditions, there is an optimal range of $\Fp$ between $6.0$ and $20$, which attains a high lift-to-drag ratio, $C_{\rm{L}}/C_{\rm D}$, in the time- and spanwise-averaged flow fields.
The controlled (attached) flows typically exhibit a LSB near the leading edge, and a turbulent boundary layer grows in the downstream of the separation bubble.
It was also shown that the smaller separation bubble is associated with the strong pressure suction peak at the leading edge, thereby significantly reducing the drag and eventually improving the lift-to-drag ratio.
As such, promoting a turbulent transition to reduce the separation bubble size near the leading edge is one of the effective strategies to suppress a separation and gain the better aerodynamic performance.
Furthermore, in those separation-controlled flow fields, the separation bubble expands and contracts in the period of $\Fp$, which periodically releases small spanwise-uniform vortex structures.
These structures are eventually diffused and merged to form a large periodic coherent vortex, which is clearly visualized in the phase- and spanwise-averaged (based on $\Fp$) flow fields.

Regarding the mechanism to maintain the attached flows, it is important to continuously inject momentum in the chordwise direction near the airfoil surface.
This is achieved by entraining the chordwise momentum from the freestream that exists above the turbulent boundary layer (chordwise momentum exchange) because the present input momentum $\Cm$ is too small to directly compensate for the lack of chordwise momentum in the separated flow, i.e., $\mathcal{O}(10^{-5}$-$10^{-3})$ of the freestream momentum, and thus, it cannot directly alter the direction of the freestream.  
The chordwise momentum exchange was evaluated by the phase decomposition (based on $\Fp$) of the Reynolds stress into periodic and turbulent components, where the turbulent component is dominant but locally enhanced around the periodic coherent vortices, irrespective of $\Fp$.
Therefore, the chordwise momentum exchange is mainly caused by the three-dimensional turbulent vortex structures, while they are also convected by the periodic coherent vortex in the downstream direction.

Next, we investigated the mechanism of the optimal actuation frequency to maintain the LSB in the separation-controlled flows, based on the analysis for the spatial growth of a wall-normal velocity fluctuation.
The inviscid linear stability analysis was conducted for the time and spanwise-averaged flow fields, assuming that the fluctuation follows the Rayleigh equation which is able to predict the KH instability.
It was shown that the linear growth regime appears near the leading edge even in the separation-controlled flows, where the most unstable frequency based on the LST corresponds to that computed from the actual LES data (taking a spatial gradient of the PSD of the wall-normal velocity component).
This finding indicates that even with a strong disturbance from the SJ (i.e., 100 times stronger than the noncontrolled or weak-input cases in the PSD of wall-normal fluctuation), a primary instability can be explained by the simple KH instability, which is associated with the LST, as long as the LSB is maintained in the separation-controlled flow.
The linear instability frequency is then identified to be $St=30$-$40$, which is generally much higher than the optimal range of the actuation frequency, $\Fp=6.0$-$20$.
Therefore, in the optimally controlled cases, the SJ introduces the linear instability mode as a harmonic mode (and its sidelobe) of the flow disturbance with $\Fp$, rather than directly introducing the disturbance of the linear instability frequency.
This indicates that if $\Fp$ is closer to the linear instability frequency, the magnitude of the linear instability mode is more strongly introduced as a harmonic mode of $\Fp$ itself, and thus the spatial growth of the linear instability mode saturates more quickly.
Subsequently, the nonlinear growth regime begins at the more upstream position.
The nonlinear growth regime starts after the linear instability mode sufficiently grows, where the modes whose frequencies are higher and lower than the linear instability frequency show a larger spatial growth rate.
The emergence of a larger spatial growth rate in the high-frequency modes ($St=\mathcal{O}(100)$) corresponds to a typical nonlinear growth regime that exhibits a turbulent transition, involving a generation of smaller scale turbulent vortices.
On the other hand, the low-frequency modes correspond to a formation of the periodic coherent vortex, which takes place in the downstream of the separation bubble.
The periodic coherent vortex convects in the downstream direction, entraining smaller turbulent vortex structures, which significantly contributes to the chordwise momentum exchange as was explained in the previous paragraph.
Interestingly, the PSD of both higher and lower frequency modes remains approximately constant until the linear instability mode sufficiently grows near the leading edge.
Therefore, it is crucial to effectively introduce both linear instability mode and lower frequency mode, which are essential for a quick and smooth turbulent transition and a formation of the periodic coherent vortex, respectively.
As such, a well-balanced input frequency to achieve the better control ability is considered to be close to but lower than the linear instability frequency, to promote both turbulent transition and generation of the periodic coherent vortex in the downstream of the separation bubble.
Indeed, $\Fp=6.0$ to $20$ are identified as the optimal range of the actuation frequency in this study, while the case with $\Fp=30$ is not able to suppress the separation with the weak input ($\Cm=2.0\dd{-5}$), which is due to the lack of input fluctuation with the low-frequency that promotes a generation of periodic coherent vortices.

Finally, we discuss the applicability of the present conclusion.
In the present flow condition, the chord Reynolds number is set to be moderate, $\mathcal{O}(10^{4}-10^{5})$, and the AoA is not significantly high such that the separated shear layer at the leading edge remains relatively close to the airfoil surface.
Eventually, the laminar flow separates from in the vicinity of the leading edge, where a fluctuation is initially amplified via the KH instability, and then a turbulent transition occurs.
Therefore, the findings in this study are available for the case such that the flow is laminar at a separation point, and fluctuation from the SJ can be directly introduced to the separated shear layer that effectively promotes a turbulent transition.
On the other hand, it would be difficult to discuss the flows at high Reynolds number ($\mathcal{O}(10^6)$) as the flow often becomes turbulence before the separation and a significant amount of input fluctuation is required to enhance a turbulent intensity effectively. 
Simultaneously, separated flows at high AoA (analogous to separated flows around a thin airfoil) would require additional discussion, where the separated shear layer stays far from the airfoil surface.
In such cases, the optimal $\Fp$ is often reported to be a lower value ($\Fp\simeq 1$), 
 which implies that a coherent vortex at low frequencies ($St\simeq 1$) plays a more critical role than enhancing a turbulent intensity by higher $\Fp$ such as $\Fp=6$.
Finally, we should emphasize that the present study aims at identifying how the SJ effectively maintains attached flows with various actuation conditions, mainly based on the analysis of time- and phase-averaged flow fields.
A transient process from the separated (noncontrolled) to attached (controlled) flow was studied in the limited studies Refs.~\onlinecite{Amitay2002a,Asada2015,Fukumoto2016} of flows in both static and dynamic stall conditions, and it would involve additional physics. We leave these discussions in the future study.

\begin{acknowledgments}
The present study was mainly supported by a Grant-in-Aid for JSPS Fellows (Grant Number 258793) and in part by Strategic Programs for Innovative Research (SPIRE) of the High Performance Computing Initiative (HPCI) (Project IDs: hp120296, hp130001, hp140207, and hp150219).
The computations in the present study were performed using the ``K'' supercomputer at the Advanced Institute of Computational Science, Riken, Japan.
\end{acknowledgments}
\appendix
\section{Validation and verification of the present flow solver and grids}\label{sec:validation}

In this appendix, we summarize results of the validation and verification study of the flow simulation.
Here, the airfoil grid (Zone 1) is mainly focused since the grid resolution around the SJ has been well verified and validated in the previous study~\cite{Okada2012} using the same flow solver;
moreover, the computational method related to the moving grid has also been adequately verified in Ref.~\onlinecite{Abe2013b}.

This section is organized as follows:
first, the grid resolution is examined for both controlled and noncontrolled cases;
next, the time step size is verified for the controlled case;
finally, the spanwise length of the computational domain is examined for the controlled case.
Table~\ref{tab:Gridpoints-app} summarizes the grids that are examined in this section.
%
%
\begin{table}
  \begin{center}
\def~{\hphantom{0}}
  \begin{tabular}{ccccc}
\hline\hline
grid name & number of total grid points  & $(N_\xi,N_\eta,N_\zeta)$ & span length  \\\hline
Coarse    & $8,598,600$   & $(562,100,153)$   &$20\%$ of the chord length  \\
Medium    & $19,068,870$  & $(795,134,179)$   &$20\%$ of the chord length  \\
Fine      & $43,498,800$  & $(1,124,180,215)$  &$20\%$ of the chord length  \\
Wide      & $35,007,030$  & $(795,246,179)$   &$36\%$ of the chord length  \\\hline\hline
  \end{tabular}
\caption{Numbers of the grid points are summarized. $(N_\xi,N_\eta,N_\zeta)$ indicate the number of grid points in each direction}\label{tab:Gridpoints-app} 
  \end{center}
\end{table}

\subsection{Validation and verification of the grid resolution}
First, we perform a series of convergence studies for a grid resolution as well as a comparison with an experimental result using the Fine, Medium, and Coarse grids given in Table~\ref{tab:Gridpoints-app}, where the number of grid points are approximately $9$ million, $19$ million, and $44$ million, respectively.
Note that the Medium grid corresponds to the grid used in this paper.

\subsubsection{Noncontrolled cases} 
\begin{figure}
  \centering
  \renewcommand{\size}{0.475}
  \ifCONDITION
  \subfloat[][$C_{\rm p}$]{\includegraphics[width=\size\textwidth,clip]{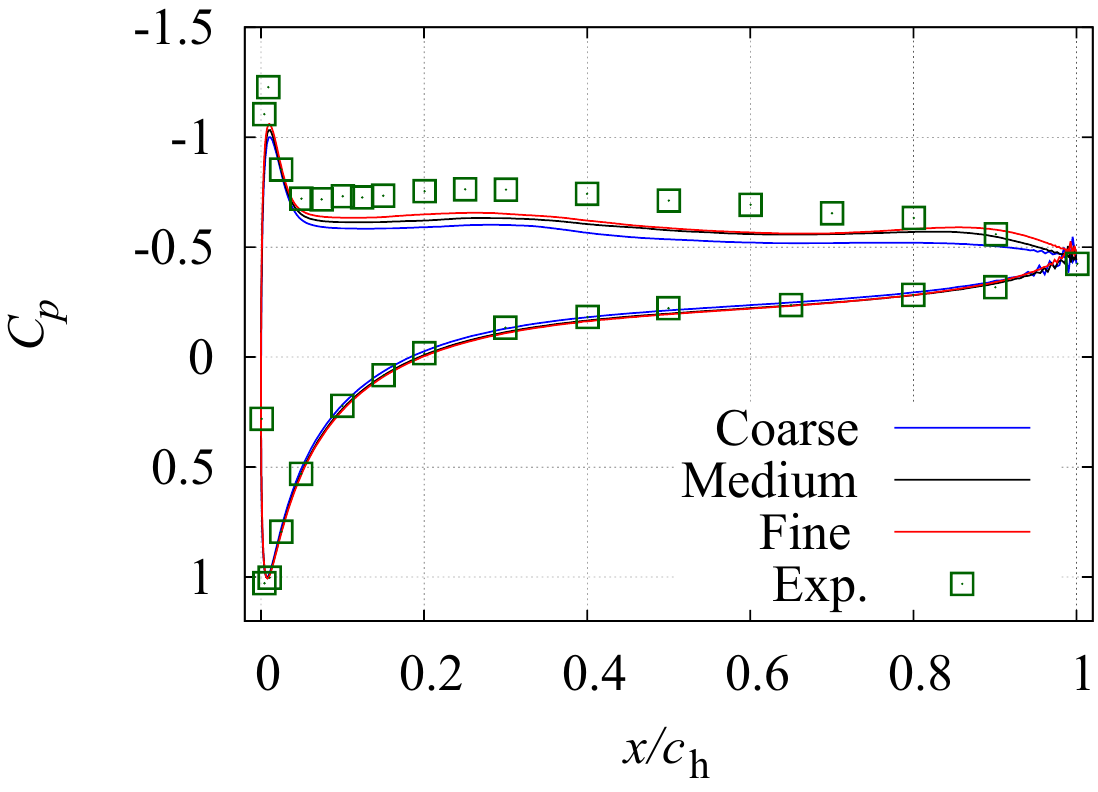}}
  \subfloat[][$C_{\rm f}$]{\includegraphics[width=\size\textwidth,clip]{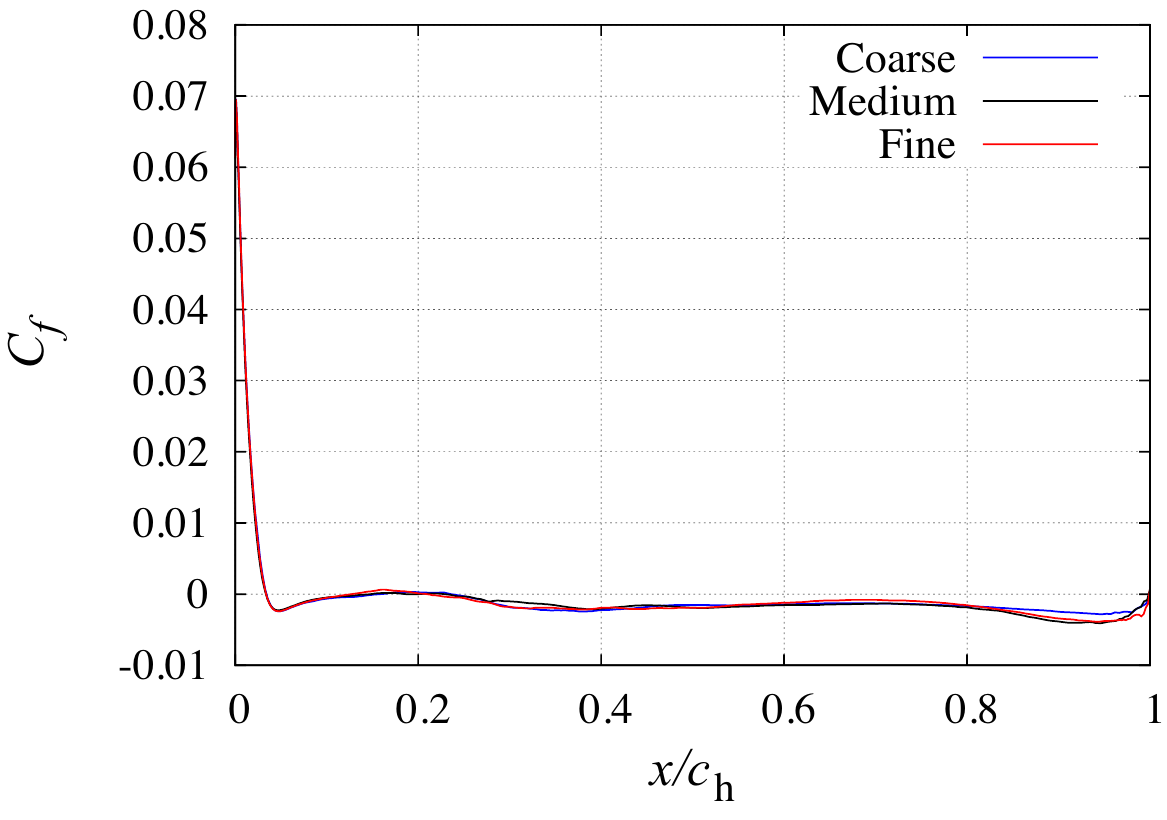}}
  \else
  \fi
 \caption{
Pressure coefficient $C_{\rm p}$ and skin friction $C_{\rm f}$ of the time- and spanwise-averaged flow fields on the Fine, Medium, and Coarse grids are visualized.
The experimental result by Ref.~\onlinecite{Asada2012} is shown in (a) $C_{\rm p}$ distribution.
In (b), $C_{\rm f}$ is the skin friction on the suction side.
 }\label{Fig_veri01}
\end{figure}

The noncontrolled flow is computed using the three different grids (Coarse, Medium, and Fine) shown in Table~\ref{tab:Gridpoints-app}.
Table~\ref{tab:veri01} summarizes the time- and spanwise-averaged aerodynamic coefficients ($C_{\rm{L}}$, $C_{\rm D}$, and $C_{\rm{L}}/C_{\rm D}$).
Both $C_{\rm{L}}$ and $C_{\rm D}$ slightly increase as the grid density increases;
however, the difference of $C_{\rm{L}}/C_{\rm D}$ between the Medium and Fine grid is approximately $2\%$ of that with the Medium (baseline) grid.
This is sufficiently small in terms of the separation-control performances listed in Fig.~\ref{fig:freq-clcd}.
Figure~\ref{Fig_veri01} shows the $C_{\rm p}$ and $C_{\rm f}$ distributions on the airfoil surface.
In $C_{\rm p}$ distribution (Fig.~\ref{Fig_veri01}(a)), the suction peak at the leading edge and the plateau pressure distribution are similarly observed for all grids.
The experimental result \cite{Asada2009} of $C_{\rm p}$ distribution is also plotted by green points, which shows good agreement with the computational results.
The $C_{\rm f}$ distributions (Fig.~\ref{Fig_veri01}(b)) are also in good agreement among three different grids although an experimental data is not available.

\begin{table}
\centering
\def~{\hphantom{0}}
\begin{tabular}{cccc}
\hline\hline
grid name &  $C_{\rm{L}}$  &  $C_{\rm D}$  &$C_{\rm{L}}/C_{\rm D}$\\\hline
Coarse &  0.410     &  0.143   &2.85 \\
Medium &  0.458     &  0.157  &2.92\\
Fine   &  0.477     &  0.159  &2.99\\\hline\hline
\end{tabular} 
\caption{Time- and spanwise-averaged aerodynamic coefficients in the noncontrolled case with the three different grids}\label{tab:veri01}
\end{table}

\subsubsection{Controlled cases} 
\begin{figure}
  \centering
\renewcommand{\size}{0.475}
\ifCONDITION
\subfloat[][$C_{\rm p}$]{\includegraphics[width=\size\textwidth,clip]{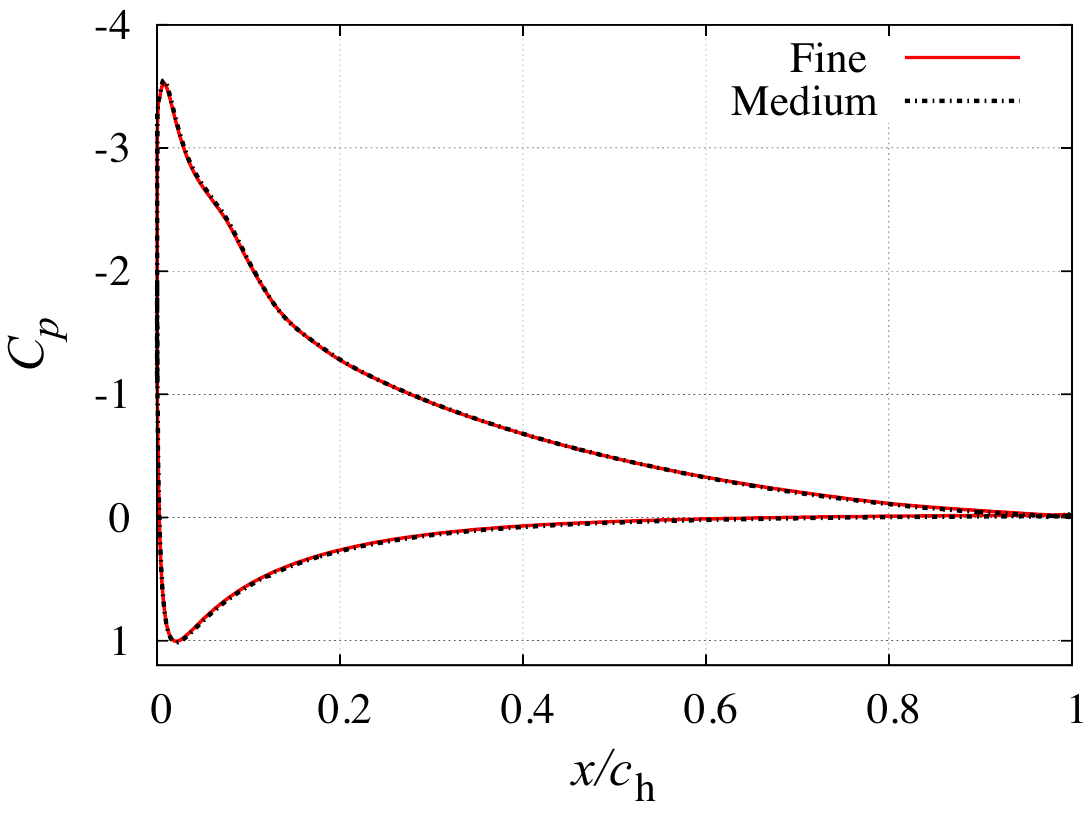}}
\subfloat[][$C_{\rm f}$]{\includegraphics[width=\size\textwidth,clip]{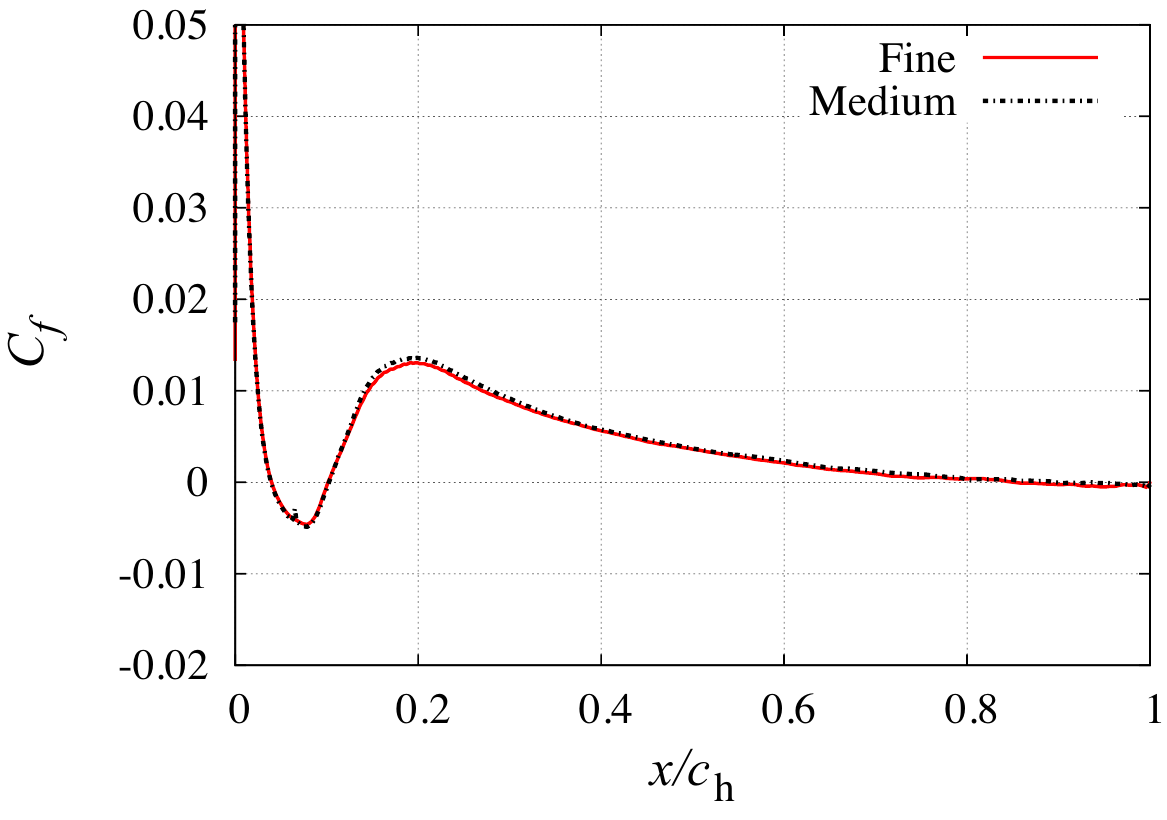}}
\else
\fi
 \caption{
Pressure coefficient $C_{\rm p}$ and skin friction $C_{\rm f}$ of the time- and spanwise-averaged flow fields on the Fine and Medium grids are visualized.
(b) shows the skin friction $C_{\rm f}$ on the suction side.
 }\label{Fig_veri01_GCcont}
\end{figure}

\begin{figure}
  \centering
\ifCONDITION
  \includegraphics[width=0.5\textwidth,clip]{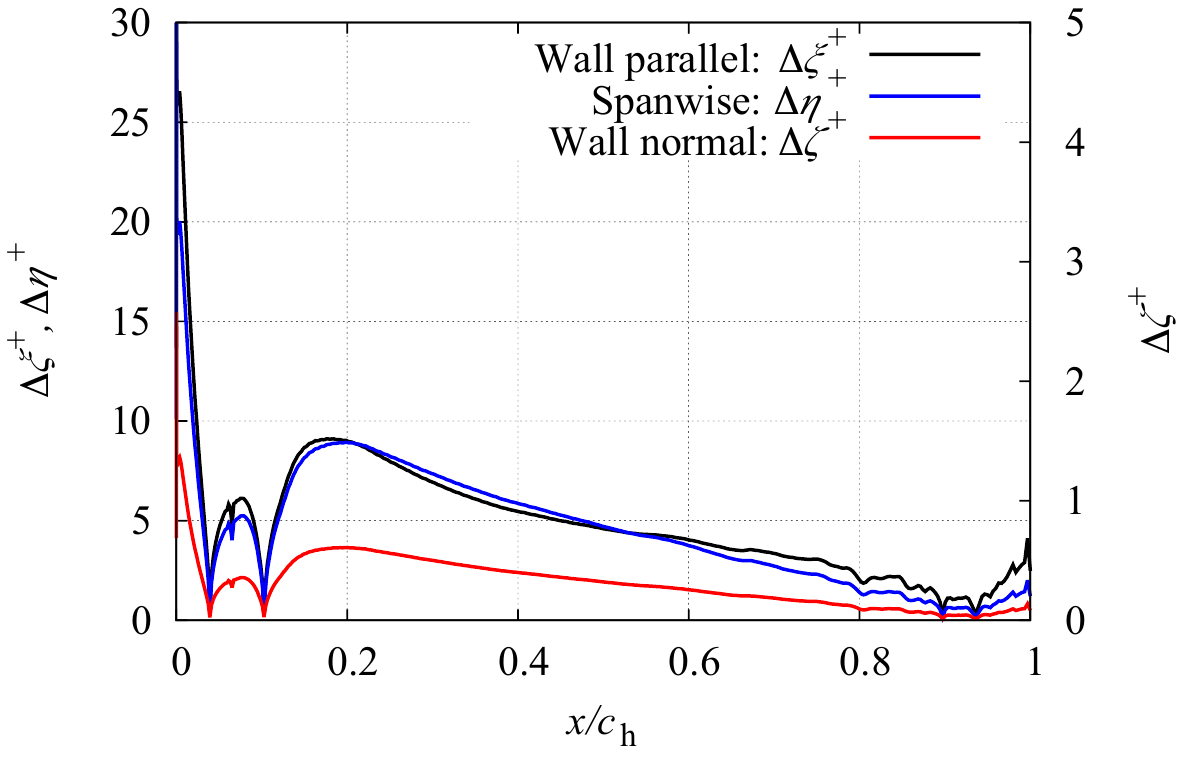}
\else
\fi
\vspace{-0.5cm}
\caption{
Visualization of the grid spacings on the suction side of the airfoil surface.
}\label{Fig_veri02}
\end{figure}

The separation-controlled flow with $\Cm=2.0\dd{-3}$ and $\Fp=6.0$ is examined by the Medium and Fine grids shown in Table~\ref{tab:Gridpoints-app}.
The present cases show attached flows with LSBs near the leading edge.
Figure~\ref{Fig_veri01_GCcont} shows the $C_{\rm p}$ and $C_{\rm f}$ distributions on the airfoil surface.
Both $C_{\rm p}$ and $C_{\rm f}$ distributions are in good agreement between the Medium and Fine grids, and thus the grid convergence on those distributions is confirmed.
The grid sizes in the wall unit are also examined on the Medium grid, which is $(\Delta\xi^+,\Delta\eta^+,\Delta\zeta^+)\leq (10,9,1)$ in the most part of the attached region (see Fig.~\ref{Fig_veri02}).
There are many studies for near-wall resolution as follows.
Ref.~\onlinecite{Nicholas2010} suggests that a flat-plate-like configuration requires $(\Delta\xi^+,\Delta\eta^+,\Delta\zeta^+)\leq (20,10,1)$ for a DNS resolution. For flows over turbine blade, Ref.~\onlinecite{Michelassi2002} suggests $(\Delta\xi^+,\Delta\eta^+,\Delta\zeta^+)\leq (10,3,0.8)$;
Ref.~\onlinecite{Sandberg2015} suggests $(\Delta\xi^+,\Delta\eta^+,\Delta\zeta^+)\leq (10,11,1.0)$;
Ref.~\onlinecite{Alhawwary2019} shows $(\Delta\xi^+,\Delta\eta^+,\Delta\zeta^+)\leq (15,20,0.3)$ for a DNS resolution;
Ref.~\onlinecite{Larsson2016} shows that $(\Delta\xi^+,\Delta\eta^+,\Delta\zeta^+)\leq (40,20,1.0)$ for a wall-resolved LES resolution.
Based on the values above, the present grid resolution can be considered sufficient for resolving near-wall turbulence as a LES, which is also close to a DNS resolution.


\begin{figure}
  \centering
\renewcommand{\size}{0.4725}
\ifCONDITION
\subfloat[][Medium grid]{\includegraphics[width=\size\textwidth,clip]{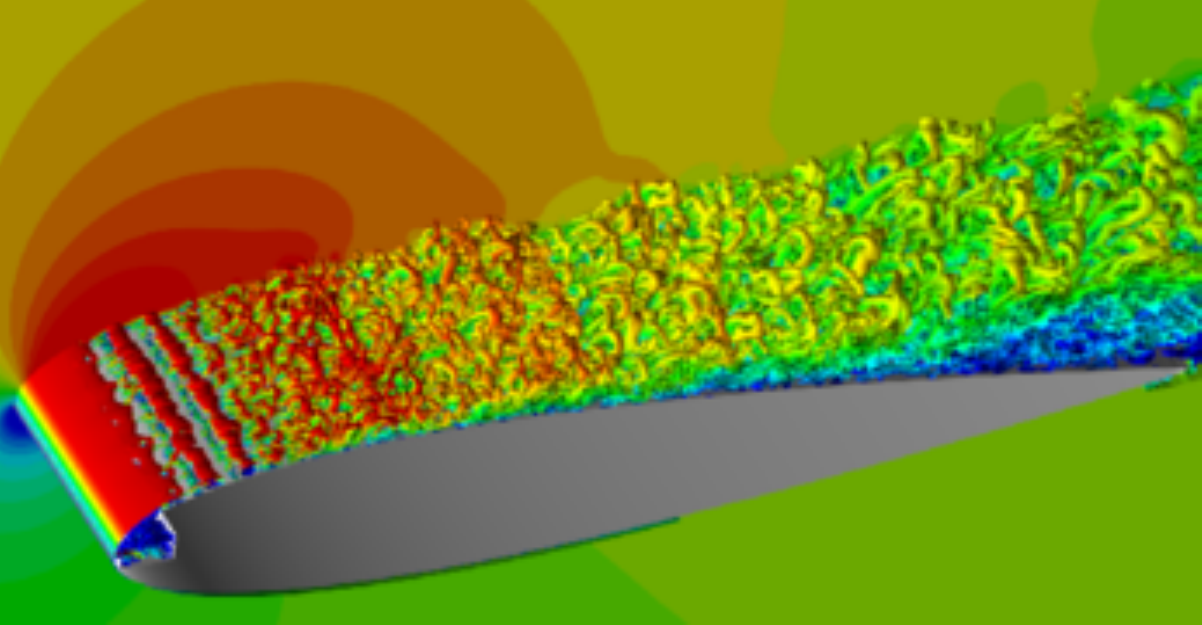}}
\subfloat[][Medium grid (zoom)]{\includegraphics[width=\size\textwidth,clip]{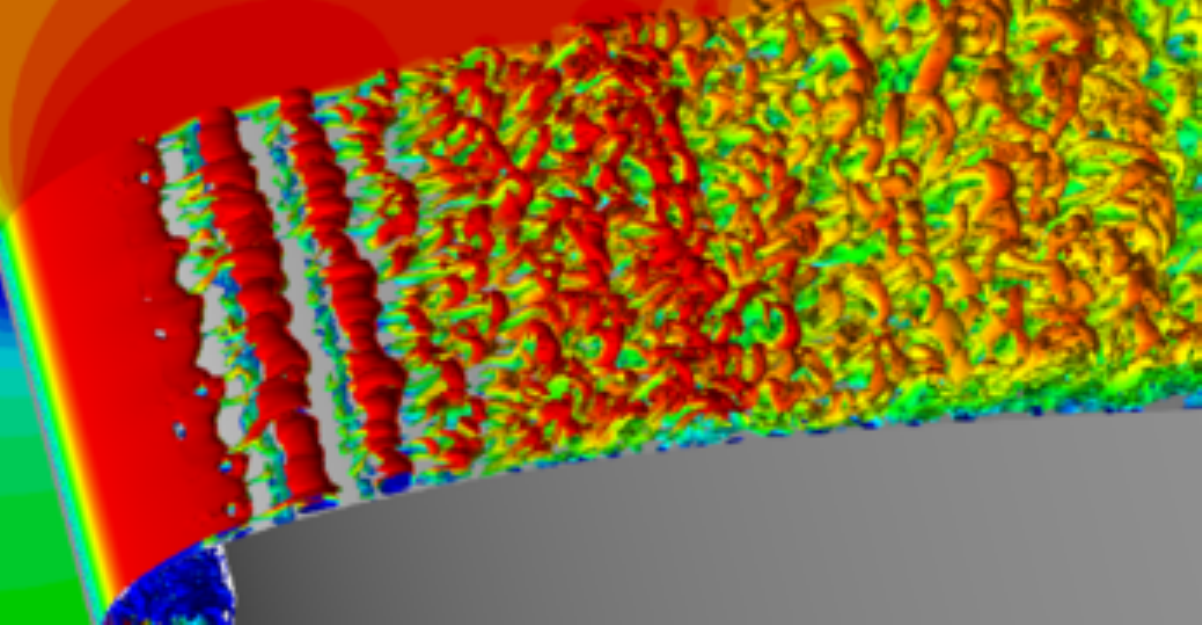}}\\
\subfloat[][Fine grid]{\includegraphics[width=\size\textwidth,clip]{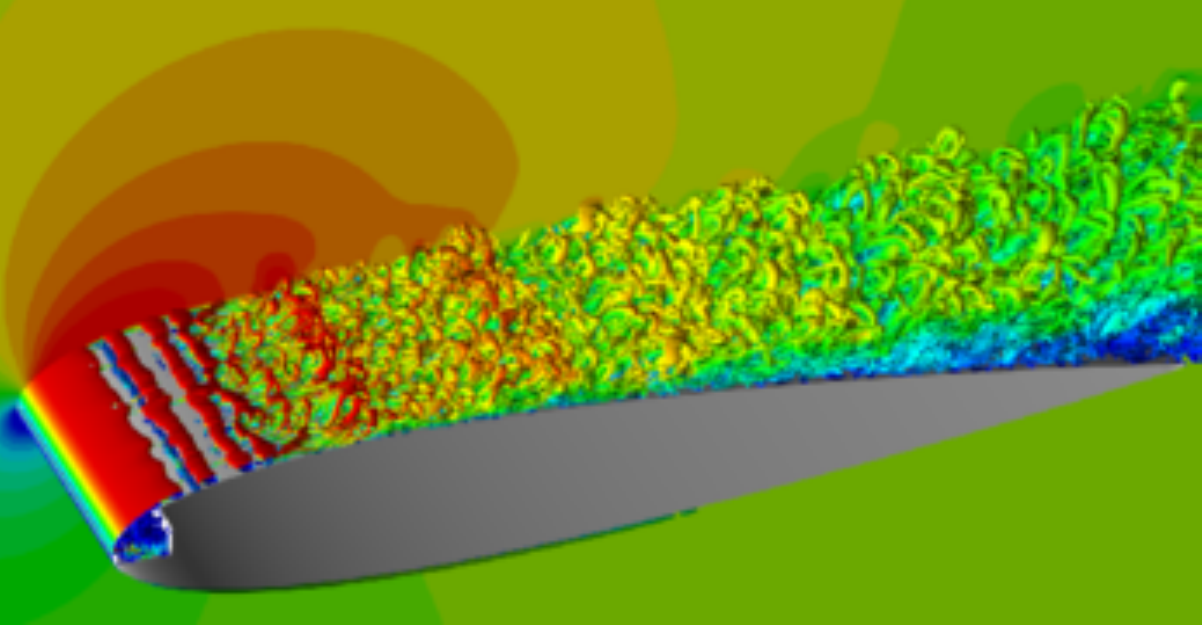}}
\subfloat[][Fine grid (zoom)]{\includegraphics[width=\size\textwidth,clip]{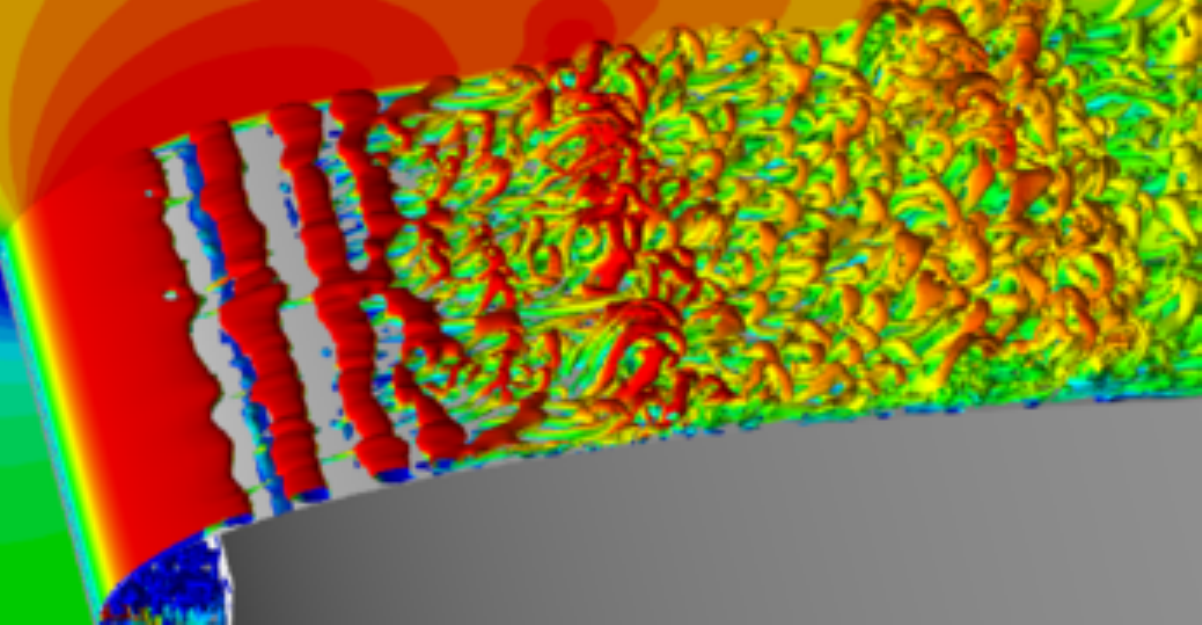}}
\else
\fi
 \caption{
Instantaneous flow fields on the Fine and Medium grids are visualized, using all the grid points.
An isosurface represents a second invariant of the velocity gradient tensor;
both the contour and isosurfaces are colored by the chordwise velocity component: $0.0\leq u/u_\infty\leq 1.5$.
 }\label{Fig_veri01_GCcont_ins}
\end{figure}

\begin{figure}
  \centering
\renewcommand{\size}{0.7}
\ifCONDITION
\subfloat[][Spatial distribution of TKE]{\includegraphics[viewport=2 65 355 190,clip,width=\size\textwidth,clip]{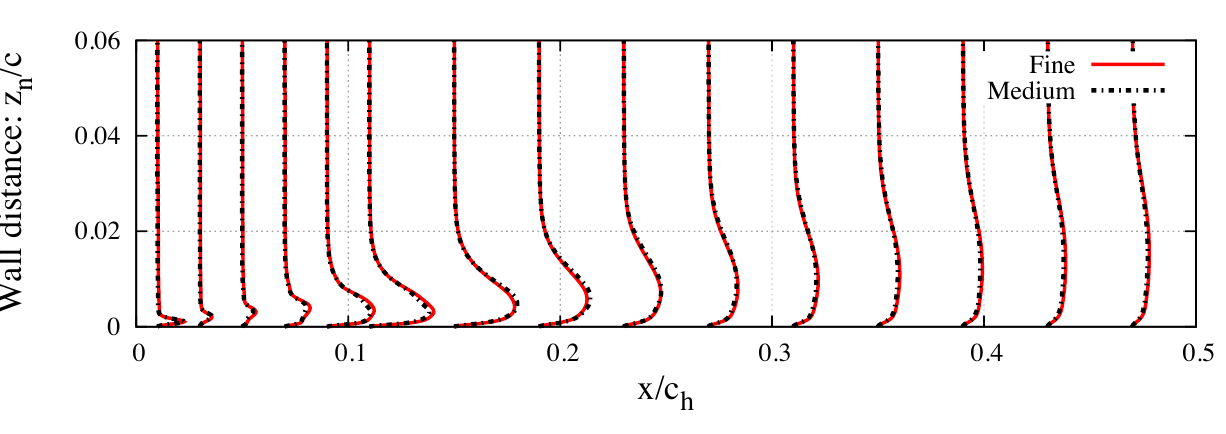}}
\renewcommand{\size}{0.475}
\subfloat[][TKE-max distribution]{\includegraphics[width=\size\textwidth,clip]{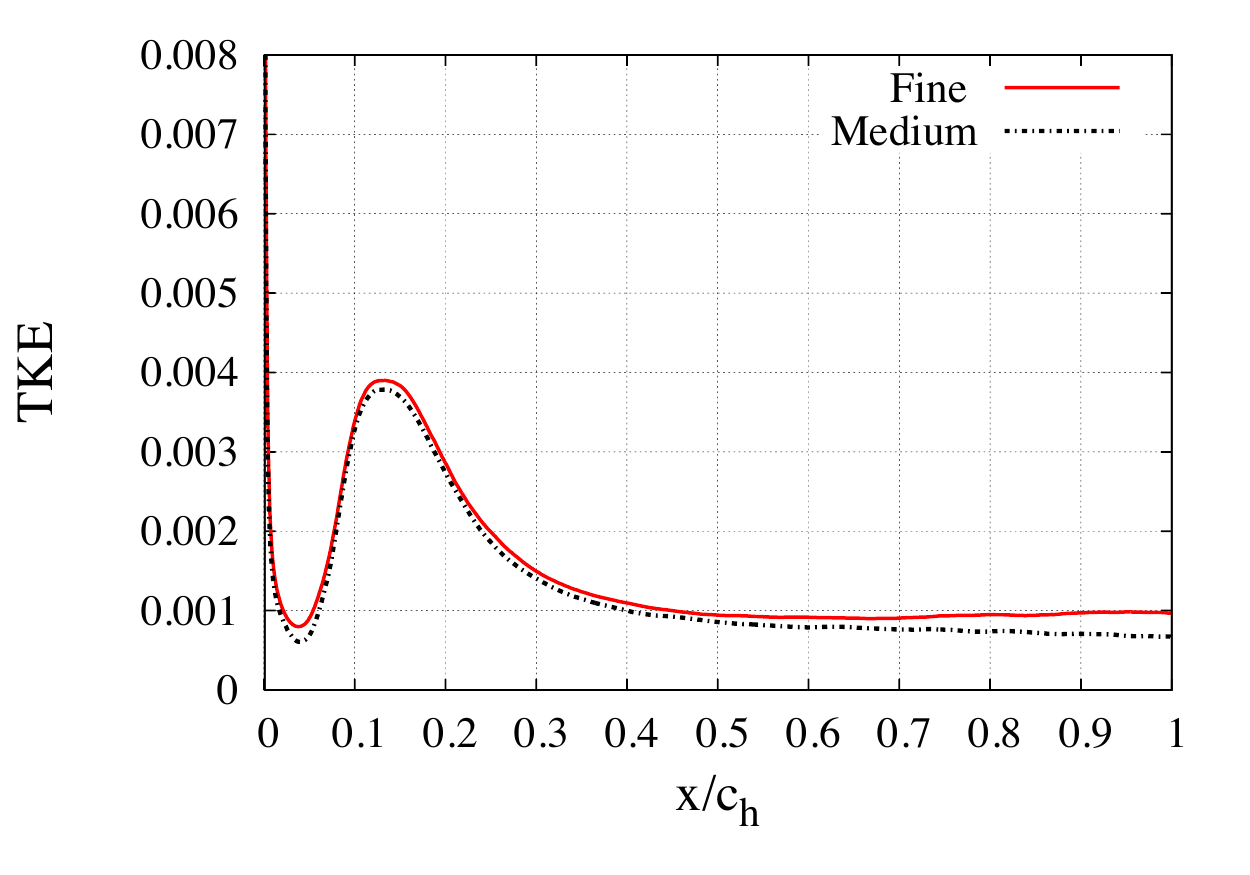}}
\else
\fi
 \caption{
$C_{\rm p}$ and $C_{\rm f}$ distribution of the time- and spanwise-averaged flow fields on the Fine and Medium grids.
 }\label{Fig_veri01_GCcontTKE}
\end{figure}

Figure~\ref{Fig_veri01_GCcont_ins} shows instantaneous flow fields of the Medium and Fine grids.
The isosurfaces show a second invariant of the velocity gradient tensor, which is colored by the chordwise velocity: $u/u_\infty$.
Both of the cases show the similar vortex structures, where the two-dimensional coherent vortices (i.e., only little fluctuation in the spanwise direction) emerge near the leading edge, which are then broken down into three-dimensional turbulent structures in the downstream.
Figure~\ref{Fig_veri01_GCcontTKE}(a) shows the wall-normal profiles of TKE on the suction side of the airfoil at $0\leq x/c_{\text{h}}\leq 0.5$.
These two profiles are almost identical especially concerning the distance between the airfoil surface and the location where the TKE takes the maximum value at each chordwise position: $x/c_{\text{h}}$.
Figure~\ref{Fig_veri01_GCcontTKE}(b) visualizes the spatial distribution of the TKE-max  distribution (the similar plots are given in Fig.~\ref{fig:TKEmax} of Sec.~\ref{subsubsec:tke}) at $0\leq x/c_{\text{h}}\leq 1.0$, where the peak values of TKE at each $x/c_{\text{h}}$ are plotted.
In the vicinity of the trailing edge, the TKE-max value of the Medium grid becomes slightly lower than that of the Fine grid;
nevertheless, a chordwise location of the peak in the TKE-max profile ($x/c_{\text{h}}\simeq 0.15$) is in good agreement between each grid.
This suggests that a turbulent transition occurs in the same chordwise location, which would be considered one of the most important phenomena in the present study.
Based on these observations, the grid density of the Medium grid can be considered sufficient for a discussion of the separation control in this paper.

\subsection{Verification of the time step size}
\begin{figure}
\centering
\ifCONDITION
\includegraphics[width=0.5\textwidth,clip]{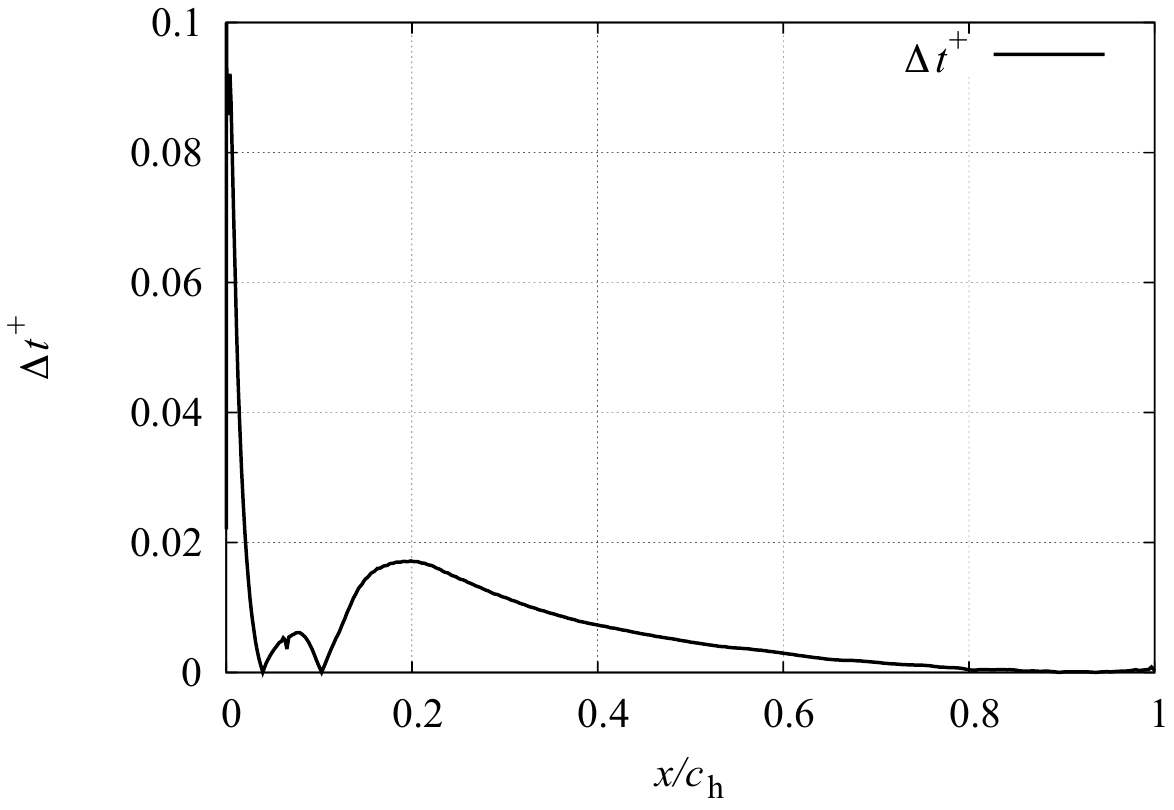}
\else
\fi
\vspace{-0.4cm}
\caption{
Computational time step normalized by the wall-unit scale (on the suction side of the airfoil) is visualized for the controlled case ($\Cm=2.0\dd{-3}, \Fp=6.0$ on the Medium grid).
}\label{Fig_veri03}
\end{figure}

The computational time step nondimensionalized by the freestream velocity and chord length is $\Delta t = 4.0\dd{-5}$ throughout this paper, where the maximum Courant number is approximately $2.0$.
The verification of this time step size is performed by the wall-unit scaling~\cite{Choi1994} for the separation-controlled flow ($\Cm=2.0\dd{-3}$ with $\Fp=6.0$).
The computational time step in the wall-unit scale is obtained as $\Delta t^+ \leq 0.1$ as shown in Fig.~\ref{Fig_veri03} for the Medium grid, which is smaller than that proposed by Ref.~\onlinecite{Choi1994} for turbulent flows, i.e., $\Delta t^+ = O(1)$.

\subsection{Verification of the span length of the grid}
\begin{figure}
  \centering
\renewcommand{\size}{0.475}
\ifCONDITION
\subfloat[][$C_{\rm p}$]{\includegraphics[width=\size\textwidth,clip]{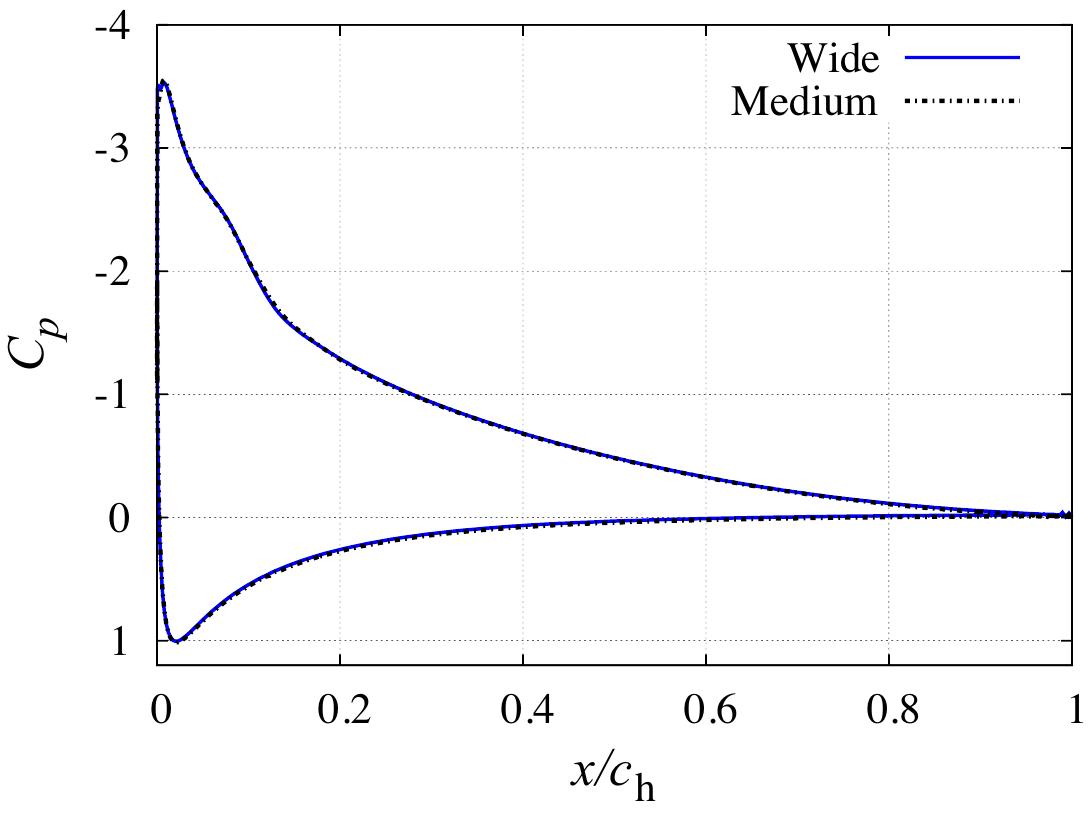}}
\subfloat[][$C_{\rm f}$]{\includegraphics[width=\size\textwidth,clip]{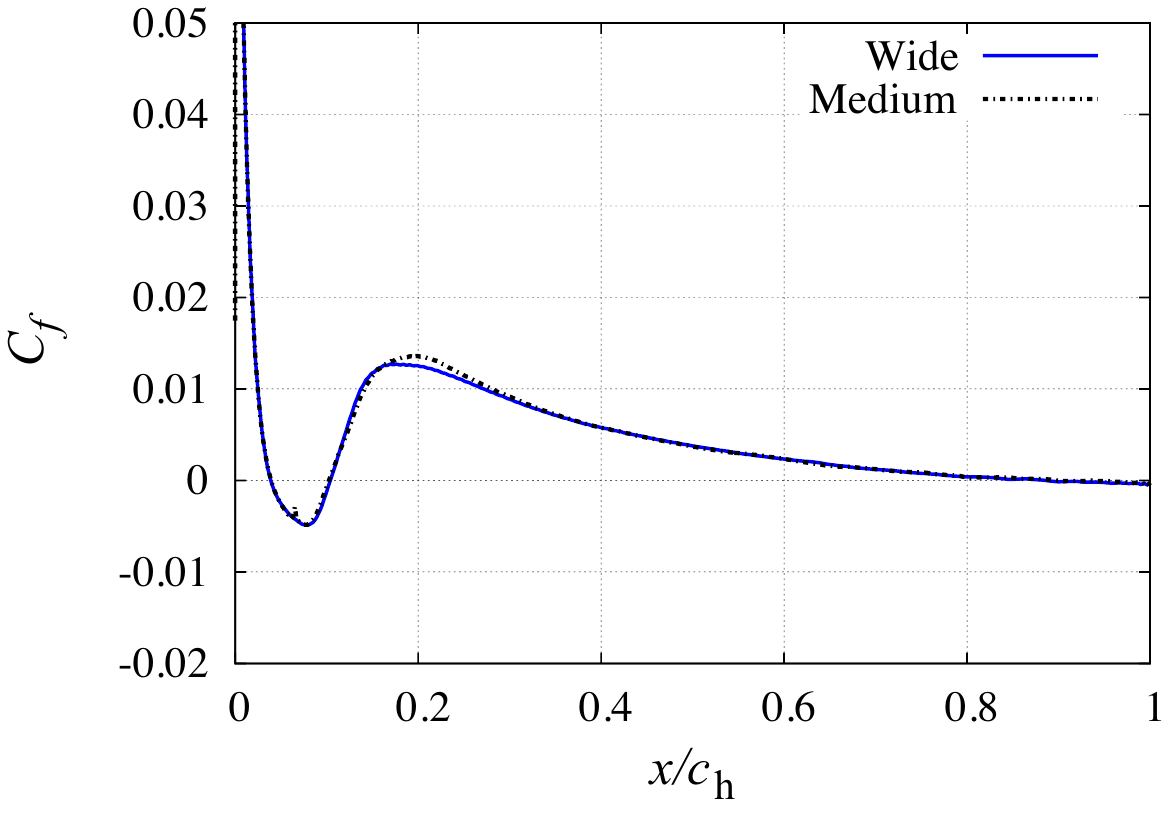}}
\else
\fi
 \caption{
$C_{\rm p}$ and $C_{\rm f}$ distribution of the time- and spanwise-averaged flow fields on the Medium and Wide grids.}\label{Fig_veri01_SPANcont}
\end{figure}

The span length of the computational grid is examined using the Medium and Wide grids shown in Table~\ref{tab:Gridpoints-app}.
The separation-controlled flow using $\Cm=2.0\dd{-3}$ with $\Fp=6.0$ is focused.
Figure~\ref{Fig_veri01_SPANcont} shows the $C_{\rm p}$ and $C_{\rm f}$ distributions on the suction side of the airfoil surface.
Both $C_{\rm p}$ and $C_{\rm f}$ distributions are in good agreement between the Medium and Wide grids, which indicates that the Medium grid provides a converged result concerning the size of a computational domain in the spanwise direction.

\begin{figure}
  \centering
\renewcommand{\size}{0.7}
\ifCONDITION
\subfloat[][Spatial distribution of TKE]{\includegraphics[viewport=2 65 350 190,clip,width=\size\textwidth,clip]{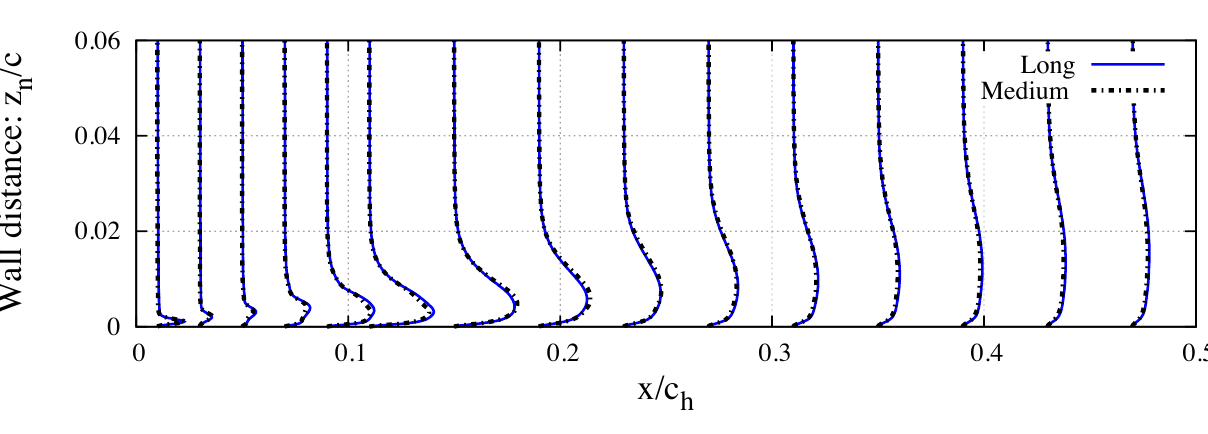}}
\renewcommand{\size}{0.475}
\subfloat[][TKE-max distribution]{\includegraphics[width=\size\textwidth,clip]{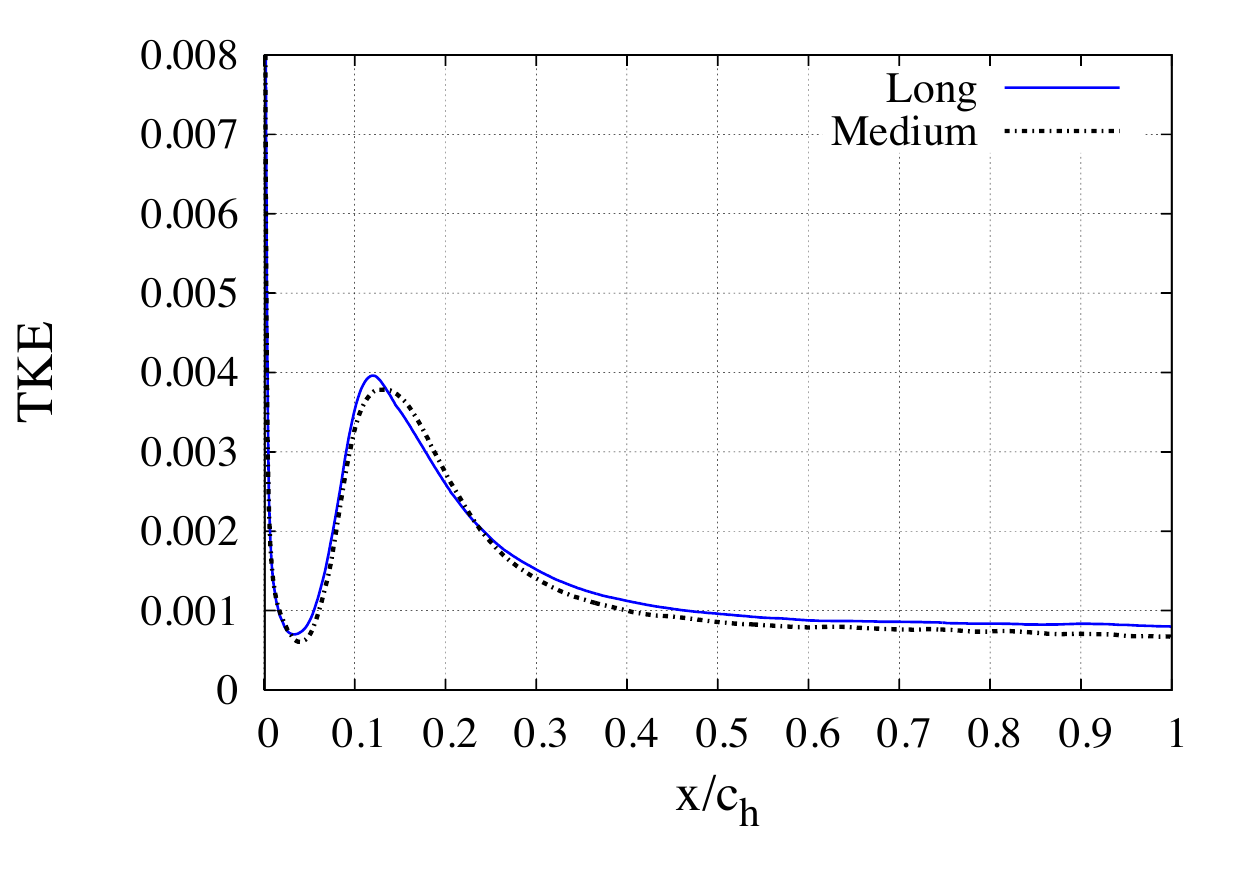}}
\else
\fi
 \caption{
The TKE distribution of the time- and spanwise-averaged flow fields on the Medium and Wide grids.}\label{Fig_veri01_SPANcontTKE}
\end{figure}

Figure~\ref{Fig_veri01_SPANcontTKE}(a) shows the wall-normal profiles of the TKE at $0\leq x/c_{\text{h}}\leq 0.5$.
These profiles show almost the same distribution regarding the distance between the airfoil surface and the location where the TKE takes the maximum value at each chordwise position: $x/c_{\text{h}}$.
Figure~\ref{Fig_veri01_SPANcontTKE}(b) visualizes the spatial distribution of the TKE-max distribution at $0\leq x/c_{\text{h}}\leq 1.0$.
Although the TKE-max values near the trailing edge and the peak at $x/c_{\text{h}}=0.15$ are slightly different between the Medium and Wide grids, the location of turbulent transition is almost the same, which ensures that the spanwise size of the Medium grid (20\% of the chord length) is sufficiently wide in the present study.


\bibliography{flab}

\begin{thebibliography}{99}%
\makeatletter
\providecommand \@ifxundefined [1]{%
 \@ifx{#1\undefined}
}%
\providecommand \@ifnum [1]{%
 \ifnum #1\expandafter \@firstoftwo
 \else \expandafter \@secondoftwo
 \fi
}%
\providecommand \@ifx [1]{%
 \ifx #1\expandafter \@firstoftwo
 \else \expandafter \@secondoftwo
 \fi
}%
\providecommand \natexlab [1]{#1}%
\providecommand \enquote  [1]{``#1''}%
\providecommand \bibnamefont  [1]{#1}%
\providecommand \bibfnamefont [1]{#1}%
\providecommand \citenamefont [1]{#1}%
\providecommand \href@noop [0]{\@secondoftwo}%
\providecommand \href [0]{\begingroup \@sanitize@url \@href}%
\providecommand \@href[1]{\@@startlink{#1}\@@href}%
\providecommand \@@href[1]{\endgroup#1\@@endlink}%
\providecommand \@sanitize@url [0]{\catcode `\\12\catcode `\$12\catcode
  `\&12\catcode `\#12\catcode `\^12\catcode `\_12\catcode `\%12\relax}%
\providecommand \@@startlink[1]{}%
\providecommand \@@endlink[0]{}%
\providecommand \url  [0]{\begingroup\@sanitize@url \@url }%
\providecommand \@url [1]{\endgroup\@href {#1}{\urlprefix }}%
\providecommand \urlprefix  [0]{URL }%
\providecommand \Eprint [0]{\href }%
\providecommand \doibase [0]{https://doi.org/}%
\providecommand \selectlanguage [0]{\@gobble}%
\providecommand \bibinfo  [0]{\@secondoftwo}%
\providecommand \bibfield  [0]{\@secondoftwo}%
\providecommand \translation [1]{[#1]}%
\providecommand \BibitemOpen [0]{}%
\providecommand \bibitemStop [0]{}%
\providecommand \bibitemNoStop [0]{.\EOS\space}%
\providecommand \EOS [0]{\spacefactor3000\relax}%
\providecommand \BibitemShut  [1]{\csname bibitem#1\endcsname}%
\let\auto@bib@innerbib\@empty
\bibitem [{\citenamefont {Pack}\ and\ \citenamefont
  {Seifert}(2001)}]{Pack2001}%
  \BibitemOpen
  \bibfield  {author} {\bibinfo {author} {\bibfnamefont {L.~G.}\ \bibnamefont
  {Pack}}\ and\ \bibinfo {author} {\bibfnamefont {A.}~\bibnamefont {Seifert}},\
  }\bibfield  {title} {\bibinfo {title} {Periodic excitation for jet vectoring
  and enhanced spreading},\ }\href@noop {} {\bibfield  {journal} {\bibinfo
  {journal} {Journal of Aircraft}\ }\textbf {\bibinfo {volume} {38}},\ \bibinfo
  {pages} {486} (\bibinfo {year} {2001})}\BibitemShut {NoStop}%
\bibitem [{\citenamefont {Smith}\ and\ \citenamefont
  {Glezer}(2002)}]{Smith2002}%
  \BibitemOpen
  \bibfield  {author} {\bibinfo {author} {\bibfnamefont {B.~L.}\ \bibnamefont
  {Smith}}\ and\ \bibinfo {author} {\bibfnamefont {A.}~\bibnamefont {Glezer}},\
  }\bibfield  {title} {\bibinfo {title} {Jet vectoring using synthetic jets},\
  }\href@noop {} {\bibfield  {journal} {\bibinfo  {journal} {Journal of Fluid
  Mechanics}\ }\textbf {\bibinfo {volume} {458}},\ \bibinfo {pages} {1}
  (\bibinfo {year} {2002})}\BibitemShut {NoStop}%
\bibitem [{\citenamefont {Chen}\ \emph {et~al.}(2000)\citenamefont {Chen},
  \citenamefont {Yao}, \citenamefont {Beeler}, \citenamefont {Bryant},\ and\
  \citenamefont {Fox}}]{Chen2000}%
  \BibitemOpen
  \bibfield  {author} {\bibinfo {author} {\bibfnamefont {F.}~\bibnamefont
  {Chen}}, \bibinfo {author} {\bibfnamefont {C.}~\bibnamefont {Yao}}, \bibinfo
  {author} {\bibfnamefont {G.}~\bibnamefont {Beeler}}, \bibinfo {author}
  {\bibfnamefont {R.}~\bibnamefont {Bryant}},\ and\ \bibinfo {author}
  {\bibfnamefont {R.}~\bibnamefont {Fox}},\ }\bibfield  {title} {\bibinfo
  {title} {Development of synthetic jet actuators for active flow control at
  nasa langley},\ }in\ \href@noop {} {\emph {\bibinfo {booktitle}
  {AIAA-2000-2405}}}\ (\bibinfo {year} {2000})\BibitemShut {NoStop}%
\bibitem [{\citenamefont {Mittal}\ \emph {et~al.}(2005)\citenamefont {Mittal},
  \citenamefont {Kotapati},\ and\ \citenamefont {III}}]{Mittal2005}%
  \BibitemOpen
  \bibfield  {author} {\bibinfo {author} {\bibfnamefont {R.}~\bibnamefont
  {Mittal}}, \bibinfo {author} {\bibfnamefont {R.~B.}\ \bibnamefont
  {Kotapati}},\ and\ \bibinfo {author} {\bibfnamefont {L.~N.~C.}\ \bibnamefont
  {III}},\ }\bibfield  {title} {\bibinfo {title} {Numerical study of resonant
  interactions and flow control in a canonical separated flow},\ }in\
  \href@noop {} {\emph {\bibinfo {booktitle} {AIAA-2005-1261}}}\ (\bibinfo
  {year} {2005})\BibitemShut {NoStop}%
\bibitem [{\citenamefont {Amitay}\ and\ \citenamefont
  {Glezer}(2002{\natexlab{a}})}]{Amitay2002}%
  \BibitemOpen
  \bibfield  {author} {\bibinfo {author} {\bibfnamefont {M.}~\bibnamefont
  {Amitay}}\ and\ \bibinfo {author} {\bibfnamefont {A.}~\bibnamefont
  {Glezer}},\ }\bibfield  {title} {\bibinfo {title} {Role of actuation
  frequency in controlled flow reattachment over a stalled airfoil},\
  }\href@noop {} {\bibfield  {journal} {\bibinfo  {journal} {AIAA Journal}\
  }\textbf {\bibinfo {volume} {40}},\ \bibinfo {pages} {209} (\bibinfo {year}
  {2002}{\natexlab{a}})}\BibitemShut {NoStop}%
\bibitem [{\citenamefont {Tensi}\ \emph {et~al.}(2002)\citenamefont {Tensi},
  \citenamefont {Bou\'e}, \citenamefont {Paill\'e},\ and\ \citenamefont
  {Dury}}]{Tensi2002}%
  \BibitemOpen
  \bibfield  {author} {\bibinfo {author} {\bibfnamefont {J.}~\bibnamefont
  {Tensi}}, \bibinfo {author} {\bibfnamefont {I.}~\bibnamefont {Bou\'e}},
  \bibinfo {author} {\bibfnamefont {F.}~\bibnamefont {Paill\'e}},\ and\
  \bibinfo {author} {\bibfnamefont {G.}~\bibnamefont {Dury}},\ }\bibfield
  {title} {\bibinfo {title} {Modification of the wake behind a circular
  cylinder by using synthetic jets},\ }\href@noop {} {\bibfield  {journal}
  {\bibinfo  {journal} {Journal of Visualization}\ }\textbf {\bibinfo {volume}
  {5}},\ \bibinfo {pages} {37} (\bibinfo {year} {2002})}\BibitemShut {NoStop}%
\bibitem [{\citenamefont {Kercher}\ \emph {et~al.}(2003)\citenamefont
  {Kercher}, \citenamefont {Lee}, \citenamefont {Brand}, \citenamefont
  {Allen},\ and\ \citenamefont {Glezer}}]{Kercher2003}%
  \BibitemOpen
  \bibfield  {author} {\bibinfo {author} {\bibfnamefont {D.~S.}\ \bibnamefont
  {Kercher}}, \bibinfo {author} {\bibfnamefont {J.~B.}\ \bibnamefont {Lee}},
  \bibinfo {author} {\bibfnamefont {O.}~\bibnamefont {Brand}}, \bibinfo
  {author} {\bibfnamefont {M.}~\bibnamefont {Allen}},\ and\ \bibinfo {author}
  {\bibfnamefont {A.}~\bibnamefont {Glezer}},\ }\bibfield  {title} {\bibinfo
  {title} {Microjet cooling devices for thermal management of electronics},\
  }\href@noop {} {\bibfield  {journal} {\bibinfo  {journal} {Journal of
  Chemical Theory Computation}\ }\textbf {\bibinfo {volume} {26}},\ \bibinfo
  {pages} {359} (\bibinfo {year} {2003})}\BibitemShut {NoStop}%
\bibitem [{\citenamefont {Mahalingam}\ and\ \citenamefont
  {Menart}(2010)}]{Mahalingam2010}%
  \BibitemOpen
  \bibfield  {author} {\bibinfo {author} {\bibfnamefont {S.}~\bibnamefont
  {Mahalingam}}\ and\ \bibinfo {author} {\bibfnamefont {J.~A.}\ \bibnamefont
  {Menart}},\ }\bibfield  {title} {\bibinfo {title} {Particle-based plasma
  simulations for an ion engine discharge chamber},\ }\href
  {https://doi.org/10.2514/1.45954} {\bibfield  {journal} {\bibinfo  {journal}
  {Journal of Propulsion and Power}\ }\textbf {\bibinfo {volume} {Vol. 26}},\
  \bibinfo {pages} {673} (\bibinfo {year} {2010})}\BibitemShut {NoStop}%
\bibitem [{\citenamefont {Chaudhari}\ \emph {et~al.}(2010)\citenamefont
  {Chaudhari}, \citenamefont {Puranik},\ and\ \citenamefont
  {Agrawal}}]{Chaudhari2010}%
  \BibitemOpen
  \bibfield  {author} {\bibinfo {author} {\bibfnamefont {M.}~\bibnamefont
  {Chaudhari}}, \bibinfo {author} {\bibfnamefont {B.}~\bibnamefont {Puranik}},\
  and\ \bibinfo {author} {\bibfnamefont {A.}~\bibnamefont {Agrawal}},\
  }\bibfield  {title} {\bibinfo {title} {Heat transfer characteristics of
  synthetic jet impingement cooling},\ }\href@noop {} {\bibfield  {journal}
  {\bibinfo  {journal} {International Journal of Heat and Mass Transfer}\
  }\textbf {\bibinfo {volume} {53}},\ \bibinfo {pages} {1057} (\bibinfo {year}
  {2010})}\BibitemShut {NoStop}%
\bibitem [{\citenamefont {Wang}\ \emph {et~al.}(2001)\citenamefont {Wang},
  \citenamefont {Wu}, \citenamefont {Wang}, \citenamefont {Igra},\ and\
  \citenamefont {Falcovitz}}]{Wang2001}%
  \BibitemOpen
  \bibfield  {author} {\bibinfo {author} {\bibfnamefont {B.}~\bibnamefont
  {Wang}}, \bibinfo {author} {\bibfnamefont {Q.}~\bibnamefont {Wu}}, \bibinfo
  {author} {\bibfnamefont {C.}~\bibnamefont {Wang}}, \bibinfo {author}
  {\bibfnamefont {O.}~\bibnamefont {Igra}},\ and\ \bibinfo {author}
  {\bibfnamefont {J.}~\bibnamefont {Falcovitz}},\ }\bibfield  {title} {\bibinfo
  {title} {Shock wave diffraction by a square cavity filled with dusty gas},\
  }\href@noop {} {\bibfield  {journal} {\bibinfo  {journal} {Shock Waves}\
  }\textbf {\bibinfo {volume} {11}},\ \bibinfo {pages} {7} (\bibinfo {year}
  {2001})}\BibitemShut {NoStop}%
\bibitem [{\citenamefont {Nishioka}\ \emph {et~al.}(1989)\citenamefont
  {Nishioka}, \citenamefont {Asai},\ and\ \citenamefont
  {Yoshida}}]{Nishioka1989}%
  \BibitemOpen
  \bibfield  {author} {\bibinfo {author} {\bibfnamefont {M.}~\bibnamefont
  {Nishioka}}, \bibinfo {author} {\bibfnamefont {M.}~\bibnamefont {Asai}},\
  and\ \bibinfo {author} {\bibfnamefont {S.}~\bibnamefont {Yoshida}},\
  }\bibfield  {title} {\bibinfo {title} {Control of flow separation by acoustic
  excitation},\ }in\ \href@noop {} {\emph {\bibinfo {booktitle}
  {AIAA-1989-973}}}\ (\bibinfo {year} {1989})\BibitemShut {NoStop}%
\bibitem [{\citenamefont {Vatsa}(2012)}]{Vatsa2012}%
  \BibitemOpen
  \bibfield  {author} {\bibinfo {author} {\bibfnamefont {V.}~\bibnamefont
  {Vatsa}},\ }\bibfield  {title} {\bibinfo {title} {Numerical simulation of
  fluidic actuators for flow control applications},\ }in\ \href@noop {} {\emph
  {\bibinfo {booktitle} {AIAA-2012-3239}}}\ (\bibinfo {year}
  {2012})\BibitemShut {NoStop}%
\bibitem [{\citenamefont {Graff}\ \emph {et~al.}(1990)\citenamefont {Graff},
  \citenamefont {Seele},\ and\ \citenamefont {Lin}}]{Graff2013}%
  \BibitemOpen
  \bibfield  {author} {\bibinfo {author} {\bibfnamefont {E.}~\bibnamefont
  {Graff}}, \bibinfo {author} {\bibfnamefont {R.}~\bibnamefont {Seele}},\ and\
  \bibinfo {author} {\bibfnamefont {J.}~\bibnamefont {Lin}},\ }\href@noop {}
  {\emph {\bibinfo {title} {Sweeping Jet Actuators - a New Design Tool for High
  Lift Generation}}},\ \bibinfo {type} {Tech. Rep.}\ (\bibinfo  {institution}
  {NASA Technical Reports},\ \bibinfo {year} {1990})\BibitemShut {NoStop}%
\bibitem [{\citenamefont {Corke}\ \emph {et~al.}(2010)\citenamefont {Corke},
  \citenamefont {Enloe},\ and\ \citenamefont {P.Wilkinson}}]{Corke2010}%
  \BibitemOpen
  \bibfield  {author} {\bibinfo {author} {\bibfnamefont {T.~C.}\ \bibnamefont
  {Corke}}, \bibinfo {author} {\bibfnamefont {C.~L.}\ \bibnamefont {Enloe}},\
  and\ \bibinfo {author} {\bibfnamefont {S.}~\bibnamefont {P.Wilkinson}},\
  }\bibfield  {title} {\bibinfo {title} {Dielectric barrier discharge plasma
  actuators for flow control},\ }\href@noop {} {\bibfield  {journal} {\bibinfo
  {journal} {Annual Review of Fluid Mechanics}\ } (\bibinfo {year}
  {2010})}\BibitemShut {NoStop}%
\bibitem [{\citenamefont {Fujii}(2014)}]{Fujii2014}%
  \BibitemOpen
  \bibfield  {author} {\bibinfo {author} {\bibfnamefont {K.}~\bibnamefont
  {Fujii}},\ }\bibfield  {title} {\bibinfo {title} {High-performance
  computing-based exploration of flow control with micro devices},\ }\href@noop
  {} {\bibfield  {journal} {\bibinfo  {journal} {Philosophical Transactions of
  Royal Society}\ }\textbf {\bibinfo {volume} {372}} (\bibinfo {year}
  {2014})}\BibitemShut {NoStop}%
\bibitem [{\citenamefont {Glezer}\ and\ \citenamefont
  {Amitay}(2002)}]{Glezer2002}%
  \BibitemOpen
  \bibfield  {author} {\bibinfo {author} {\bibfnamefont {A.}~\bibnamefont
  {Glezer}}\ and\ \bibinfo {author} {\bibfnamefont {M.}~\bibnamefont
  {Amitay}},\ }\bibfield  {title} {\bibinfo {title} {Synthetic jets},\
  }\href@noop {} {\bibfield  {journal} {\bibinfo  {journal} {Annual Review of
  Fluid Mechanics}\ }\textbf {\bibinfo {volume} {34}},\ \bibinfo {pages} {503}
  (\bibinfo {year} {2002})}\BibitemShut {NoStop}%
\bibitem [{\citenamefont {Greenblatt}\ \emph {et~al.}(2008)\citenamefont
  {Greenblatt}, \citenamefont {Goksel}, \citenamefont {Schule}, \citenamefont
  {Romann},\ and\ \citenamefont {Paschereit}}]{Greenblatt2008}%
  \BibitemOpen
  \bibfield  {author} {\bibinfo {author} {\bibfnamefont {D.}~\bibnamefont
  {Greenblatt}}, \bibinfo {author} {\bibfnamefont {B.}~\bibnamefont {Goksel}},
  \bibinfo {author} {\bibfnamefont {C.~Y.}\ \bibnamefont {Schule}}, \bibinfo
  {author} {\bibfnamefont {D.}~\bibnamefont {Romann}},\ and\ \bibinfo {author}
  {\bibfnamefont {C.~O.}\ \bibnamefont {Paschereit}},\ }\bibfield  {title}
  {\bibinfo {title} {Dielectric barrier discharge flow control at very low
  flight reynolds numbers},\ }\href@noop {} {\bibfield  {journal} {\bibinfo
  {journal} {AIAA Journal}\ }\textbf {\bibinfo {volume} {46}},\ \bibinfo
  {pages} {1528} (\bibinfo {year} {2008})}\BibitemShut {NoStop}%
\bibitem [{\citenamefont {Nagib}\ and\ \citenamefont
  {Kiedaisch}(2009)}]{Nagib2004}%
  \BibitemOpen
  \bibfield  {author} {\bibinfo {author} {\bibfnamefont {H.~M.}\ \bibnamefont
  {Nagib}}\ and\ \bibinfo {author} {\bibfnamefont {J.~W.}\ \bibnamefont
  {Kiedaisch}},\ }\bibfield  {title} {\bibinfo {title} {First-in-flight
  full-scale applications of active flow control: The {XV}-15 tiltrotor
  download reduction},\ }in\ \href@noop {} {\emph {\bibinfo {booktitle}
  {RTO-MP-AVT-111}}}\ (\bibinfo {year} {2009})\BibitemShut {NoStop}%
\bibitem [{\citenamefont {Martin}\ \emph {et~al.}(2014)\citenamefont {Martin},
  \citenamefont {Ovemeyer}, \citenamefont {Tanner},\ and\ \citenamefont
  {Wilson}}]{Martin2014}%
  \BibitemOpen
  \bibfield  {author} {\bibinfo {author} {\bibfnamefont {P.~B.}\ \bibnamefont
  {Martin}}, \bibinfo {author} {\bibfnamefont {A.~D.}\ \bibnamefont
  {Ovemeyer}}, \bibinfo {author} {\bibfnamefont {P.~E.}\ \bibnamefont
  {Tanner}},\ and\ \bibinfo {author} {\bibfnamefont {J.~S.}\ \bibnamefont
  {Wilson}},\ }\bibfield  {title} {\bibinfo {title} {Helicopter fuselage active
  flow control in the presence of a rotor},\ }in\ \href@noop {} {\emph
  {\bibinfo {booktitle} {AIAA-2014}}}\ (\bibinfo {year} {2014})\BibitemShut
  {NoStop}%
\bibitem [{\citenamefont {Enloe}\ \emph {et~al.}(2004)\citenamefont {Enloe},
  \citenamefont {McLaughlin}, \citenamefont {VanDyken}, \citenamefont
  {Kachner}, \citenamefont {Jumper},\ and\ \citenamefont {Corke}}]{Enloe2004}%
  \BibitemOpen
  \bibfield  {author} {\bibinfo {author} {\bibfnamefont {C.~L.}\ \bibnamefont
  {Enloe}}, \bibinfo {author} {\bibfnamefont {T.~E.}\ \bibnamefont
  {McLaughlin}}, \bibinfo {author} {\bibfnamefont {R.~D.}\ \bibnamefont
  {VanDyken}}, \bibinfo {author} {\bibfnamefont {K.~D.}\ \bibnamefont
  {Kachner}}, \bibinfo {author} {\bibfnamefont {E.~J.}\ \bibnamefont
  {Jumper}},\ and\ \bibinfo {author} {\bibfnamefont {T.~C.}\ \bibnamefont
  {Corke}},\ }\bibfield  {title} {\bibinfo {title} {Mechanisms and responses of
  a single dielectric barrier plasma actuator: Geometric effect},\ }\href@noop
  {} {\bibfield  {journal} {\bibinfo  {journal} {AIAA Journal}\ }\textbf
  {\bibinfo {volume} {42}},\ \bibinfo {pages} {595} (\bibinfo {year}
  {2004})}\BibitemShut {NoStop}%
\bibitem [{\citenamefont {Visbal}\ \emph {et~al.}(2006)\citenamefont {Visbal},
  \citenamefont {Gaitonde},\ and\ \citenamefont {Roy}}]{Visbal2006}%
  \BibitemOpen
  \bibfield  {author} {\bibinfo {author} {\bibfnamefont {M.~R.}\ \bibnamefont
  {Visbal}}, \bibinfo {author} {\bibfnamefont {D.~V.}\ \bibnamefont
  {Gaitonde}},\ and\ \bibinfo {author} {\bibfnamefont {S.}~\bibnamefont
  {Roy}},\ }\bibfield  {title} {\bibinfo {title} {Control of transitional and
  turbulent flows using plasma-based actuators},\ }in\ \href@noop {} {\emph
  {\bibinfo {booktitle} {AIAA-2006-3230}}}\ (\bibinfo {year}
  {2006})\BibitemShut {NoStop}%
\bibitem [{\citenamefont {Patel}\ \emph {et~al.}(2007)\citenamefont {Patel},
  \citenamefont {Ng}, \citenamefont {Vasudevan}, \citenamefont {Corke},\ and\
  \citenamefont {He}}]{Patel2007a}%
  \BibitemOpen
  \bibfield  {author} {\bibinfo {author} {\bibfnamefont {M.~P.}\ \bibnamefont
  {Patel}}, \bibinfo {author} {\bibfnamefont {T.~T.}\ \bibnamefont {Ng}},
  \bibinfo {author} {\bibfnamefont {S.}~\bibnamefont {Vasudevan}}, \bibinfo
  {author} {\bibfnamefont {T.~C.}\ \bibnamefont {Corke}},\ and\ \bibinfo
  {author} {\bibfnamefont {C.}~\bibnamefont {He}},\ }\bibfield  {title}
  {\bibinfo {title} {Plasma actuators for hingeless aerodynamic control of an
  unmanned air vehicle},\ }\href {https://doi.org/10.2514/1.25368} {\bibfield
  {journal} {\bibinfo  {journal} {Journal of Aircraft}\ }\textbf {\bibinfo
  {volume} {Vol. 44}},\ \bibinfo {pages} {1264} (\bibinfo {year}
  {2007})}\BibitemShut {NoStop}%
\bibitem [{\citenamefont {Sidorenko}\ \emph {et~al.}(2007)\citenamefont
  {Sidorenko}, \citenamefont {Zanin}, \citenamefont {Postnikov}, \citenamefont
  {Budovsky}, \citenamefont {Starikovskii}, \citenamefont {Roupassov},
  \citenamefont {Zavialov}, \citenamefont {Malmuth}, \citenamefont
  {Smereczniak},\ and\ \citenamefont {Silkey}}]{Sidorenko2007}%
  \BibitemOpen
  \bibfield  {author} {\bibinfo {author} {\bibfnamefont {A.~A.}\ \bibnamefont
  {Sidorenko}}, \bibinfo {author} {\bibfnamefont {B.~Y.}\ \bibnamefont
  {Zanin}}, \bibinfo {author} {\bibfnamefont {B.~V.}\ \bibnamefont
  {Postnikov}}, \bibinfo {author} {\bibfnamefont {A.~D.}\ \bibnamefont
  {Budovsky}}, \bibinfo {author} {\bibfnamefont {A.~Y.}\ \bibnamefont
  {Starikovskii}}, \bibinfo {author} {\bibfnamefont {D.~V.}\ \bibnamefont
  {Roupassov}}, \bibinfo {author} {\bibfnamefont {I.~N.}\ \bibnamefont
  {Zavialov}}, \bibinfo {author} {\bibfnamefont {N.~D.}\ \bibnamefont
  {Malmuth}}, \bibinfo {author} {\bibfnamefont {P.}~\bibnamefont
  {Smereczniak}},\ and\ \bibinfo {author} {\bibfnamefont {J.~S.}\ \bibnamefont
  {Silkey}},\ }\bibfield  {title} {\bibinfo {title} {Pulsed discharge actuators
  for rectangular wings separation control},\ }in\ \href@noop {} {\emph
  {\bibinfo {booktitle} {AIAA-2007-941}}}\ (\bibinfo {year} {2007})\BibitemShut
  {NoStop}%
\bibitem [{\citenamefont {Wang}\ \emph {et~al.}(2013)\citenamefont {Wang},
  \citenamefont {Choi}, \citenamefont {Feng}, \citenamefont {Jukes},\ and\
  \citenamefont {Whalley}}]{WangChoi2013}%
  \BibitemOpen
  \bibfield  {author} {\bibinfo {author} {\bibfnamefont {J.-J.}\ \bibnamefont
  {Wang}}, \bibinfo {author} {\bibfnamefont {K.-S.}\ \bibnamefont {Choi}},
  \bibinfo {author} {\bibfnamefont {L.-H.}\ \bibnamefont {Feng}}, \bibinfo
  {author} {\bibfnamefont {T.~N.}\ \bibnamefont {Jukes}},\ and\ \bibinfo
  {author} {\bibfnamefont {R.~D.}\ \bibnamefont {Whalley}},\ }\bibfield
  {title} {\bibinfo {title} {Recent developments in {DBD} plasma flow
  control},\ }\href@noop {} {\bibfield  {journal} {\bibinfo  {journal} {Prog.
  Aerosp. Sci.}\ }\textbf {\bibinfo {volume} {62}},\ \bibinfo {pages} {52}
  (\bibinfo {year} {2013})}\BibitemShut {NoStop}%
\bibitem [{\citenamefont {Sato}\ \emph {et~al.}(2015)\citenamefont {Sato},
  \citenamefont {Nonomura}, \citenamefont {Okada}, \citenamefont {Asada},
  \citenamefont {Aono}, \citenamefont {Yakeno}, \citenamefont {Abe},\ and\
  \citenamefont {Fujii}}]{Sato2015PoF}%
  \BibitemOpen
  \bibfield  {author} {\bibinfo {author} {\bibfnamefont {M.}~\bibnamefont
  {Sato}}, \bibinfo {author} {\bibfnamefont {T.}~\bibnamefont {Nonomura}},
  \bibinfo {author} {\bibfnamefont {K.}~\bibnamefont {Okada}}, \bibinfo
  {author} {\bibfnamefont {K.}~\bibnamefont {Asada}}, \bibinfo {author}
  {\bibfnamefont {H.}~\bibnamefont {Aono}}, \bibinfo {author} {\bibfnamefont
  {A.}~\bibnamefont {Yakeno}}, \bibinfo {author} {\bibfnamefont
  {Y.}~\bibnamefont {Abe}},\ and\ \bibinfo {author} {\bibfnamefont
  {K.}~\bibnamefont {Fujii}},\ }\bibfield  {title} {\bibinfo {title}
  {Mechanisms for laminar separated-flow control using dbd plasma actuator at
  low reynolds number},\ }\href@noop {} {\bibfield  {journal} {\bibinfo
  {journal} {Physics of Fluids}\ }\textbf {\bibinfo {volume} {27}},\ \bibinfo
  {pages} {117101} (\bibinfo {year} {2015})}\BibitemShut {NoStop}%
\bibitem [{\citenamefont {Yakeno}\ \emph {et~al.}(2015)\citenamefont {Yakeno},
  \citenamefont {Kawai}, \citenamefont {Nonomura},\ and\ \citenamefont
  {Fujii}}]{Yakeno2015}%
  \BibitemOpen
  \bibfield  {author} {\bibinfo {author} {\bibfnamefont {A.}~\bibnamefont
  {Yakeno}}, \bibinfo {author} {\bibfnamefont {S.}~\bibnamefont {Kawai}},
  \bibinfo {author} {\bibfnamefont {T.}~\bibnamefont {Nonomura}},\ and\
  \bibinfo {author} {\bibfnamefont {K.}~\bibnamefont {Fujii}},\ }\bibfield
  {title} {\bibinfo {title} {Separation control based on turbulence transition
  around a two-dimensional hump at different reynolds numbers},\ }\href@noop {}
  {\bibfield  {journal} {\bibinfo  {journal} {International Journal of Heat and
  Fluid Flow}\ }\textbf {\bibinfo {volume} {55}},\ \bibinfo {pages} {52}
  (\bibinfo {year} {2015})}\BibitemShut {NoStop}%
\bibitem [{\citenamefont {Yakeno}\ \emph {et~al.}(2017)\citenamefont {Yakeno},
  \citenamefont {Abe}, \citenamefont {Kawai}, \citenamefont {Nonomura},\ and\
  \citenamefont {Fujii}}]{Yakeno2017}%
  \BibitemOpen
  \bibfield  {author} {\bibinfo {author} {\bibfnamefont {A.}~\bibnamefont
  {Yakeno}}, \bibinfo {author} {\bibfnamefont {Y.}~\bibnamefont {Abe}},
  \bibinfo {author} {\bibfnamefont {S.}~\bibnamefont {Kawai}}, \bibinfo
  {author} {\bibfnamefont {T.}~\bibnamefont {Nonomura}},\ and\ \bibinfo
  {author} {\bibfnamefont {K.}~\bibnamefont {Fujii}},\ }\bibfield  {title}
  {\bibinfo {title} {Spanwise modulation effects of local body force on
  downstream turbulence growth around two-dimensional hump},\ }\href@noop {}
  {\bibfield  {journal} {\bibinfo  {journal} {International Journal of Heat and
  Fluid Flow}\ }\textbf {\bibinfo {volume} {63}},\ \bibinfo {pages} {108}
  (\bibinfo {year} {2017})}\BibitemShut {NoStop}%
\bibitem [{\citenamefont {Yarusevych}\ and\ \citenamefont
  {Kotsonis}(2017{\natexlab{a}})}]{Yarusevych2017}%
  \BibitemOpen
  \bibfield  {author} {\bibinfo {author} {\bibfnamefont {S.}~\bibnamefont
  {Yarusevych}}\ and\ \bibinfo {author} {\bibfnamefont {M.}~\bibnamefont
  {Kotsonis}},\ }\bibfield  {title} {\bibinfo {title} {Effect of local {DBD}
  plasma actuation on transition in a laminar separation bubble},\ }\href@noop
  {} {\bibfield  {journal} {\bibinfo  {journal} {Flow Turbulence Combustion}\
  }\textbf {\bibinfo {volume} {98}},\ \bibinfo {pages} {195} (\bibinfo {year}
  {2017}{\natexlab{a}})}\BibitemShut {NoStop}%
\bibitem [{\citenamefont {Ziad\'e}\ \emph {et~al.}(2018)\citenamefont
  {Ziad\'e}, \citenamefont {Feerob},\ and\ \citenamefont
  {Sullivanc}}]{Ziade2018}%
  \BibitemOpen
  \bibfield  {author} {\bibinfo {author} {\bibfnamefont {P.}~\bibnamefont
  {Ziad\'e}}, \bibinfo {author} {\bibfnamefont {M.~A.}\ \bibnamefont
  {Feerob}},\ and\ \bibinfo {author} {\bibfnamefont {P.~E.}\ \bibnamefont
  {Sullivanc}},\ }\bibfield  {title} {\bibinfo {title} {A numerical study on
  the influence of cavity shape on synthetic jet performance},\ }\href@noop {}
  {\bibfield  {journal} {\bibinfo  {journal} {International Journal of Heat and
  Fluid Flow}\ }\textbf {\bibinfo {volume} {74}},\ \bibinfo {pages} {187}
  (\bibinfo {year} {2018})}\BibitemShut {NoStop}%
\bibitem [{\citenamefont {Benton}\ and\ \citenamefont
  {Visbal}(2019)}]{Benton2019}%
  \BibitemOpen
  \bibfield  {author} {\bibinfo {author} {\bibfnamefont {S.~I.}\ \bibnamefont
  {Benton}}\ and\ \bibinfo {author} {\bibfnamefont {M.~R.}\ \bibnamefont
  {Visbal}},\ }\bibfield  {title} {\bibinfo {title} {Extending the reynolds
  number range of high-frequency control of dynamic stall},\ }\href@noop {}
  {\bibfield  {journal} {\bibinfo  {journal} {AIAA Journal}\ }\textbf {\bibinfo
  {volume} {57(7)}},\ \bibinfo {pages} {2676} (\bibinfo {year}
  {2019})}\BibitemShut {NoStop}%
\bibitem [{\citenamefont {Sato}\ \emph {et~al.}(2020)\citenamefont {Sato},
  \citenamefont {Okada}, \citenamefont {Asada}, \citenamefont {Aono},
  \citenamefont {Nonomura},\ and\ \citenamefont {Fujii}}]{Sato2020}%
  \BibitemOpen
  \bibfield  {author} {\bibinfo {author} {\bibfnamefont {M.}~\bibnamefont
  {Sato}}, \bibinfo {author} {\bibfnamefont {K.}~\bibnamefont {Okada}},
  \bibinfo {author} {\bibfnamefont {K.}~\bibnamefont {Asada}}, \bibinfo
  {author} {\bibfnamefont {H.}~\bibnamefont {Aono}}, \bibinfo {author}
  {\bibfnamefont {T.}~\bibnamefont {Nonomura}},\ and\ \bibinfo {author}
  {\bibfnamefont {K.}~\bibnamefont {Fujii}},\ }\bibfield  {title} {\bibinfo
  {title} {Unified mechanisms for separation control around airfoil using
  plasma actuator with burst actuation over {R}eynolds number range of
  10$^3$-10$^6$},\ }\href@noop {} {\bibfield  {journal} {\bibinfo  {journal}
  {Physics of Fluids}\ }\textbf {\bibinfo {volume} {32(2)}} (\bibinfo {year}
  {2020})}\BibitemShut {NoStop}%
\bibitem [{\citenamefont {Rizzetta}\ \emph {et~al.}(1999)\citenamefont
  {Rizzetta}, \citenamefont {Visbal},\ and\ \citenamefont
  {Stanek}}]{Rizzetta1999a}%
  \BibitemOpen
  \bibfield  {author} {\bibinfo {author} {\bibfnamefont {D.~P.}\ \bibnamefont
  {Rizzetta}}, \bibinfo {author} {\bibfnamefont {M.~R.}\ \bibnamefont
  {Visbal}},\ and\ \bibinfo {author} {\bibfnamefont {M.~J.}\ \bibnamefont
  {Stanek}},\ }\bibfield  {title} {\bibinfo {title} {Numerical investigation of
  synthetic-jet flowfields},\ }\href@noop {} {\bibfield  {journal} {\bibinfo
  {journal} {AIAA Journal}\ }\textbf {\bibinfo {volume} {37}},\ \bibinfo
  {pages} {919} (\bibinfo {year} {1999})}\BibitemShut {NoStop}%
\bibitem [{\citenamefont {You}\ and\ \citenamefont {Moin}(2008)}]{You2008}%
  \BibitemOpen
  \bibfield  {author} {\bibinfo {author} {\bibfnamefont {D.}~\bibnamefont
  {You}}\ and\ \bibinfo {author} {\bibfnamefont {P.}~\bibnamefont {Moin}},\
  }\bibfield  {title} {\bibinfo {title} {Active control of flow separation over
  an airfoil using synthetic jets},\ }\href@noop {} {\bibfield  {journal}
  {\bibinfo  {journal} {Journal of Fluids and Structures}\ }\textbf {\bibinfo
  {volume} {24}},\ \bibinfo {pages} {1349} (\bibinfo {year}
  {2008})}\BibitemShut {NoStop}%
\bibitem [{\citenamefont {Crowther}\ and\ \citenamefont
  {Gomes}(2008)}]{Crowther2008}%
  \BibitemOpen
  \bibfield  {author} {\bibinfo {author} {\bibfnamefont {W.~J.}\ \bibnamefont
  {Crowther}}\ and\ \bibinfo {author} {\bibnamefont {Gomes}},\ }\bibfield
  {title} {\bibinfo {title} {An evaluation of the mass and power scaling of
  synthetic jet actuator flow control technology for civil transport aircraft
  applications},\ }in\ \href {https://doi.org/doi:10.1243/09596518JSCE519}
  {\emph {\bibinfo {booktitle} {Proceedings of the Institution of Mechanical
  Engineers, Part I (Journal of Systems and Control Engineering)}}}\ (\bibinfo
  {year} {2008})\BibitemShut {NoStop}%
\bibitem [{\citenamefont {Ingard}\ and\ \citenamefont
  {Labate}(1950)}]{Ingard1950}%
  \BibitemOpen
  \bibfield  {author} {\bibinfo {author} {\bibfnamefont {U.}~\bibnamefont
  {Ingard}}\ and\ \bibinfo {author} {\bibfnamefont {S.}~\bibnamefont
  {Labate}},\ }\bibfield  {title} {\bibinfo {title} {Acoustic circulation
  effects and the nonlinear impedance of orifices},\ }\href@noop {} {\bibfield
  {journal} {\bibinfo  {journal} {Journal of the Acoustical Society of
  America}\ }\textbf {\bibinfo {volume} {22}},\ \bibinfo {pages} {211}
  (\bibinfo {year} {1950})}\BibitemShut {NoStop}%
\bibitem [{\citenamefont {Dauphinee}(1957)}]{Dauphinee1957}%
  \BibitemOpen
  \bibfield  {author} {\bibinfo {author} {\bibfnamefont {T.~M.}\ \bibnamefont
  {Dauphinee}},\ }\bibfield  {title} {\bibinfo {title} {Acoustic air pump},\
  }\href@noop {} {\bibfield  {journal} {\bibinfo  {journal} {Review of
  Scientific Instruments}\ }\textbf {\bibinfo {volume} {28}},\ \bibinfo {pages}
  {456} (\bibinfo {year} {1957})}\BibitemShut {NoStop}%
\bibitem [{\citenamefont {Smith}\ and\ \citenamefont
  {Glezer}(1997)}]{Smith1997}%
  \BibitemOpen
  \bibfield  {author} {\bibinfo {author} {\bibfnamefont {B.~L.}\ \bibnamefont
  {Smith}}\ and\ \bibinfo {author} {\bibfnamefont {A.}~\bibnamefont {Glezer}},\
  }\bibfield  {title} {\bibinfo {title} {Vectoring and small-scale motions
  effected in free shear flows using synthetic jet actuators},\ }in\ \href@noop
  {} {\emph {\bibinfo {booktitle} {AIAA-1997-213}}}\ (\bibinfo {year}
  {1997})\BibitemShut {NoStop}%
\bibitem [{\citenamefont {Smith}\ and\ \citenamefont
  {Glezer}(1998)}]{Smith1998}%
  \BibitemOpen
  \bibfield  {author} {\bibinfo {author} {\bibfnamefont {B.~L.}\ \bibnamefont
  {Smith}}\ and\ \bibinfo {author} {\bibfnamefont {A.}~\bibnamefont {Glezer}},\
  }\bibfield  {title} {\bibinfo {title} {The formation and evolution of
  synthetic jets},\ }\href@noop {} {\bibfield  {journal} {\bibinfo  {journal}
  {Physics of Fluids}\ }\textbf {\bibinfo {volume} {10}},\ \bibinfo {pages}
  {2281} (\bibinfo {year} {1998})}\BibitemShut {NoStop}%
\bibitem [{\citenamefont {Kral}\ \emph {et~al.}(1997)\citenamefont {Kral},
  \citenamefont {Donovan}, \citenamefont {Cain},\ and\ \citenamefont
  {Cary}}]{Kral1997}%
  \BibitemOpen
  \bibfield  {author} {\bibinfo {author} {\bibfnamefont {L.~D.}\ \bibnamefont
  {Kral}}, \bibinfo {author} {\bibfnamefont {J.~F.}\ \bibnamefont {Donovan}},
  \bibinfo {author} {\bibfnamefont {A.~B.}\ \bibnamefont {Cain}},\ and\
  \bibinfo {author} {\bibfnamefont {A.~W.}\ \bibnamefont {Cary}},\ }\bibfield
  {title} {\bibinfo {title} {Numerical simulation of synthetic jet actuators},\
  }in\ \href@noop {} {\emph {\bibinfo {booktitle} {AIAA-1997-1824}}}\ (\bibinfo
  {year} {1997})\BibitemShut {NoStop}%
\bibitem [{\citenamefont {Okada}\ \emph
  {et~al.}(2012{\natexlab{a}})\citenamefont {Okada}, \citenamefont {Oyama},
  \citenamefont {Fujii},\ and\ \citenamefont {Miyaji}}]{Okada2012}%
  \BibitemOpen
  \bibfield  {author} {\bibinfo {author} {\bibfnamefont {K.}~\bibnamefont
  {Okada}}, \bibinfo {author} {\bibfnamefont {A.}~\bibnamefont {Oyama}},
  \bibinfo {author} {\bibfnamefont {K.}~\bibnamefont {Fujii}},\ and\ \bibinfo
  {author} {\bibfnamefont {K.}~\bibnamefont {Miyaji}},\ }\bibfield  {title}
  {\bibinfo {title} {Computational study of effects of nondimensional
  parameters on synthetic jets},\ }\href@noop {} {\bibfield  {journal}
  {\bibinfo  {journal} {Transactions of the Japan Society for Aeronautical and
  Space Sciences}\ }\textbf {\bibinfo {volume} {55}},\ \bibinfo {pages} {1}
  (\bibinfo {year} {2012}{\natexlab{a}})}\BibitemShut {NoStop}%
\bibitem [{\citenamefont {Okada}\ \emph
  {et~al.}(2012{\natexlab{b}})\citenamefont {Okada}, \citenamefont {Nonomura},
  \citenamefont {Fujii},\ and\ \citenamefont {Miyaji}}]{Okada2012a}%
  \BibitemOpen
  \bibfield  {author} {\bibinfo {author} {\bibfnamefont {K.}~\bibnamefont
  {Okada}}, \bibinfo {author} {\bibfnamefont {T.}~\bibnamefont {Nonomura}},
  \bibinfo {author} {\bibfnamefont {K.}~\bibnamefont {Fujii}},\ and\ \bibinfo
  {author} {\bibfnamefont {K.}~\bibnamefont {Miyaji}},\ }\bibfield  {title}
  {\bibinfo {title} {Computational analysis of vortex structures induced by a
  synthetic jet to control separated flows},\ }\href
  {https://doi.org/10.1260/1756-8250.4.1-2.47} {\bibfield  {journal} {\bibinfo
  {journal} {International Journal of Flow Control}\ }\textbf {\bibinfo
  {volume} {4}},\ \bibinfo {pages} {47} (\bibinfo {year}
  {2012}{\natexlab{b}})}\BibitemShut {NoStop}%
\bibitem [{\citenamefont {Seifert}\ \emph {et~al.}(1996)\citenamefont
  {Seifert}, \citenamefont {Darabi},\ and\ \citenamefont
  {Wygnanski}}]{Seifert1996}%
  \BibitemOpen
  \bibfield  {author} {\bibinfo {author} {\bibfnamefont {A.}~\bibnamefont
  {Seifert}}, \bibinfo {author} {\bibfnamefont {A.}~\bibnamefont {Darabi}},\
  and\ \bibinfo {author} {\bibfnamefont {I.}~\bibnamefont {Wygnanski}},\
  }\bibfield  {title} {\bibinfo {title} {Delay of airfoil stall by periodic
  excitation},\ }\href@noop {} {\bibfield  {journal} {\bibinfo  {journal}
  {Journal of Aircraft}\ }\textbf {\bibinfo {volume} {33}},\ \bibinfo {pages}
  {691} (\bibinfo {year} {1996})}\BibitemShut {NoStop}%
\bibitem [{\citenamefont {Donovan}\ \emph {et~al.}(1998)\citenamefont
  {Donovan}, \citenamefont {Kral},\ and\ \citenamefont {Cary}}]{Donovan1998}%
  \BibitemOpen
  \bibfield  {author} {\bibinfo {author} {\bibfnamefont {J.~F.}\ \bibnamefont
  {Donovan}}, \bibinfo {author} {\bibfnamefont {L.~D.}\ \bibnamefont {Kral}},\
  and\ \bibinfo {author} {\bibfnamefont {A.~W.}\ \bibnamefont {Cary}},\
  }\bibfield  {title} {\bibinfo {title} {Active flow control applied to an
  airfoil},\ }in\ \href@noop {} {\emph {\bibinfo {booktitle} {AIAA-1998-210}}}\
  (\bibinfo {year} {1998})\BibitemShut {NoStop}%
\bibitem [{\citenamefont {Feero}\ \emph {et~al.}(2015)\citenamefont {Feero},
  \citenamefont {Goodfellow}, \citenamefont {Lavoie},\ and\ \citenamefont
  {Sullivan}}]{Feero2015}%
  \BibitemOpen
  \bibfield  {author} {\bibinfo {author} {\bibfnamefont {M.~A.}\ \bibnamefont
  {Feero}}, \bibinfo {author} {\bibfnamefont {S.~D.}\ \bibnamefont
  {Goodfellow}}, \bibinfo {author} {\bibfnamefont {P.}~\bibnamefont {Lavoie}},\
  and\ \bibinfo {author} {\bibfnamefont {P.~E.}\ \bibnamefont {Sullivan}},\
  }\bibfield  {title} {\bibinfo {title} {Flow reattachment using synthetic jet
  actuation on a low-reynolds-number airfoil},\ }\href@noop {} {\bibfield
  {journal} {\bibinfo  {journal} {AIAA Journal}\ }\textbf {\bibinfo {volume}
  {53(7)}} (\bibinfo {year} {2015})}\BibitemShut {NoStop}%
\bibitem [{\citenamefont {Zhang}\ and\ \citenamefont
  {Samtaney}(2015)}]{Zhang2015}%
  \BibitemOpen
  \bibfield  {author} {\bibinfo {author} {\bibfnamefont {W.}~\bibnamefont
  {Zhang}}\ and\ \bibinfo {author} {\bibfnamefont {R.}~\bibnamefont
  {Samtaney}},\ }\bibfield  {title} {\bibinfo {title} {A direct numerical
  simulation investigation of the synthetic jet frequency effects on separation
  control of low-{R}e flow past an airfoil},\ }\href@noop {} {\bibfield
  {journal} {\bibinfo  {journal} {Physics of Fluids}\ }\textbf {\bibinfo
  {volume} {27}} (\bibinfo {year} {2015})}\BibitemShut {NoStop}%
\bibitem [{\citenamefont {Glezer}(1999)}]{Glezer1999}%
  \BibitemOpen
  \bibfield  {author} {\bibinfo {author} {\bibfnamefont {A.}~\bibnamefont
  {Glezer}},\ }\href@noop {} {\emph {\bibinfo {title} {Shear flow control using
  synthetic jet fluidic actuator technology}}},\ \bibinfo {type} {Tech. Rep.}\
  (\bibinfo  {institution} {Georgia Institute of Technology},\ \bibinfo {year}
  {1999})\BibitemShut {NoStop}%
\bibitem [{\citenamefont {Glezer}\ \emph {et~al.}(2005)\citenamefont {Glezer},
  \citenamefont {Amitay},\ and\ \citenamefont {Honohan}}]{Glezer2005}%
  \BibitemOpen
  \bibfield  {author} {\bibinfo {author} {\bibfnamefont {A.}~\bibnamefont
  {Glezer}}, \bibinfo {author} {\bibfnamefont {M.}~\bibnamefont {Amitay}},\
  and\ \bibinfo {author} {\bibfnamefont {A.~M.}\ \bibnamefont {Honohan}},\
  }\bibfield  {title} {\bibinfo {title} {Aspects of low- and high-frequency
  actuation for aerodynamic flow control},\ }\href@noop {} {\bibfield
  {journal} {\bibinfo  {journal} {AIAA Journal}\ }\textbf {\bibinfo {volume}
  {43}},\ \bibinfo {pages} {1501} (\bibinfo {year} {2005})}\BibitemShut
  {NoStop}%
\bibitem [{\citenamefont {Yoshioka}\ \emph {et~al.}(2001)\citenamefont
  {Yoshioka}, \citenamefont {Obi},\ and\ \citenamefont
  {Masuda}}]{Yoshioka2001}%
  \BibitemOpen
  \bibfield  {author} {\bibinfo {author} {\bibfnamefont {S.}~\bibnamefont
  {Yoshioka}}, \bibinfo {author} {\bibfnamefont {S.}~\bibnamefont {Obi}},\ and\
  \bibinfo {author} {\bibfnamefont {S.}~\bibnamefont {Masuda}},\ }\bibfield
  {title} {\bibinfo {title} {Turbulence statistics of peridically perturbed
  separated flow over backward-facing step},\ }\href@noop {} {\bibfield
  {journal} {\bibinfo  {journal} {International Journal of Heat and Fluid
  Flow}\ }\textbf {\bibinfo {volume} {22}},\ \bibinfo {pages} {393} (\bibinfo
  {year} {2001})}\BibitemShut {NoStop}%
\bibitem [{\citenamefont {Dandois}\ \emph {et~al.}(2007)\citenamefont
  {Dandois}, \citenamefont {Garnier},\ and\ \citenamefont
  {Sagaut}}]{Dandois2007}%
  \BibitemOpen
  \bibfield  {author} {\bibinfo {author} {\bibfnamefont {J.}~\bibnamefont
  {Dandois}}, \bibinfo {author} {\bibfnamefont {E.}~\bibnamefont {Garnier}},\
  and\ \bibinfo {author} {\bibfnamefont {P.}~\bibnamefont {Sagaut}},\
  }\bibfield  {title} {\bibinfo {title} {Numerical simulation of active
  separation control by a synthetic jet},\ }\href@noop {} {\bibfield  {journal}
  {\bibinfo  {journal} {Journal of Fluid Mechanics}\ }\textbf {\bibinfo
  {volume} {574}},\ \bibinfo {pages} {25} (\bibinfo {year} {2007})}\BibitemShut
  {NoStop}%
\bibitem [{\citenamefont {Corke}\ \emph {et~al.}(2004)\citenamefont {Corke},
  \citenamefont {He},\ and\ \citenamefont {Patelz}}]{Corke2004}%
  \BibitemOpen
  \bibfield  {author} {\bibinfo {author} {\bibfnamefont {T.~C.}\ \bibnamefont
  {Corke}}, \bibinfo {author} {\bibfnamefont {C.}~\bibnamefont {He}},\ and\
  \bibinfo {author} {\bibfnamefont {M.~P.}\ \bibnamefont {Patelz}},\ }\bibfield
   {title} {\bibinfo {title} {Plasma flaps and slats: An application of
  weakly-ionized plasma actuators},\ }in\ \href@noop {} {\emph {\bibinfo
  {booktitle} {AIAA-2004-2127}}}\ (\bibinfo {year} {2004})\BibitemShut
  {NoStop}%
\bibitem [{\citenamefont {Aono}\ \emph {et~al.}(2017)\citenamefont {Aono},
  \citenamefont {Kawai}, \citenamefont {Nonomura}, \citenamefont {Sato},
  \citenamefont {Fujii},\ and\ \citenamefont {Okada}}]{Aono2017}%
  \BibitemOpen
  \bibfield  {author} {\bibinfo {author} {\bibfnamefont {H.}~\bibnamefont
  {Aono}}, \bibinfo {author} {\bibfnamefont {S.}~\bibnamefont {Kawai}},
  \bibinfo {author} {\bibfnamefont {T.}~\bibnamefont {Nonomura}}, \bibinfo
  {author} {\bibfnamefont {M.}~\bibnamefont {Sato}}, \bibinfo {author}
  {\bibfnamefont {K.}~\bibnamefont {Fujii}},\ and\ \bibinfo {author}
  {\bibfnamefont {K.}~\bibnamefont {Okada}},\ }\bibfield  {title} {\bibinfo
  {title} {Plasma-actuator burst-mode frequency effects on leading-edge
  flow-separation control at reynolds number $2.6\times 10^5$},\ }\href@noop {}
  {\bibfield  {journal} {\bibinfo  {journal} {AIAA Journal}\ }\textbf {\bibinfo
  {volume} {55}},\ \bibinfo {pages} {3789} (\bibinfo {year}
  {2017})}\BibitemShut {NoStop}%
\bibitem [{\citenamefont {Sekimoto}\ \emph {et~al.}(2017)\citenamefont
  {Sekimoto}, \citenamefont {Nonomura},\ and\ \citenamefont
  {Fujii}}]{Sekimoto2017}%
  \BibitemOpen
  \bibfield  {author} {\bibinfo {author} {\bibfnamefont {S.}~\bibnamefont
  {Sekimoto}}, \bibinfo {author} {\bibfnamefont {T.}~\bibnamefont {Nonomura}},\
  and\ \bibinfo {author} {\bibfnamefont {K.}~\bibnamefont {Fujii}},\ }\bibfield
   {title} {\bibinfo {title} {Burst-mode frequency effects of dielectric
  barrier discharge plasma actuator for separation control},\ }\href@noop {}
  {\bibfield  {journal} {\bibinfo  {journal} {AIAA Journal}\ }\textbf {\bibinfo
  {volume} {55}},\ \bibinfo {pages} {1385} (\bibinfo {year}
  {2017})}\BibitemShut {NoStop}%
\bibitem [{\citenamefont {Wu}\ and\ \citenamefont {Squires}(1998)}]{Wu1998}%
  \BibitemOpen
  \bibfield  {author} {\bibinfo {author} {\bibfnamefont {X.}~\bibnamefont
  {Wu}}\ and\ \bibinfo {author} {\bibfnamefont {K.~D.}\ \bibnamefont
  {Squires}},\ }\bibfield  {title} {\bibinfo {title} {Numerical investigation
  of the turbulent boundary layer over a bump},\ }\href@noop {} {\bibfield
  {journal} {\bibinfo  {journal} {Journal of Fluid Mechanics}\ }\textbf
  {\bibinfo {volume} {362}},\ \bibinfo {pages} {229} (\bibinfo {year}
  {1998})}\BibitemShut {NoStop}%
\bibitem [{\citenamefont {Gaster}(1966)}]{Gaster1966}%
  \BibitemOpen
  \bibfield  {author} {\bibinfo {author} {\bibfnamefont {M.}~\bibnamefont
  {Gaster}},\ }\bibfield  {title} {\bibinfo {title} {The structure and
  behaviour of laminar separation bubbles},\ }in\ \href@noop {} {\emph
  {\bibinfo {booktitle} {AGARD CP-4}}}\ (\bibinfo {year} {1966})\BibitemShut
  {NoStop}%
\bibitem [{\citenamefont {Watmuff}(1999)}]{Watmuff1999}%
  \BibitemOpen
  \bibfield  {author} {\bibinfo {author} {\bibfnamefont {J.~H.}\ \bibnamefont
  {Watmuff}},\ }\bibfield  {title} {\bibinfo {title} {Evolution of a wave
  packet into vortex loops in a laminar separation bubble},\ }\href@noop {}
  {\bibfield  {journal} {\bibinfo  {journal} {Journal of Fluid Mechanics}\
  }\textbf {\bibinfo {volume} {397}},\ \bibinfo {pages} {119} (\bibinfo {year}
  {1999})}\BibitemShut {NoStop}%
\bibitem [{\citenamefont {Spalart}(2000)}]{Spalart2000}%
  \BibitemOpen
  \bibfield  {author} {\bibinfo {author} {\bibfnamefont {P.~R.}\ \bibnamefont
  {Spalart}},\ }\bibfield  {title} {\bibinfo {title} {Strategies for turbulence
  modeling and simulations},\ }\href@noop {} {\bibfield  {journal} {\bibinfo
  {journal} {International Journal of Heat and Fluid Flow}\ }\textbf {\bibinfo
  {volume} {21}},\ \bibinfo {pages} {252} (\bibinfo {year} {2000})}\BibitemShut
  {NoStop}%
\bibitem [{\citenamefont {Rist}\ and\ \citenamefont
  {Maucher}(2002)}]{Rist2002}%
  \BibitemOpen
  \bibfield  {author} {\bibinfo {author} {\bibfnamefont {U.}~\bibnamefont
  {Rist}}\ and\ \bibinfo {author} {\bibfnamefont {U.}~\bibnamefont {Maucher}},\
  }\bibfield  {title} {\bibinfo {title} {Investigations of time-growing
  instabilities in laminar separation bubbles},\ }\href@noop {} {\bibfield
  {journal} {\bibinfo  {journal} {European Journal of Mechanics - B/Fluids}\
  }\textbf {\bibinfo {volume} {21}},\ \bibinfo {pages} {495} (\bibinfo {year}
  {2002})}\BibitemShut {NoStop}%
\bibitem [{\citenamefont {Rist}\ and\ \citenamefont
  {Augustin}(2006)}]{Rist2006}%
  \BibitemOpen
  \bibfield  {author} {\bibinfo {author} {\bibfnamefont {U.}~\bibnamefont
  {Rist}}\ and\ \bibinfo {author} {\bibfnamefont {K.}~\bibnamefont
  {Augustin}},\ }\bibfield  {title} {\bibinfo {title} {Control of laminar
  separation bubbles using instability waves},\ }\href@noop {} {\bibfield
  {journal} {\bibinfo  {journal} {AIAA Journal}\ }\textbf {\bibinfo {volume}
  {44}},\ \bibinfo {pages} {2217} (\bibinfo {year} {2006})}\BibitemShut
  {NoStop}%
\bibitem [{\citenamefont {Marxen}\ and\ \citenamefont
  {Henningson}(2011)}]{Marxen2011}%
  \BibitemOpen
  \bibfield  {author} {\bibinfo {author} {\bibfnamefont {O.}~\bibnamefont
  {Marxen}}\ and\ \bibinfo {author} {\bibfnamefont {D.~S.}\ \bibnamefont
  {Henningson}},\ }\bibfield  {title} {\bibinfo {title} {The effect of
  small-amplitude convective disturbances on the size and bursting of a laminar
  separation bubble},\ }\href@noop {} {\bibfield  {journal} {\bibinfo
  {journal} {Journal of Fluid Mechanics}\ }\textbf {\bibinfo {volume} {671}},\
  \bibinfo {pages} {1} (\bibinfo {year} {2011})}\BibitemShut {NoStop}%
\bibitem [{\citenamefont {Ahuja}(1984)}]{Ahuja1984}%
  \BibitemOpen
  \bibfield  {author} {\bibinfo {author} {\bibfnamefont {K.~K.}\ \bibnamefont
  {Ahuja}},\ }\bibfield  {title} {\bibinfo {title} {Control of flow separation
  by sound},\ }in\ \href@noop {} {\emph {\bibinfo {booktitle}
  {AIAA-1984-2298}}}\ (\bibinfo {year} {1984})\BibitemShut {NoStop}%
\bibitem [{\citenamefont {Zaman}\ \emph {et~al.}(1987)\citenamefont {Zaman},
  \citenamefont {Bar-Sever},\ and\ \citenamefont {Mangalam}}]{Zaman1987}%
  \BibitemOpen
  \bibfield  {author} {\bibinfo {author} {\bibfnamefont {K.~B. M.~Q.}\
  \bibnamefont {Zaman}}, \bibinfo {author} {\bibfnamefont {A.}~\bibnamefont
  {Bar-Sever}},\ and\ \bibinfo {author} {\bibfnamefont {S.~M.}\ \bibnamefont
  {Mangalam}},\ }\bibfield  {title} {\bibinfo {title} {Effect acoustic
  excitation flow over low-{R}e airfoil},\ }\href@noop {} {\bibfield  {journal}
  {\bibinfo  {journal} {Journal of Fluid Mechanics}\ }\textbf {\bibinfo
  {volume} {182}},\ \bibinfo {pages} {127} (\bibinfo {year}
  {1987})}\BibitemShut {NoStop}%
\bibitem [{\citenamefont {Marxen}\ \emph {et~al.}(2004)\citenamefont {Marxen},
  \citenamefont {Rist},\ and\ \citenamefont {Wagner}}]{Marxen2004}%
  \BibitemOpen
  \bibfield  {author} {\bibinfo {author} {\bibfnamefont {O.}~\bibnamefont
  {Marxen}}, \bibinfo {author} {\bibfnamefont {U.}~\bibnamefont {Rist}},\ and\
  \bibinfo {author} {\bibfnamefont {S.}~\bibnamefont {Wagner}},\ }\bibfield
  {title} {\bibinfo {title} {Effect of spanwise-modulated disturbances on
  transition in a separated boundary layer},\ }\href@noop {} {\bibfield
  {journal} {\bibinfo  {journal} {AIAA Journal}\ }\textbf {\bibinfo {volume}
  {42}},\ \bibinfo {pages} {937} (\bibinfo {year} {2004})}\BibitemShut
  {NoStop}%
\bibitem [{\citenamefont {Yarusevych}\ \emph {et~al.}(2009)\citenamefont
  {Yarusevych}, \citenamefont {Sullivan},\ and\ \citenamefont
  {Kawall}}]{Yarusevych2009}%
  \BibitemOpen
  \bibfield  {author} {\bibinfo {author} {\bibfnamefont {S.}~\bibnamefont
  {Yarusevych}}, \bibinfo {author} {\bibfnamefont {P.~E.}\ \bibnamefont
  {Sullivan}},\ and\ \bibinfo {author} {\bibfnamefont {J.~G.}\ \bibnamefont
  {Kawall}},\ }\bibfield  {title} {\bibinfo {title} {On vortex shedding from an
  airfoil in low-reynolds-number flows},\ }\href@noop {} {\bibfield  {journal}
  {\bibinfo  {journal} {Journal of Fluid Mechanics}\ }\textbf {\bibinfo
  {volume} {632}},\ \bibinfo {pages} {245} (\bibinfo {year}
  {2009})}\BibitemShut {NoStop}%
\bibitem [{\citenamefont {Avdis}\ \emph {et~al.}(2009)\citenamefont {Avdis},
  \citenamefont {Lardeau},\ and\ \citenamefont {Leschziner}}]{Avdis2009}%
  \BibitemOpen
  \bibfield  {author} {\bibinfo {author} {\bibfnamefont {A.}~\bibnamefont
  {Avdis}}, \bibinfo {author} {\bibfnamefont {S.}~\bibnamefont {Lardeau}},\
  and\ \bibinfo {author} {\bibfnamefont {M.}~\bibnamefont {Leschziner}},\
  }\bibfield  {title} {\bibinfo {title} {Large eddy simulation of separated
  flow over a two-dimensional hump with and without control by means of a
  synthetic slot-jet},\ }\href@noop {} {\bibfield  {journal} {\bibinfo
  {journal} {Flow Turbulence Combust}\ }\textbf {\bibinfo {volume} {83}},\
  \bibinfo {pages} {343} (\bibinfo {year} {2009})}\BibitemShut {NoStop}%
\bibitem [{\citenamefont {Boutilier}\ and\ \citenamefont
  {Yarusevych}(2012)}]{Boutilier2012b}%
  \BibitemOpen
  \bibfield  {author} {\bibinfo {author} {\bibfnamefont {M.~S.~H.}\
  \bibnamefont {Boutilier}}\ and\ \bibinfo {author} {\bibfnamefont
  {S.}~\bibnamefont {Yarusevych}},\ }\bibfield  {title} {\bibinfo {title}
  {Parametric study of separation and transition characteristicsover an airfoil
  at low reynolds numbers},\ }\href@noop {} {\bibfield  {journal} {\bibinfo
  {journal} {Experiments in Fluids}\ }\textbf {\bibinfo {volume} {52}},\
  \bibinfo {pages} {1491} (\bibinfo {year} {2012})}\BibitemShut {NoStop}%
\bibitem [{\citenamefont {Marxen}\ \emph {et~al.}(2015)\citenamefont {Marxen},
  \citenamefont {Kotapati}, \citenamefont {Mittal},\ and\ \citenamefont
  {Zaki}}]{Marxen2015}%
  \BibitemOpen
  \bibfield  {author} {\bibinfo {author} {\bibfnamefont {O.}~\bibnamefont
  {Marxen}}, \bibinfo {author} {\bibfnamefont {R.~B.}\ \bibnamefont
  {Kotapati}}, \bibinfo {author} {\bibfnamefont {R.}~\bibnamefont {Mittal}},\
  and\ \bibinfo {author} {\bibfnamefont {T.}~\bibnamefont {Zaki}},\ }\bibfield
  {title} {\bibinfo {title} {Stability analysis of separated flows subject to
  control by zero-net-mass-flux jet},\ }\href@noop {} {\bibfield  {journal}
  {\bibinfo  {journal} {Physics of Fluids}\ }\textbf {\bibinfo {volume} {27}},\
  \bibinfo {pages} {68} (\bibinfo {year} {2015})}\BibitemShut {NoStop}%
\bibitem [{\citenamefont {Yarusevych}\ and\ \citenamefont
  {Kotsonis}(2017{\natexlab{b}})}]{Yarusevych2017JFM}%
  \BibitemOpen
  \bibfield  {author} {\bibinfo {author} {\bibfnamefont {S.}~\bibnamefont
  {Yarusevych}}\ and\ \bibinfo {author} {\bibfnamefont {M.}~\bibnamefont
  {Kotsonis}},\ }\bibfield  {title} {\bibinfo {title} {Steady and transient
  response of a laminar separation bubble to controlled disturbances},\
  }\href@noop {} {\bibfield  {journal} {\bibinfo  {journal} {Journal of Fluid
  Mechanics}\ }\textbf {\bibinfo {volume} {813}},\ \bibinfo {pages} {955}
  (\bibinfo {year} {2017}{\natexlab{b}})}\BibitemShut {NoStop}%
\bibitem [{\citenamefont {Kurelek}\ \emph {et~al.}(2019)\citenamefont
  {Kurelek}, \citenamefont {Yarusevych},\ and\ \citenamefont
  {Kotsonis}}]{Yarusevych2019}%
  \BibitemOpen
  \bibfield  {author} {\bibinfo {author} {\bibfnamefont {J.~W.}\ \bibnamefont
  {Kurelek}}, \bibinfo {author} {\bibfnamefont {S.}~\bibnamefont
  {Yarusevych}},\ and\ \bibinfo {author} {\bibfnamefont {M.}~\bibnamefont
  {Kotsonis}},\ }\bibfield  {title} {\bibinfo {title} {Vortex merging in a
  laminar separation bubble under natural and forced conditions},\ }\href@noop
  {} {\bibfield  {journal} {\bibinfo  {journal} {Physical Review Fluids}\
  }\textbf {\bibinfo {volume} {4}} (\bibinfo {year} {2019})}\BibitemShut
  {NoStop}%
\bibitem [{\citenamefont {Postl}\ \emph {et~al.}(2011)\citenamefont {Postl},
  \citenamefont {Balzer},\ and\ \citenamefont {Fasel}}]{Postl2011}%
  \BibitemOpen
  \bibfield  {author} {\bibinfo {author} {\bibfnamefont {D.}~\bibnamefont
  {Postl}}, \bibinfo {author} {\bibfnamefont {W.}~\bibnamefont {Balzer}},\ and\
  \bibinfo {author} {\bibfnamefont {H.~F.}\ \bibnamefont {Fasel}},\ }\bibfield
  {title} {\bibinfo {title} {Control of laminar separation using pulsed vortex
  generator jets: direct numerical simulations},\ }\href@noop {} {\bibfield
  {journal} {\bibinfo  {journal} {Journal of Fluid Mechanics}\ }\textbf
  {\bibinfo {volume} {676}},\ \bibinfo {pages} {81} (\bibinfo {year}
  {2011})}\BibitemShut {NoStop}%
\bibitem [{\citenamefont {Dejoan}\ and\ \citenamefont
  {Leschziner}(2004)}]{Dejoan2004}%
  \BibitemOpen
  \bibfield  {author} {\bibinfo {author} {\bibfnamefont {A.}~\bibnamefont
  {Dejoan}}\ and\ \bibinfo {author} {\bibfnamefont {M.~A.}\ \bibnamefont
  {Leschziner}},\ }\bibfield  {title} {\bibinfo {title} {Large eddy simulation
  of periodically perturbed separated flow over a backward-facing step},\
  }\href@noop {} {\bibfield  {journal} {\bibinfo  {journal} {International
  Journal of Heat and Fluid Flow}\ }\textbf {\bibinfo {volume} {25}},\ \bibinfo
  {pages} {581} (\bibinfo {year} {2004})}\BibitemShut {NoStop}%
\bibitem [{\citenamefont {Lambert}\ and\ \citenamefont
  {Yarusevych}(2019)}]{Lambert2019}%
  \BibitemOpen
  \bibfield  {author} {\bibinfo {author} {\bibfnamefont {A.}~\bibnamefont
  {Lambert}}\ and\ \bibinfo {author} {\bibfnamefont {S.}~\bibnamefont
  {Yarusevych}},\ }\bibfield  {title} {\bibinfo {title} {Effect of angle of
  attack on vortex dynamics in laminar separation bubbles},\ }\href@noop {}
  {\bibfield  {journal} {\bibinfo  {journal} {Physics of Fluids}\ }\textbf
  {\bibinfo {volume} {31}},\ \bibinfo {pages} {064105} (\bibinfo {year}
  {2019})}\BibitemShut {NoStop}%
\bibitem [{\citenamefont {Lin}\ and\ \citenamefont {Pauley}(1996)}]{Lin1996}%
  \BibitemOpen
  \bibfield  {author} {\bibinfo {author} {\bibfnamefont {J.~C.~M.}\
  \bibnamefont {Lin}}\ and\ \bibinfo {author} {\bibfnamefont {L.~L.}\
  \bibnamefont {Pauley}},\ }\bibfield  {title} {\bibinfo {title}
  {Low-reynolds-number separation on an airfoil},\ }\href@noop {} {\bibfield
  {journal} {\bibinfo  {journal} {AIAA Journal}\ }\textbf {\bibinfo {volume}
  {34}},\ \bibinfo {pages} {1570} (\bibinfo {year} {1996})}\BibitemShut
  {NoStop}%
\bibitem [{\citenamefont {Asada}\ \emph {et~al.}(2015)\citenamefont {Asada},
  \citenamefont {Nonomura}, \citenamefont {Aono}, \citenamefont {Sato},
  \citenamefont {Okada},\ and\ \citenamefont {Fujii}}]{Asada2015}%
  \BibitemOpen
  \bibfield  {author} {\bibinfo {author} {\bibfnamefont {K.}~\bibnamefont
  {Asada}}, \bibinfo {author} {\bibfnamefont {T.}~\bibnamefont {Nonomura}},
  \bibinfo {author} {\bibfnamefont {H.}~\bibnamefont {Aono}}, \bibinfo {author}
  {\bibfnamefont {M.}~\bibnamefont {Sato}}, \bibinfo {author} {\bibfnamefont
  {K.}~\bibnamefont {Okada}},\ and\ \bibinfo {author} {\bibfnamefont
  {K.}~\bibnamefont {Fujii}},\ }\bibfield  {title} {\bibinfo {title} {{LES} of
  transient flows controlled by {DBD} plasma actuator over a stalled airfoil},\
  }\href@noop {} {\bibfield  {journal} {\bibinfo  {journal} {International
  Journal of Computational Fluid Dynamics}\ }\textbf {\bibinfo {volume} {29}},\
  \bibinfo {pages} {215} (\bibinfo {year} {2015})}\BibitemShut {NoStop}%
\bibitem [{\citenamefont {Fukumoto}\ \emph {et~al.}(2016)\citenamefont
  {Fukumoto}, \citenamefont {Aono}, \citenamefont {Watanabe}, \citenamefont
  {Tanaka}, \citenamefont {Matsuda}, \citenamefont {Osako}, \citenamefont
  {Nonomura}, \citenamefont {Oyama},\ and\ \citenamefont
  {Fujii}}]{Fukumoto2016}%
  \BibitemOpen
  \bibfield  {author} {\bibinfo {author} {\bibfnamefont {H.}~\bibnamefont
  {Fukumoto}}, \bibinfo {author} {\bibfnamefont {H.}~\bibnamefont {Aono}},
  \bibinfo {author} {\bibfnamefont {T.}~\bibnamefont {Watanabe}}, \bibinfo
  {author} {\bibfnamefont {M.}~\bibnamefont {Tanaka}}, \bibinfo {author}
  {\bibfnamefont {H.}~\bibnamefont {Matsuda}}, \bibinfo {author} {\bibfnamefont
  {T.}~\bibnamefont {Osako}}, \bibinfo {author} {\bibfnamefont
  {T.}~\bibnamefont {Nonomura}}, \bibinfo {author} {\bibfnamefont
  {A.}~\bibnamefont {Oyama}},\ and\ \bibinfo {author} {\bibfnamefont
  {K.}~\bibnamefont {Fujii}},\ }\bibfield  {title} {\bibinfo {title} {Control
  of dynamic flowfield around a pitching {NACA}633-618 airfoil by a {DBD}
  plasma actuator},\ }\href@noop {} {\bibfield  {journal} {\bibinfo  {journal}
  {International Journal of Heat and Fluid Flow}\ }\textbf {\bibinfo {volume}
  {62}},\ \bibinfo {pages} {10} (\bibinfo {year} {2016})}\BibitemShut {NoStop}%
\bibitem [{\citenamefont {Asada}\ \emph {et~al.}(2009)\citenamefont {Asada},
  \citenamefont {Ninomiya}, \citenamefont {Oyama},\ and\ \citenamefont
  {Fujii}}]{Asada2009}%
  \BibitemOpen
  \bibfield  {author} {\bibinfo {author} {\bibfnamefont {K.}~\bibnamefont
  {Asada}}, \bibinfo {author} {\bibfnamefont {Y.}~\bibnamefont {Ninomiya}},
  \bibinfo {author} {\bibfnamefont {A.}~\bibnamefont {Oyama}},\ and\ \bibinfo
  {author} {\bibfnamefont {K.}~\bibnamefont {Fujii}},\ }\bibfield  {title}
  {\bibinfo {title} {Airfoil flow experiment on the duty cycle of dbd plasma
  actuator},\ }in\ \href@noop {} {\emph {\bibinfo {booktitle}
  {AIAA-2009-531}}}\ (\bibinfo {year} {2009})\BibitemShut {NoStop}%
\bibitem [{\citenamefont {Greenblatt}\ and\ \citenamefont
  {Wygnanski}(2000)}]{Greenblatt2000}%
  \BibitemOpen
  \bibfield  {author} {\bibinfo {author} {\bibfnamefont {D.}~\bibnamefont
  {Greenblatt}}\ and\ \bibinfo {author} {\bibfnamefont {I.~J.}\ \bibnamefont
  {Wygnanski}},\ }\bibfield  {title} {\bibinfo {title} {The control of flow
  separation by periodic excitation},\ }\href@noop {} {\bibfield  {journal}
  {\bibinfo  {journal} {Progress in Aerospace Sciences}\ }\textbf {\bibinfo
  {volume} {36}},\ \bibinfo {pages} {487} (\bibinfo {year} {2000})}\BibitemShut
  {NoStop}%
\bibitem [{\citenamefont {Fujii}\ \emph {et~al.}(1990)\citenamefont {Fujii},
  \citenamefont {Endo},\ and\ \citenamefont {Yasuhara}}]{Fujii1990lans3d}%
  \BibitemOpen
  \bibfield  {author} {\bibinfo {author} {\bibfnamefont {K.}~\bibnamefont
  {Fujii}}, \bibinfo {author} {\bibfnamefont {H.}~\bibnamefont {Endo}},\ and\
  \bibinfo {author} {\bibfnamefont {M.}~\bibnamefont {Yasuhara}},\ }\href@noop
  {} {\emph {\bibinfo {title} {Activities of Computational Fluid Dynamics in
  Japan: Compressible Flow Simulations, High Performance Computing Research and
  Practice in Japan}}}\ (\bibinfo  {publisher} {JOHN WILEY \& SONS},\ \bibinfo
  {year} {1990})\BibitemShut {NoStop}%
\bibitem [{\citenamefont {Fujii}(2005)}]{Fujii2005}%
  \BibitemOpen
  \bibfield  {author} {\bibinfo {author} {\bibfnamefont {K.}~\bibnamefont
  {Fujii}},\ }\bibfield  {title} {\bibinfo {title} {Progress and future
  prospects of {CFD} in aerospace-wind tunnel and beyond},\ }\href@noop {}
  {\bibfield  {journal} {\bibinfo  {journal} {Progress in Aerospace Sciences}\
  }\textbf {\bibinfo {volume} {41}},\ \bibinfo {pages} {455} (\bibinfo {year}
  {2005})}\BibitemShut {NoStop}%
\bibitem [{\citenamefont {Fujii}(2008)}]{Fujii2008}%
  \BibitemOpen
  \bibfield  {author} {\bibinfo {author} {\bibfnamefont {K.}~\bibnamefont
  {Fujii}},\ }\bibfield  {title} {\bibinfo {title} {{CFD} contributions to
  high-speed shock-related problems},\ }\href@noop {} {\bibfield  {journal}
  {\bibinfo  {journal} {Shock Waves}\ }\textbf {\bibinfo {volume} {18}},\
  \bibinfo {pages} {145} (\bibinfo {year} {2008})}\BibitemShut {NoStop}%
\bibitem [{\citenamefont {Lele}(1992)}]{Lele1992}%
  \BibitemOpen
  \bibfield  {author} {\bibinfo {author} {\bibfnamefont {S.~K.}\ \bibnamefont
  {Lele}},\ }\bibfield  {title} {\bibinfo {title} {Compact finite difference
  schemes with spectral-like resolution.},\ }\href
  {https://doi.org/10.1016/0021-9991(92)90324-R} {\bibfield  {journal}
  {\bibinfo  {journal} {Journal of Computational Physics}\ }\textbf {\bibinfo
  {volume} {103}},\ \bibinfo {pages} {16} (\bibinfo {year} {1992})}\BibitemShut
  {NoStop}%
\bibitem [{\citenamefont {Abe}\ \emph {et~al.}(2014)\citenamefont {Abe},
  \citenamefont {Nonomura}, \citenamefont {Iizuka},\ and\ \citenamefont
  {Fujii}}]{Abe2013b}%
  \BibitemOpen
  \bibfield  {author} {\bibinfo {author} {\bibfnamefont {Y.}~\bibnamefont
  {Abe}}, \bibinfo {author} {\bibfnamefont {T.}~\bibnamefont {Nonomura}},
  \bibinfo {author} {\bibfnamefont {N.}~\bibnamefont {Iizuka}},\ and\ \bibinfo
  {author} {\bibfnamefont {K.}~\bibnamefont {Fujii}},\ }\bibfield  {title}
  {\bibinfo {title} {Geometric interpretations and spatial symmetry property of
  metrics in the conservative form for high-order finite-difference schemes on
  moving and deforming grids},\ }\href@noop {} {\bibfield  {journal} {\bibinfo
  {journal} {Journal of Computational Physics}\ }\textbf {\bibinfo {volume}
  {260}},\ \bibinfo {pages} {163} (\bibinfo {year} {2014})}\BibitemShut
  {NoStop}%
\bibitem [{\citenamefont {Gaitonde}\ and\ \citenamefont
  {Visbal}(2000)}]{Gaitonde2000}%
  \BibitemOpen
  \bibfield  {author} {\bibinfo {author} {\bibfnamefont {D.~V.}\ \bibnamefont
  {Gaitonde}}\ and\ \bibinfo {author} {\bibfnamefont {M.~R.}\ \bibnamefont
  {Visbal}},\ }\bibfield  {title} {\bibinfo {title} {Pad\'{e}-type higher-order
  boundary filters for the navier-stokes equations},\ }\href@noop {} {\bibfield
   {journal} {\bibinfo  {journal} {AIAA Journal}\ }\textbf {\bibinfo {volume}
  {38}},\ \bibinfo {pages} {2103} (\bibinfo {year} {2000})}\BibitemShut
  {NoStop}%
\bibitem [{\citenamefont {Visbal}\ and\ \citenamefont
  {Rizzetta}(2002)}]{Visbal2002}%
  \BibitemOpen
  \bibfield  {author} {\bibinfo {author} {\bibfnamefont {M.~R.}\ \bibnamefont
  {Visbal}}\ and\ \bibinfo {author} {\bibfnamefont {D.~P.}\ \bibnamefont
  {Rizzetta}},\ }\bibfield  {title} {\bibinfo {title} {Large-eddy simulationn
  on general geometries using compact differencing and filtering schemes},\
  }in\ \href@noop {} {\emph {\bibinfo {booktitle} {AIAA-2002-288}}}\ (\bibinfo
  {year} {2002})\BibitemShut {NoStop}%
\bibitem [{\citenamefont {Nishida}\ and\ \citenamefont
  {Nonomura}(2009)}]{Nishida2009}%
  \BibitemOpen
  \bibfield  {author} {\bibinfo {author} {\bibfnamefont {H.}~\bibnamefont
  {Nishida}}\ and\ \bibinfo {author} {\bibfnamefont {T.}~\bibnamefont
  {Nonomura}},\ }\bibfield  {title} {\bibinfo {title} {Adi-sgs scheme on ideal
  magnetohydrodynamics},\ }\href@noop {} {\bibfield  {journal} {\bibinfo
  {journal} {Journal of Computational Physics}\ }\textbf {\bibinfo {volume}
  {228}},\ \bibinfo {pages} {3182} (\bibinfo {year} {2009})}\BibitemShut
  {NoStop}%
\bibitem [{\citenamefont {Fujii}(1995)}]{Fujii1995a}%
  \BibitemOpen
  \bibfield  {author} {\bibinfo {author} {\bibfnamefont {K.}~\bibnamefont
  {Fujii}},\ }\bibfield  {title} {\bibinfo {title} {Unified zonal method based
  on the fortified solution algorithm},\ }\href@noop {} {\bibfield  {journal}
  {\bibinfo  {journal} {Journal of Computational Physics}\ }\textbf {\bibinfo
  {volume} {118}},\ \bibinfo {pages} {92} (\bibinfo {year} {1995})}\BibitemShut
  {NoStop}%
\bibitem [{\citenamefont {Melville}\ \emph {et~al.}(1997)\citenamefont
  {Melville}, \citenamefont {Moiton},\ and\ \citenamefont
  {Rizzetta}}]{Melville1997}%
  \BibitemOpen
  \bibfield  {author} {\bibinfo {author} {\bibfnamefont {R.~B.}\ \bibnamefont
  {Melville}}, \bibinfo {author} {\bibfnamefont {S.~A.}\ \bibnamefont
  {Moiton}},\ and\ \bibinfo {author} {\bibfnamefont {D.~P.}\ \bibnamefont
  {Rizzetta}},\ }\bibfield  {title} {\bibinfo {title} {Implementation of a
  fully-implicit, aeroelastic navier-stokes solver},\ }in\ \href@noop {} {\emph
  {\bibinfo {booktitle} {AIAA-1997-2039}}}\ (\bibinfo {year}
  {1997})\BibitemShut {NoStop}%
\bibitem [{\citenamefont {Larsson}\ \emph {et~al.}(2016)\citenamefont
  {Larsson}, \citenamefont {Kawai}, \citenamefont {Bodart},\ and\ \citenamefont
  {Bermejo-Moreno}}]{Larsson2016}%
  \BibitemOpen
  \bibfield  {author} {\bibinfo {author} {\bibfnamefont {J.}~\bibnamefont
  {Larsson}}, \bibinfo {author} {\bibfnamefont {S.}~\bibnamefont {Kawai}},
  \bibinfo {author} {\bibfnamefont {J.}~\bibnamefont {Bodart}},\ and\ \bibinfo
  {author} {\bibfnamefont {I.}~\bibnamefont {Bermejo-Moreno}},\ }\bibfield
  {title} {\bibinfo {title} {Large eddy simulation with modeled wall-stress:
  recent progress and future directions},\ }\href@noop {} {\bibfield  {journal}
  {\bibinfo  {journal} {Mech. Eng. Reviews}\ }\textbf {\bibinfo {volume} {3}},\
  \bibinfo {pages} {15} (\bibinfo {year} {2016})}\BibitemShut {NoStop}%
\bibitem [{\citenamefont {Kawai}\ and\ \citenamefont
  {Fujii}(2008)}]{Kawai2008}%
  \BibitemOpen
  \bibfield  {author} {\bibinfo {author} {\bibfnamefont {S.}~\bibnamefont
  {Kawai}}\ and\ \bibinfo {author} {\bibfnamefont {K.}~\bibnamefont {Fujii}},\
  }\bibfield  {title} {\bibinfo {title} {Compact scheme with filtering for
  large-eddy simulation of transitional boundary layer},\ }\href@noop {}
  {\bibfield  {journal} {\bibinfo  {journal} {AIAA Journal}\ }\textbf {\bibinfo
  {volume} {46}},\ \bibinfo {pages} {690} (\bibinfo {year} {2008})}\BibitemShut
  {NoStop}%
\bibitem [{\citenamefont {Choi}\ and\ \citenamefont {Moin}(1994)}]{Choi1994}%
  \BibitemOpen
  \bibfield  {author} {\bibinfo {author} {\bibfnamefont {H.}~\bibnamefont
  {Choi}}\ and\ \bibinfo {author} {\bibfnamefont {P.}~\bibnamefont {Moin}},\
  }\bibfield  {title} {\bibinfo {title} {Effects of the computational time step
  on numerical solutions of turbulent flow},\ }\href@noop {} {\bibfield
  {journal} {\bibinfo  {journal} {Journal of Computational Physics}\ }\textbf
  {\bibinfo {volume} {113}},\ \bibinfo {pages} {1} (\bibinfo {year}
  {1994})}\BibitemShut {NoStop}%
\bibitem [{\citenamefont {Frigo}\ and\ \citenamefont
  {Johnson}(2005)}]{Frigo2006}%
  \BibitemOpen
  \bibfield  {author} {\bibinfo {author} {\bibfnamefont {M.}~\bibnamefont
  {Frigo}}\ and\ \bibinfo {author} {\bibfnamefont {S.~G.}\ \bibnamefont
  {Johnson}},\ }\bibfield  {title} {\bibinfo {title} {The design and
  implementation of {FFTW3}},\ }\href@noop {} {\bibfield  {journal} {\bibinfo
  {journal} {Proceedings of the IEEE}\ }\textbf {\bibinfo {volume} {93}},\
  \bibinfo {pages} {216} (\bibinfo {year} {2005})},\ \bibinfo {note} {special
  issue on ``Program Generation, Optimization, and Platform
  Adaptation''}\BibitemShut {NoStop}%
\bibitem [{\citenamefont {Simoni}\ \emph {et~al.}(2014)\citenamefont {Simoni},
  \citenamefont {Ubaldi},\ and\ \citenamefont {Zunino}}]{Simoni2014}%
  \BibitemOpen
  \bibfield  {author} {\bibinfo {author} {\bibfnamefont {D.}~\bibnamefont
  {Simoni}}, \bibinfo {author} {\bibfnamefont {M.}~\bibnamefont {Ubaldi}},\
  and\ \bibinfo {author} {\bibfnamefont {P.}~\bibnamefont {Zunino}},\
  }\bibfield  {title} {\bibinfo {title} {Experimental investigation of flow
  instabilities in a laminar separation bubble},\ }\href@noop {} {\bibfield
  {journal} {\bibinfo  {journal} {Journal of Thermal Science}\ }\textbf
  {\bibinfo {volume} {23}},\ \bibinfo {pages} {203} (\bibinfo {year}
  {2014})}\BibitemShut {NoStop}%
\bibitem [{\citenamefont {Alam}\ and\ \citenamefont
  {Sandham}(2000)}]{Alam2000}%
  \BibitemOpen
  \bibfield  {author} {\bibinfo {author} {\bibfnamefont {M.}~\bibnamefont
  {Alam}}\ and\ \bibinfo {author} {\bibfnamefont {N.}~\bibnamefont {Sandham}},\
  }\bibfield  {title} {\bibinfo {title} {Direct numerical simulation of short
  laminar separation bubbles with turbulent reattachment},\ }\href@noop {}
  {\bibfield  {journal} {\bibinfo  {journal} {Journal of Fluid Mechanics}\
  }\textbf {\bibinfo {volume} {403}},\ \bibinfo {pages} {223} (\bibinfo {year}
  {2000})}\BibitemShut {NoStop}%
\bibitem [{\citenamefont {Rowley}\ \emph {et~al.}(2002)\citenamefont {Rowley},
  \citenamefont {Colonius},\ and\ \citenamefont {Basu.}}]{Rowley2002}%
  \BibitemOpen
  \bibfield  {author} {\bibinfo {author} {\bibfnamefont {C.~W.}\ \bibnamefont
  {Rowley}}, \bibinfo {author} {\bibfnamefont {T.}~\bibnamefont {Colonius}},\
  and\ \bibinfo {author} {\bibfnamefont {A.~J.}\ \bibnamefont {Basu.}},\
  }\bibfield  {title} {\bibinfo {title} {On self-sustained oscillations in
  two-dimensional compressible flow over rectangular cavities},\ }\href
  {https://doi.org/10.1017/S0022112001007534} {\bibfield  {journal} {\bibinfo
  {journal} {Journal of Fluid Mechanics}\ }\textbf {\bibinfo {volume} {455}},\
  \bibinfo {pages} {315} (\bibinfo {year} {2002})}\BibitemShut {NoStop}%
\bibitem [{\citenamefont {Amitay}\ and\ \citenamefont
  {Glezer}(2002{\natexlab{b}})}]{Amitay2002a}%
  \BibitemOpen
  \bibfield  {author} {\bibinfo {author} {\bibfnamefont {M.}~\bibnamefont
  {Amitay}}\ and\ \bibinfo {author} {\bibfnamefont {A.}~\bibnamefont
  {Glezer}},\ }\bibfield  {title} {\bibinfo {title} {Controlled transients of
  flow reattachment over stalled airfoils},\ }\href@noop {} {\bibfield
  {journal} {\bibinfo  {journal} {International Journal of Heat and Fluid
  Flow}\ }\textbf {\bibinfo {volume} {23}},\ \bibinfo {pages} {690} (\bibinfo
  {year} {2002}{\natexlab{b}})}\BibitemShut {NoStop}%
\bibitem [{\citenamefont {Asada}\ and\ \citenamefont
  {Fujii}(2012)}]{Asada2012}%
  \BibitemOpen
  \bibfield  {author} {\bibinfo {author} {\bibfnamefont {K.}~\bibnamefont
  {Asada}}\ and\ \bibinfo {author} {\bibfnamefont {K.}~\bibnamefont {Fujii}},\
  }\bibfield  {title} {\bibinfo {title} {Burst frequency effect of dbd plasma
  actuator on the control of separated flow over an airfoil},\ }in\ \href@noop
  {} {\emph {\bibinfo {booktitle} {AIAA 2012-3054}}}\ (\bibinfo {year}
  {2012})\BibitemShut {NoStop}%
\bibitem [{\citenamefont {Nicholas}\ \emph {et~al.}(2010)\citenamefont
  {Nicholas}, \citenamefont {Rizzetta},\ and\ \citenamefont
  {Fureby}}]{Nicholas2010}%
  \BibitemOpen
  \bibfield  {author} {\bibinfo {author} {\bibfnamefont {J.~G.}\ \bibnamefont
  {Nicholas}}, \bibinfo {author} {\bibfnamefont {D.~P.}\ \bibnamefont
  {Rizzetta}},\ and\ \bibinfo {author} {\bibfnamefont {C.}~\bibnamefont
  {Fureby}},\ }\bibfield  {title} {\bibinfo {title} {Large-eddy simulation:
  Current capabilities, recommended practices, and future research},\
  }\href@noop {} {\bibfield  {journal} {\bibinfo  {journal} {AIAA Journal}\
  }\textbf {\bibinfo {volume} {48}} (\bibinfo {year} {2010})}\BibitemShut
  {NoStop}%
\bibitem [{\citenamefont {Michelassi}\ \emph {et~al.}(2002)\citenamefont
  {Michelassi}, \citenamefont {Wissink},\ and\ \citenamefont
  {Rodi}}]{Michelassi2002}%
  \BibitemOpen
  \bibfield  {author} {\bibinfo {author} {\bibfnamefont {V.}~\bibnamefont
  {Michelassi}}, \bibinfo {author} {\bibfnamefont {J.}~\bibnamefont
  {Wissink}},\ and\ \bibinfo {author} {\bibfnamefont {W.}~\bibnamefont
  {Rodi}},\ }\bibfield  {title} {\bibinfo {title} {Analysis of dns and les of
  flow in a low pressure turbine cascade with incoming wakes and comparison
  with experiments},\ }\href@noop {} {\bibfield  {journal} {\bibinfo  {journal}
  {Flow, Turbulence and Combustion}\ }\textbf {\bibinfo {volume} {69}},\
  \bibinfo {pages} {295} (\bibinfo {year} {2002})}\BibitemShut {NoStop}%
\bibitem [{\citenamefont {Sandberg}\ \emph {et~al.}(2015)\citenamefont
  {Sandberg}, \citenamefont {Michelassi}, \citenamefont {Pichler},
  \citenamefont {Chen},\ and\ \citenamefont {Johnstone}}]{Sandberg2015}%
  \BibitemOpen
  \bibfield  {author} {\bibinfo {author} {\bibfnamefont {R.~D.}\ \bibnamefont
  {Sandberg}}, \bibinfo {author} {\bibfnamefont {V.}~\bibnamefont
  {Michelassi}}, \bibinfo {author} {\bibfnamefont {R.}~\bibnamefont {Pichler}},
  \bibinfo {author} {\bibfnamefont {L.}~\bibnamefont {Chen}},\ and\ \bibinfo
  {author} {\bibfnamefont {R.}~\bibnamefont {Johnstone}},\ }\bibfield  {title}
  {\bibinfo {title} {Compressible direct numerical simulation of low-pressure
  turbines part {I}: Methodology},\ }\href@noop {} {\bibfield  {journal}
  {\bibinfo  {journal} {Journal of Turbomachinery}\ }\textbf {\bibinfo {volume}
  {137}} (\bibinfo {year} {2015})},\ \bibinfo {note} {051011},\ \Eprint
  {https://arxiv.org/abs/https://asmedigitalcollection.asme.org/turbomachinery/article-pdf/137/5/051011/6302098/turbo\_137\_05\_051011.pdf}
  {https://asmedigitalcollection.asme.org/turbomachinery/article-pdf/137/5/051011/6302098/turbo\_137\_05\_051011.pdf}
  \BibitemShut {NoStop}%
\bibitem [{\citenamefont {Alhawwary}\ and\ \citenamefont
  {Wang}(2019)}]{Alhawwary2019}%
  \BibitemOpen
  \bibfield  {author} {\bibinfo {author} {\bibfnamefont {M.}~\bibnamefont
  {Alhawwary}}\ and\ \bibinfo {author} {\bibfnamefont {Z.}~\bibnamefont
  {Wang}},\ }\bibfield  {title} {\bibinfo {title} {On the mesh resolution of
  industrial les based on the dns of flow over the t106c turbine},\ }\href@noop
  {} {\bibfield  {journal} {\bibinfo  {journal} {Advances in Aerodynamics}\
  }\textbf {\bibinfo {volume} {1}} (\bibinfo {year} {2019})}\BibitemShut
  {NoStop}%
\end{thebibliography}%

\end{document}